\documentclass[twocolumn,showpacs,preprintnumbers,amsmath,amssymb,prd]{revtex4}
\usepackage{jabbrv}

\usepackage{subfigure,dcolumn,jabbrv}
\usepackage{graphicx}
\usepackage{dcolumn}
\usepackage{bm}
\usepackage{CJK}
\usepackage{url}
\usepackage{color}
\usepackage{booktabs}
\usepackage{bookmark}
\usepackage{natbib}
\usepackage{cleveref}
\usepackage{hyperref}
\usepackage{tikz}

\begin{document}

\title{Bayesian Quantification of Observability and Equation of State of Twin Stars}

\author{Xavier Grundler$^{1}$\footnote{xgrundler@leomail.tamuc.edu} and Bao-An Li$^{1}$\footnote{Corresponding Author: Bao-An.Li@etamu.edu}}
\affiliation{$^1$Department of Physics and Astronomy, East Texas A$\&$M University, Commerce, TX 75429-3011, USA}

\begin{abstract}
The possibility of discovering twin stars, two neutron stars (NSs) with the same mass but different radii, is usually studied in forward modelings by using a restricted number of NS matter equation of state (EOS) encapsulating a first-order phase transition from hadronic to quark matter (QM). Informing our likelihood function with the NS radius data from GW170817 and using a meta-model with 9-parameters capable of mimicking most NS EOSs available in the literature, we conduct a Bayesian quantification of the observability and underlying EOSs of twin stars. Of the accepted EOSs, between 12-18\% yield twin stars, depending on the restrictions we place on the second branch. The possibility of twin stars remains robust even under recent observational constraints. We show that many of these twin star scenarios are observable with currently available levels of accuracy in measuring NS radii. We also present the marginalized posterior probability density functions (PDFs) of every EOS parameter for each of four mass-radius correlation topologies. We find that the inferred EOS depends sensitively on not only whether twin stars are present, but also the category of twin stars, indicating that the observation of twin stars would provide a strong constraint on the underlying EOS. In particular, for two coexisting hybrid stars having QM cores at different densities, the PDF for QM speed of sound squared $c_{\rm qm}^2$ has two peaks, one below and another above the conformal limit $c_{\rm qm}^2=1/3$ predicted by perturbative QCD.  
\end{abstract}

\maketitle

\section{Introduction}\label{intro}

Neutron stars (NS) provide a natural laboratory to study high-density nuclear matter at several times saturation density ($\rho_0 = 0.16$ fm$^{-3}$). At such high densities, it is possible hadronic matter (HM) will undergo a phase transition to deconfined quark matter (QM). The nature and signatures of this transition are among the major questions for both the astrophysics and nuclear physics communities and remains unsolved. An especially interesting phenomenon that would immediately indicate that a phase transition occurs at densities appearing in NSs is the detection of twin stars, which are two NSs with the same mass but different radii. To our best knowledge, the possibility of twin stars was first suggested in Ref. \cite{gerlach_1968}, and further examined in Refs. \cite{Kampfer:1981yr, Glendenning:1998ag, Schertler_2000}. After observations of pulsars around 2 $M_\odot$, Ref. \cite{Blaschke:2013ana} for the first time demonstrated physically motivated EOS could simultaneously fulfill this mass observation and create twin star configurations, see also Refs. \cite{Benic:2014jia, Kaltenborn:2017hus, Alvarez-Castillo:2018pve} among previous works on twin stars and their EOSs.

The existence of twin stars requires a distinct feature in the mass-radius relation of another disconnected branch of stable degenerate stars called a "third family" \cite{gerlach_1968} after white dwarfs and regular neutron stars. In this paper, we refer to this as a second branch in the NS MR curves, as these hybrid stars have a quark core with a mantle composed of mostly neutrons, protons, electrons and muons.

Most often, twin star studies have used a Maxwell construction to create a strong, first-order phase transition between HM and QM matter \cite{Alford:2013aca, Christian_2018, Christian_2022, Christian_2024, Christian_2025, Li_2023, Pal_2025, ZhangLi_twin, Albino:2024ymc, Alvarez_Castillo_2016}, but they have also been found using the Glendenning construction with a mixed phase \cite{Monta_a_2019, Glendenning:1998ag, Schertler_2000} or with effects of pasta structures in a mixed phase \cite{Alvarez-Castillo:2014dva, Ayriyan:2017nby}. While features of the MR curve of spherical and static stars are commonly used to discuss the possible detection of twin stars, others have examined the tidal deformability as well \cite{Miao_2020, acastillo_2025, Christian_2020, Li_NICERtwin_2025} or the effects of rotation on twin stars \cite{Zdunik_2006, Bejger:2016emu}.

The papers referenced and many others have studied the twin star phenomenon, see e.g. Refs. \cite{Laskos_Patkos_2025, huang2025, Gorda_2023, Li_2024, Carlomagno:2023nrc, Chanlaridis:2024rov, Jimenez:2024hib, Veselsky:2024bnf}, but they usually only select a few representative EOSs or else fix the HM EOS and vary the QM EOS, or vice versa. Some recent studies have conducted a search for twin stars in a larger parameter space varying both the HM and QM EOSs, see, e.g., Ref. \cite{chunhuang_2025twin}, using a speed of sound model. Here, we examine the existence of twin stars within a 9-dimensional meta-model, capable of mimicking most HM and QM EOSs existing in the literature. Conducting a physics-informed Bayesian analysis (in contrast to agnostic Bayesian analyses, e.g. Ref. \cite{Brandes:2024wpq}) informed by indications of the binary NS merger event GW170817 and recent analyses of NICER observations, we determine the one-dimensional (marginalized) posterior probability density functions (PDFs) of each parameter, and categorize the EOSs according to the topologies of their resulting MR curves in order to determine which sub-space of the EOS parameters' priors yield twin stars that meet NS observational data. We also calculate the range $\Delta M$ over which twin stars can be found and how large the radius difference $\Delta R$ is between twin stars. By this, we show the EOS characteristics which allow the formation of twin stars. For discussion on the physical mechanism and timescale for the formation of twin stars, see, e.g., Refs. \cite{Chanlaridis:2024rov, Bejger:2016emu}.

In the following section, we outline our NS EOS model and categorization of NS MR curves. In Sections \ref{results} and \ref{nicer}, we present our results, and we have concluding remarks in Section \ref{conclusion}.

\section{Methodology}\label{method}
For completeness and ease of reference we briefly outline the EOS meta-model used in this work as well as the Bayesian method used. For a more thorough discussion, see some of our previous work \cite{zhang2018combined,Zhang:2019fog,Zhang:2021xdt,xie2019bayesian,xie2020bayesian,xie2021bayesian,Zhang:2023wqj,Zhang:2024npg,Xie:2024mxu,Li:2024imk,LiEPJA,universeReview,XieLi_phasetrans,hybridPrecision}. We also describe the adopted categorization of NS MR curves.

\subsection{NS EOS Meta-Model}
In order to find the MR curve, we need pressure as a function of energy density, the NS EOS, to solve the TOV equations \cite{tolman1939,oppenheimer1939massive}. We create hybrid EOS by coupling a HM EOS to a QM EOS with a first order phase transition under a Maxwell construction. The EOS of HM consisting of neutrons, protons, electrons and muons (\textit{npe$\mu$} matter) at $\beta$-equilibrium is constructed by parameterizing the binding energy per nucleon, $E(\rho,\delta)$, as a function of nucleon density, $\rho = \rho_n + \rho_p$, and isospin asymetry, $\delta = (\rho_n - \rho_p) / \rho$ according to the empirical isospin-parabolic law of neutron-rich matter verified by essentially all nuclear many-body theories \cite{bombaci1991asymmetric}
\begin{equation}\label{eos}
    E(\rho,\delta)=E_0(\rho)+E_{\rm{sym}}(\rho)\cdot \delta ^{2} +\mathcal{O}(\delta^4).
\end{equation}
Here, $E_0(\rho)$ is the EOS of symmetric nuclear matter (SNM), and $E_{\rm sym}(\rho)$ is the symmetry energy. These are parameterized as
\begin{eqnarray}\label{E0para}
  E_{0}(\rho)&=&E_0(\rho_0)+\frac{K_0}{2}(\frac{\rho-\rho_0}{3\rho_0})^2+\frac{J_0}{6}(\frac{\rho-\rho_0}{3\rho_0})^3,\\
  E_{\rm{sym}}(\rho)&=&E_{\rm{sym}}(\rho_0)+L(\frac{\rho-\rho_0}{3\rho_0})+\frac{K_{\rm{sym}}}{2}(\frac{\rho-\rho_0}{3\rho_0})^2\nonumber\\
  &+&\frac{J_{\rm{sym}}}{6}(\frac{\rho-\rho_0}{3\rho_0})^3\label{Esympara},
\end{eqnarray}
with $E_0(\rho_0) = 16 MeV$. The form of the parameterizations is inspired by a Taylor expansion, but they are \textit{not} expansions of any particular energy-density functionals (EDFs) known in advance in Bayesian analyses. While the coefficients approach asymptotically the correct derivatives for some EDFs near $\rho_0$, for the Bayesian analysis they serve only as parameterizations, and so do not suffer from any issue of convergence at supra-saturation densities often associated with Taylor expansions. The coefficients are defined as
\begin{eqnarray}
    K_0&=&9\rho_0^2[\partial^2 E_0(\rho)/\partial\rho^2]|_{\rho=\rho_0},\\\
    J_0&=&27\rho_0^3[\partial^3 E_0(\rho)/\partial\rho^3]|_{\rho=\rho_0},\\\
    L&=&3\rho_0[\partial E_{\rm{sym}}(\rho)/\partial\rho]|_{\rho=\rho_0},\\\ K_{\rm{sym}}&=&9\rho_0^2[\partial^2 E_{\rm{sym}}(\rho)/\partial\rho^2]|_{\rho=\rho_0},\\\
    J_{\rm{sym}}&=&27\rho_0^3[\partial^3 E_{\rm{sym}}(\rho)/\partial\rho^3]|_{\rho=\rho_0}.
\end{eqnarray}
These are the incompressibility and skewness of SNM and the slope, curvature, and skewness of the symmetry energy, respectively, and $E_{\rm sym} (\rho_0)$ is the magnitude of nuclear symmetry energy at saturation density.

The pressure can then be calculated from
\begin{equation}\label{pressure}
    P(\rho, \delta) = \rho^2 \frac{\rm{d}\varepsilon_{\rm{HM}}(\rho,\delta)/\rho}{\rm{d}\rho},
\end{equation}
where $\varepsilon_{\rm{HM}}(\rho, \delta) = \rho [E(\rho,\delta)+M_N]+ \varepsilon_l(\rho, \delta)$ is the energy density of HM matter, and $\varepsilon_l(\rho,\delta)$ is the energy density of leptons, which can be found with the non-interacting Fermi gas model \cite{oppenheimer1939massive}. After applying the charge neutrality and $\beta$ equilibrium conditions, the density profile 
$\delta(\rho)$ of isospin asymmetry can be obtained. The pressure of $npe\mu$ matter in Eq. (\ref{pressure}) then becomes barotropic.

For the QM, we adopt the constant sound speed (CSS) model \cite{Alford:2013aca}
\begin{equation}\label{css}
    \varepsilon(p)= \begin{cases}\varepsilon_{\mathrm{HM}}(p) & p<p_{t} \\ \varepsilon_{\mathrm{HM}}\left(p_{t}\right)+\Delta \varepsilon+c_{\mathrm{qm}}^{-2}\left(p-p_{t}\right) & p>p_{t}\end{cases}
\end{equation}
where $\varepsilon_{\rm HM} (p)$ is the EOS of HM described above, and $p_t$ is the pressure at the phase transition. The parameters we use are the transition density, $\rho_t/\rho_0$, which determines the transition pressure, $p_t$, the energy density discontinuity at the hadron-quark interface, $\Delta \varepsilon / \varepsilon_t$, which describes the strength of the phase transition, and the speed of sound squared in QM, $c_{\rm qm}^2$, which describes the stiffness of QM. By initializing randomly all nine parameters in their prior ranges, we can mimic most or all other proposed NS EOS in the literature.

An equivalent constant speed of sound model was proposed by Zdunik and Haensel \cite{zdunik2013}, who found that a constant speed of sound in QM well-approximated numerical calculations using the Nambu--Jona-Lasinio (NJL) model. Various other models have an approximately constant speed of sound in QM at densities relevant to NS, such as a bag model \cite{Zdunik:2000xx} and nonlocal NJL models \cite{nlNJL_cssContrera, nlNJL_cssShahrbaf}. Thus, the CSS model provides a simple approximation to analyze the effects of a first-order phase transition to QM, useful as a reference for more complicated models.

Finally, we use the popular Negele-Vautherin (NV) EOS \cite{negele1973neutron} and Baym-Pethick-Sutherland (BPS) EOS \cite{baym1971ground} for the inner and outer crusts respectively. This is connected to the core EOS described above at the point when the outer core EOS becomes thermodynamically unstable \cite{lattimer:2006xb, kubis2007nuclear, Xu:2009vi}.

\begin{figure*}
\resizebox{\linewidth}{!}{
\begin{tikzpicture}
    \node at (2.5,4.5) {Absent};

    \draw[ultra thick, ->] (0,0) -- (5,0) node[below] {$R$};
    \draw[ultra thick, ->] (0,0) -- (0,4.5) node[above] {$M$};

    \draw[ultra thick] (4,0.25) to[out=160,in=180+100] (2,3.25);
    \draw[thick,dashed,red] (2,3.25) to[out=180+100,in=180+190] (0.5,2.25);

    \draw[ultra thick] (4.5,0.5) to[out=140,in=180+90] (3.25,3) to[out=90,in=180+180] (2.5,4);
    \draw[thick,dashed] (2.5,4) to[out=180,in=180+225] (1.5,3.5);
\end{tikzpicture}
\begin{tikzpicture}
    \node at (2.5,4.5) {Both};

    \draw[ultra thick, ->] (0,0) -- (5,0) node[below] {$R$};
    \draw[ultra thick, ->] (0,0) -- (0,4.5) node[above] {$M$};

    \draw[ultra thick] (4.5,0.5) to[out=160,in=180+120] (3.25,2);
    \draw[ultra thick,red] (3.25,2) to[out=120,in=180+180] (2.5,2.75);
    \draw[thick,red,dashed] (2.5,2.75) to[out=180,in=180+180] (2,2.5);
    \draw[ultra thick,red] (2,2.5) to[out=180,in=180+180] (1.25,3.5);
    \draw[thick,red,dashed] (1.25,3.5) to[out=180,in=180+225] (0.5,3);
\end{tikzpicture}
\begin{tikzpicture}
    \node at (2.5,4.5) {Connected};

    \draw[ultra thick, ->] (0,0) -- (5,0) node[below] {$R$};
    \draw[ultra thick, ->] (0,0) -- (0,4.5) node[above] {$M$};

    \draw[ultra thick] (4.5,0.5) to[out=160,in=180+120] (3,2.5);
    \draw[ultra thick,red] (3,2.5) to[out=120,in=180+180] (2,4);
    \draw[thick,red,dashed] (2,4) to[out=180,in=180+245] (1.25,3.25);
\end{tikzpicture}
\begin{tikzpicture}
    \node at (2.5,4.5) {Disconnected};

    \draw[ultra thick, ->] (0,0) -- (5,0) node[below] {$R$};
    \draw[ultra thick, ->] (0,0) -- (0,4.5) node[above] {$M$};

    \draw[ultra thick] (4.5,0.5) to[out=160,in=180+120] (3,2.5);
    \draw[thick,red,dashed] (3,2.5) to[out=180+120,in=180+180] (2.5,1.5);
    \draw[ultra thick,red] (2.5,1.5) to[out=180,in=180+180] (1.25,3.5);
    \draw[thick,red,dashed] (1.25,3.5) to[out=180,in=180+245] (0.5,2.75);

    \draw[very thick,|-|] (4,2.5) -- (4,1.5) node[midway,xshift=0.4cm] {$\Delta M$};
    \draw[very thick,|-|] (3,2.5) -- (1.85,2.5) node[midway,yshift=0.3cm] {$\Delta R$};
\end{tikzpicture}
}
\caption{Modified slightly from Fig. 2 in Ref. \cite{Alford:2013aca}, these diagrams show an exaggerated MR curve for each category. The change from black to red marks the appearance of QM in the core. Dashed lines represent unstable configurations. Shown in the right panel are $\Delta M$, which is the difference between the maximum mass on the first branch and the minimum mass of the second branch, and $\Delta R$, which is the maximum radius difference between twins. 
The $\Delta M$ and $\Delta R$ together are used to measure the observability of twin stars \cite{ZhangLi_twin}.}
\label{fig:cat}
\end{figure*}

\subsection{Bayesian Analysis}\label{bayes}
Bayes' theorem reads
\begin{equation}
    P(\mathcal{M}|D) = \frac{P(D|\mathcal{M}) P(\mathcal{M})}{\int P(D|\mathcal{M}) P(\mathcal{M}) \,d\mathcal{M}},
\end{equation}
where $P(\mathcal{M}|D)$ is the posterior probability, $P(D|\mathcal{M})$ is the likelihood, $P(\mathcal{M})$ is the prior, and the denominator is a normalizing constant. We take the most agnostic approach to our prior to avoid biased assumptions by keeping the prior uniform within current nuclear and astrophysics bounds. The limits are shown in Table \ref{tab-prior} \cite{LiEPJA}, note that units are in $c = 1$ throughout this paper.

\begin{table}[htbp]
    \centering
    \begin{tabular}{lccccccc}
        \hline\hline
        Parameters&~~~~Lower limit  &~~~~Upper limit \\
        \hline\\
        \vspace{0.2cm}
        $K_0$ (MeV) & 220 & 260 \\
        $J_0$ (MeV) & -300 & 300 \\
        $K_{\mathrm{sym}}$ (MeV) & -400 & 100 \\
        $J_{\mathrm{sym}}$ (MeV) & -200 & 800 \\
        $L$ (MeV) & 30 & 90 \\
        $E_{\mathrm{sym}}(\rho_0)$ (MeV) & 28.5 & 34.9 \\
        $\rho_t/\rho_0$ & 1.0 & 6.0 \\
        $\Delta\varepsilon/\varepsilon_t$ & 0.2 & 1.0 \\
        $c_{\rm{qm}}^2$ & 0.0 & 1.0 \\
        \hline
    \end{tabular}
    \caption{Prior ranges of the nine EOS parameters.}\label{tab-prior}
\end{table}

The likelihood is composed of three functions, as follows
\begin{equation}\label{ll}
    P(D|\mathcal{M}) = P_{\rm filter} \times P_{\rm mass,max} \times P_R.
\end{equation}
The first term, $P_{\rm filter}$, is a step-function that guarantees (i) the crust-core transition is positive, (ii) thermodynamic stability, $dP/d\varepsilon \ge 0$, and (iii) causality is not violated. The second term, $P_{\rm mass,max}$, is also a step-function that guarantees the EOS can generate a NS at least as massive as 1.97 $M_\odot$ \cite{Antoniadis:2013pzd}. This is based on the mass observations by Antoniadis et al. of PSR J0348+0432 with $M = 2.01 \pm 0.04\;M_\odot$ \cite{Antoniadis:2013pzd}. This condition was used in deriving radius constraints from GW170817 \cite{abbott2018gw170817}. In recent years, more massive NS have been observed, for example NICER measurements of PSR J0740+6620 with a mass of $M = 2.08 \pm 0.07\;M_\odot$ \cite{Miller:2021qha, Dittmann:2024mbo}. Other measurements have greater uncertainty, such as the fastest spinning NS PSR J0952-0607 with $M = 2.35 \pm 0.17\;M_\odot$ \cite{Romani:2022jhd}. In our calculations, we mainly use $M_{\rm TOV} \ge 1.97\;M_\odot$ for the likelihood, where $M_{\rm TOV}$ is the maximum mass supported by a given EOS, choosing this as a conservative estimate. To test the effects of a larger mass constraint, we do one analysis with $M_{\rm TOV} \ge 2.08\;M_\odot$.

The last term, $P_R$, measures how well the model satisfies NS mass-radius data. Since our goal was to look at the difference in the PDFs across the categories, we did not want to use a large number of data points as that would already tightly constrain the possible EOS. Also, some of these data are in tension with each other and their accuracy still debated, with reanalysis not uncommon. Considering that, we chose for most of our calculations to use only the LIGO/VIRGO data from GW170817, which found that the radius for a canonical NS with mass around 1.4 M$_{\odot}$ was $R_{1.4} = 11.9 \pm 0.875$ km at one sigma \cite{abbott2018gw170817}. To evaluate the likelihood of a given EOS fulfilling the mass-radius data for $N$ data points, we use
\begin{equation}
    P_R = \prod_{j=1}^{N}\frac{1}{\sqrt{2\pi}\sigma_{\mathrm{obs},j}}\exp\left[-\frac{(R_{\mathrm{th},j}-R_{\mathrm{obs},j})^{2}}{2\sigma_{\mathrm{obs},j}^{2}}\right]
\end{equation}
where $R_{\rm th}$ is the theoretical radius predicted by the EOS and $R_{\rm obs}$ and $\sigma_{\rm obs}$ are the observational data. This can be either twin, so we use whichever prediction is closer to the observed data for an EOS. We note that the quoted precision of $R_{1.4}$ from analyzing gravitational waves emitted by GW170817 is higher than that from both earlier analyses of low-mass x-ray binaries from XMM-Newton-Chandra and the recent results from NICER, see, e.g., Refs.\cite{Li:2019xxz,Li:2025uaw} for reviews.

This radius value for a canonical NS from GW170817 is consistent with most NICER analyses, so it provides a good baseline. There is a growing number of simultaneous mass-radius measurements provided by the efforts of NICER as summarized in Tab. \ref{tab:nicerdata}. It is beyond the scope of this work to analyze which NICER analyses to include and how each measurement can help constrain the nuclear matter EOS. However, for comparison to the LIGO/VIRGO data, we will do one calculation using PSR J0740+6620 with $R_{2.1} = 12.92^{+2.09}_{-1.13}$ \cite{Dittmann:2024mbo} and one using PSR J0437+4715 $R_{1.4} = 11.36^{+0.95}_{-0.63}$ \cite{Choudhury:2024xbk}. For better comparison to the LIGO/VIRGO data where the mass is given exactly, we take the optimistic scenario that we know the mass of these pulsars precisely. For the asymmetric uncertainty interval, we follow Ref. \cite{Tsang:2020prc} by using two Gaussian distributions connected at the most probable value. These two measurements show the constraining effects of MR data of massive NS and the indication in some recent observations of smaller radius values.

\begin{table*}[htbp]
    \centering
    \setlength{\tabcolsep}{0.5em} 
    {\renewcommand{\arraystretch}{1.2}
    \begin{tabular}{|c|c|c|c|c|}
        \hline
        Name & Mass ($M_\odot$) & Radius (km) & Model & Reference \\
        \hline\hline
        PSR J0740+6620 & $2.072^{+0.067}_{-0.066}$ & $12.39^{+1.30}_{-0.98}$ & ST-U & T. E. Riley et al. 2021 \cite{Riley:2021pdl}\\
        PSR J0740+6620 & $2.08 \pm 0.07$ & $13.7^{+2.6}_{-1.5}$ & Two Circular Spots & M. C. Miller et al. 2021 \cite{Miller:2021qha}\\
        PSR J0740+6620 & $2.08 \pm 0.07$ & $12.92^{+2.09}_{-1.13}$ & Two Circular Spots & A. J. Dittmann et al. 2024 \cite{Dittmann:2024mbo}\\
        PSR J0740+6620 & $2.073^{+0.069}_{-0.069}$ & $12.49^{+1.28}_{-0.88}$ & ST-U & T. Salmi et al. 2024 \cite{Salmi:2024aum}\\
        \hline
        PSR J0030+0451 & $1.34^{+0.15}_{-0.16}$ & $12.71^{+1.14}_{-1.19}$ & ST+PST & T. E. Riley et al. 2019 \cite{Riley:2019yda}\\
        PSR J0030+0451 & $1.44^{+0.15}_{-0.14}$ & $13.02^{+1.24}_{-1.06}$ & Three Oval Spots & M. C. Miller et al. 2019 \cite{Miller:2019cac}\\
        PSR J0030+0451 & $1.40^{+0.13}_{-0.12}$ & $11.71^{+0.88}_{-0.83}$ & ST+PDT & S. Vinciguerra et al. 2024 \cite{Vinciguerra:2023qxq}\\
        PSR J0030+0451 & $1.70^{+0.18}_{-0.19}$ & $14.44^{+0.88}_{-1.05}$ & PDT-U & S. Vinciguerra et al. 2024 \cite{Vinciguerra:2023qxq}\\
        \hline
        PSR J0437+4715 & $1.418 \pm 0.037$ & $11.36^{+0.95}_{-0.63}$ & CST+PDT & D. Choudhury et al. 2024 \cite{Choudhury:2024xbk}\\
        \hline
        PSR J0614+3329 & $1.44^{+0.06}_{-0.07}$ & $10.29^{+1.01}_{-0.86}$ & ST+PDT & L. Mauviard et al. 2025 \cite{Mauviard:2025dmd}\\
        \hline
        PSR J1231+1411 & $1.04^{+0.05}_{-0.03}$ & $12.6 \pm 0.3$ & PDT-U & T. Salmi et al. 2024 \cite{Salmi:2024bss}\\
        \hline
    \end{tabular}}
    \caption{The headline result for each NICER analysis, median values and equal-tailed 68\% confidence intervals.}
    \label{tab:nicerdata}
\end{table*}

Other analyses incorporate current theoretical calculations of the low- and high-density behavior of the nuclear matter equation of state from chiral effective field theory ($\chi$-EFT) and perturbative Quantum Chromodynamics (pQCD), respectively, see, e.g., Refs. \cite{Lattimer:2023rpe, Carvalho:2025qie, Mondal:2022cva}. We incorporate $\chi$-EFT constraints in the values of our parameters' priors, rather than the likelihood as argued for in Ref. \cite{Brandes:2024wpq}. If we used a step-function to check an EOS was consistent with $\chi$-EFT in the likelihood, we do not expect this would be significantly different from excluding these combinations with the prior. Typically for pQCD constraints, the asymptotic behavior is calculated perturbatively at densities $\sim40\rho_0$ before non-perturbative effects appear, and then a given EOS must allow for an interpolation between its highest applicable density and this ultra-high-density behavior that respects both causality and thermodynamic stability, see, e.g. Refs. \cite{Albino:2024ymc, Komoltsev:2022prl, Kurkela:2024prl} for recent discussion and use of these constraints. Because these constraints are so far above NS densities, we chose not to include them in this study. For more detailed discussions on the physics reasons for making such choice, see, e.g., Refs. \cite{Somasundaram:2022ztm,Zhou:2023zrm}.

The Metropolis-Hastings algorithm \cite{metropolis1953,hastings1970}
is used in the Markov chain Monte Carlo (MCMC) sampling of the PDFs of EOS parameters. While some previous studies have focused solely on the twin star parameter space or parts of it, one of our aims is to find the relative frequency of twin star solutions that satisfy the above astrophysical constraints compared to non-twin star EOS. We expect, however, for the parameters to have different maximum \textit{a posteriori} (MaP) values for different categories. Because of this, we cannot use some of the more refined samplers such as those provided by the Python package \textsc{emcee} \cite{emcee} that will converge faster by refining the priors toward the more probable values. We need to explore the entire parameter space in order to find the solutions for every category. Therefore, we use a uniform random number generator as our sampler. This will still produce valid results with the Metropolis-Hastings algorithm although the walkers will be slow in exploring this high-dimensionality, multi-solution parameter space. To ensure proper analysis, we used 12 walkers, threw away the first 30,000 steps as burn-in, and accepted/rejected the following 300,000 steps for the posteriors for each walker.  We ran a total of six separate analyses to determine the effects of various constraints as summarized in Tab. \ref{tab:runs}.

\begin{table}[htbp]
    \centering
    \begin{tabular}{|c|c|c|c|}
        \hline
        Run & MR-data & $M_{\rm TOV}$ & Length \\
        \hline
        1 & $R_{1.4} = 11.9\pm0.875$ km & $\ge 1.97\;M_\odot$ & short\\
        2 & $R_{1.4} = 11.9\pm0.875$ km & $\ge 1.97\;M_\odot$ & mid-length\\
        3 & $R_{1.4} = 11.9\pm0.875$ km & $\ge 1.97\;M_\odot$ & long\\
        4 & $R_{2.1} = 12.92_{-1.13}^{+2.09}$ km & $\ge 1.97\;M_\odot$ & mid-length\\
        5 & $R_{1.4} = 11.9\pm0.875$ km & $\ge 2.08\;M_\odot$ & mid-length\\
        6 & $R_{1.4} = 11.36_{-0.63}^{+0.95}$ km & $\ge 1.97\;M_\odot$ & mid-length\\
        \hline
    \end{tabular}
    \caption{Summary of the different constraints for each of six Bayesian analyses.}
    \label{tab:runs}
\end{table}

\subsection{Categorizing NS MR curves and their EOSs}
We adopt the categorization scheme proposed by Alford, Han, and Prakash \cite{Alford:2013aca}. They classified NS MR curves (thus the underlying EOSs) into four-categories as we summarize below. 
\begin{enumerate}
    \item[A] (Absent) The phase transition does not occur at neutron star densities, or the appearance of QM destabilizes the star with no second branch so that no QM is present in even the most massive NS.
    \item[B] (Both) The phase transition does not cause immediate instability, but there is a later instable region followed by a second stable branch.
    \item[C] (Connected) The hybrid branch, QM core with HM mantle, is connected without any instability from the phase transition.
    \item[D] (Disconnected) The phase transition immediately destabilizes the star, but the NS restabilizes on a second branch.
    \item[E] (Everything) This category contains all of the above.
\end{enumerate}
Fig. \ref{fig:cat} demonstrates each category visually with a diagram adapted from Fig. 2 in Ref. \cite{Alford:2013aca}. As the color changes from black to red, QM appears in the core of the NS, which immediately destabilizes the star in the Absent and Disconnected categories. The dips and cusps in the curves are greatly exaggerated purposely here. The second all-black curve for the Absent category was added because we consider the possibility that the phase transition is not reached even in the most massive NS. Note that both categories Both and Disconnected yield twin star solutions. The fifth category (Everything) listed above is inclusive for all MR curves. It is used for normalizing the relative fractions of different categories. The underlying EOS parameters' PDFs for each category are then examined separately. Shown also in the right panel of Fig. \ref{fig:cat} are the $\Delta M$ and $\Delta R$ used to measure the observability of twin stars \cite{ZhangLi_twin}.

In order to determine the stability of the NS on a possible second branch, we use the Method 2-A of Ref. \cite{stableMethod}, recently reexamined in Ref. \cite{Alford_2017} and the stopping criteria of Ref. \cite{Essick:2024olf}. In Ref. \cite{stableMethod}, the authors explain that a NS at low central pressures on the first branch is known to be stable until it hits a maximum on the MR curve. The curve then undergoes a counterclockwise turn in which the fundamental mode becomes unstable. An additional mode becomes unstable for each subsequent counterclockwise turn in the MR curve, but a mode restabilizes for every clockwise turn. Thus, we must use the TOV equations to get the MR curve by specifying an ever increasing central pressure as a boundary condition to those equations. In order for there to be a stable second branch, and thus a twin star scenario, we must reach the first peak (where mass stops increasing with central pressure), a minimum (where the mass resumes increasing with central pressure), and at that minimum the radius must decrease from the previous central pressure.

If such a second branch is found, we then check whether the Seidov stability condition, which is \cite{seidov},
\begin{equation}\label{eq:seidov}
    \frac{\Delta \varepsilon}{\varepsilon_t} \le \frac{1}{2} + \frac{3}{2} \frac{p_t}{\varepsilon_t}
\end{equation}
is met. If the $\Delta \varepsilon / \varepsilon_t$ is small enough to satisfy the condition, then there will be a hybrid star on the first branch, so we place the EOS in the Both category. If the condition is not satisfied, then the appearance of QM immediately destabilizes the star, and we place the EOS in the Disconnected category. If no second branch is found, then we check if the densities reached before the NS becomes unstable were above $\rho_t$. If it was, then QM was present so the EOS is put in the Connected category. If no QM appeared, then the EOS is categorized as Absent.

As the MR curve is calculated, we needed to determine how long, i.e. what size range of central pressures, was sufficient to determine that there exists a physical second branch. We solve the TOV equations at intervals of about 2.75 Mev/fm$^3$ in pressure. If the mass increases for only one or a few steps, especially if by only minuscule amounts, then it is hard to determine if $dM/dp_c$ is truly positive. Further, part of the goal of our research was measuring the observability of twin stars. If only a small range of central pressures exist on the second branch, then it is unlikely that any natural process could be so exact as to actually produce a twin star. The exact range needed, however, is unknown to us, so we examined three different possibilities to determine the effects of requiring a longer second branch.
\begin{enumerate}
    \item Short: 50 TOV solutions with mass increasing with central pressure, corresponding to a range of about 140 MeV/fm$^3$
    \item Mid-length: 100 solutions, corresponding to a range of about 275 MeV/fm$^3$
    \item Long: 150 solutions, corresponding to a range of about 415 MeV/fm$^3$
\end{enumerate}
While we did not find this explicitly reported anywhere, Fig. 2 of Ref. \cite{Christian_2018} indicates a range of nearly 1000 MeV/fm$^3$ in the central pressure on the second branch. Visually, that graph appeared to have a longer second branch than many commonly shown MR-curves, so we allow for shorter third family branch.

Each scenario (short, mid-length, and long) was conducted as an independent Bayesian analysis. The existence of a second branch in each scenario was then evaluated on whether it had the required range of central pressures. The Connected and Absent EOS are not categorized based on the length of a non-existent second branch. However, the Connected and Absent categories can be affected by this because what was determined as Both under a shorter requirement could now be classified as a Connected EOS under a longer requirement, and what was determined as a Disconnected EOS could now be classified as an Absent EOS. The goal of the three Bayesian analyses was to demonstrate the effects of differing this criterion of what constitutes a second branch.

\begin{table}[htbp]
    \centering
    \begin{tabular}{|c|c|c|c|}
        \hline
         & Short & Mid-length & Long \\
        \hline
        Absent & 114,322 & 119,179 & 121,744\\
        Both & 85,108 & 76,265 & 67,180\\
        Connected & 868,612 & 894,892 & 922,880\\
        Disconnected & 125,063 & 99,227 & 75,380\\
        Everything & 1,193,105 & 1,189,563 & 1,187,184\\
        \% Twin & 17.62 & 14.75 & 12.01\\
        \hline
    \end{tabular}
    \caption{Accepted steps in each category from 12 MCMC walkers each taking 300,000 steps.}
    \label{tab:twinCount}
\end{table}

\section{Results and Discussions}\label{results}
\begin{figure*}
    \centering
    \addtolength{\tabcolsep}{-0.8em}
    \begin{tabular}{ccc}
        \includegraphics[width=0.35\linewidth]{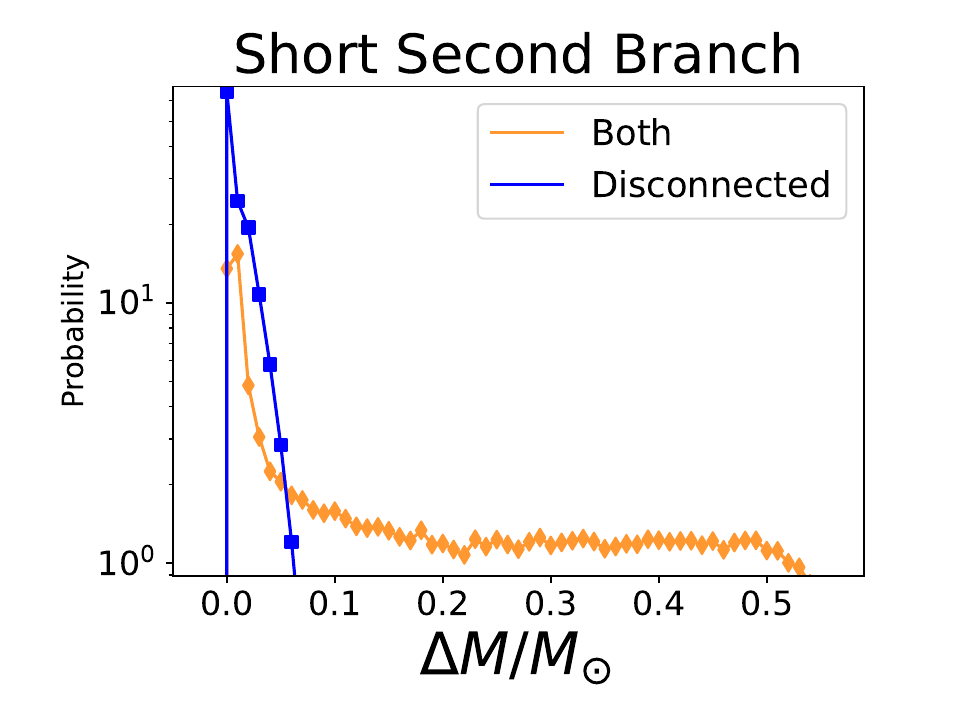} &
        \includegraphics[width=0.35\linewidth]{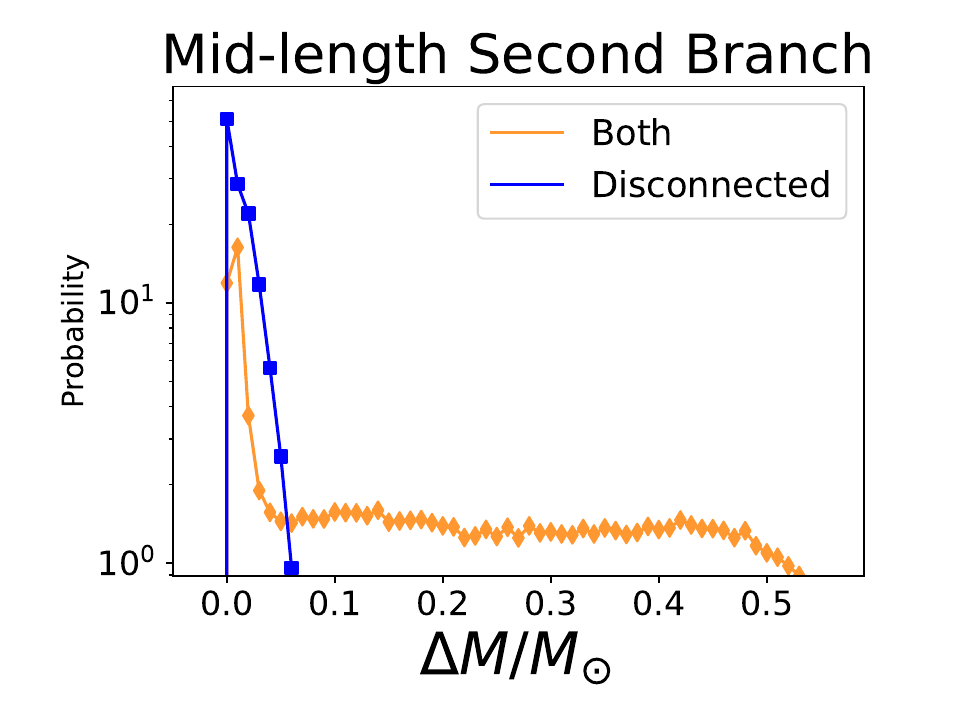} &
        \includegraphics[width=0.35\linewidth]{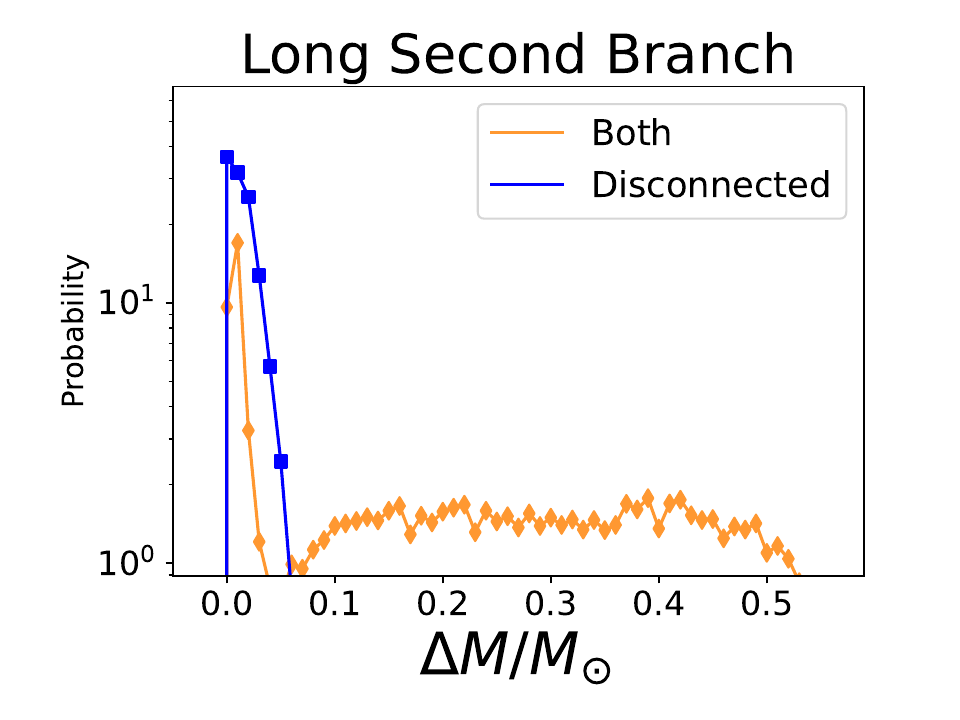} \\
        \includegraphics[width=0.35\linewidth]{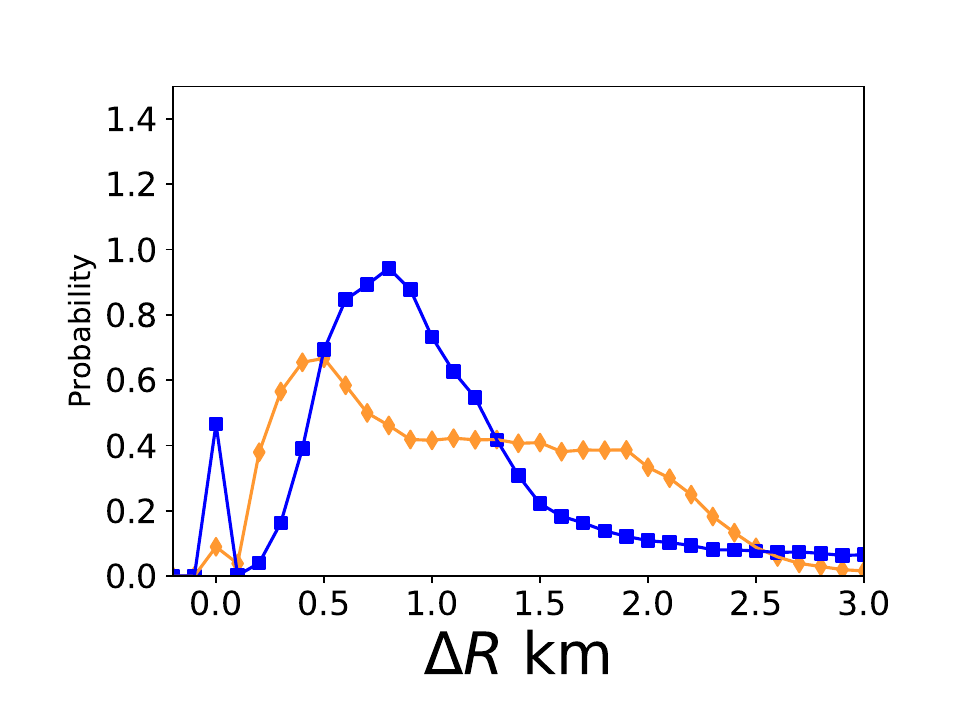} &
        \includegraphics[width=0.35\linewidth]{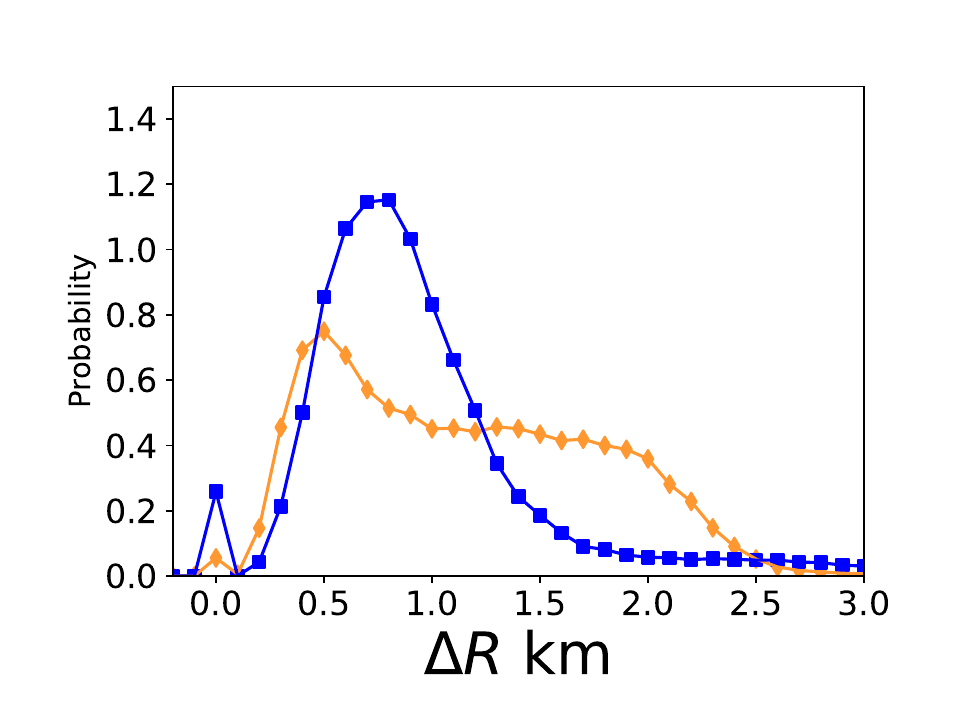} &
        \includegraphics[width=0.35\linewidth]{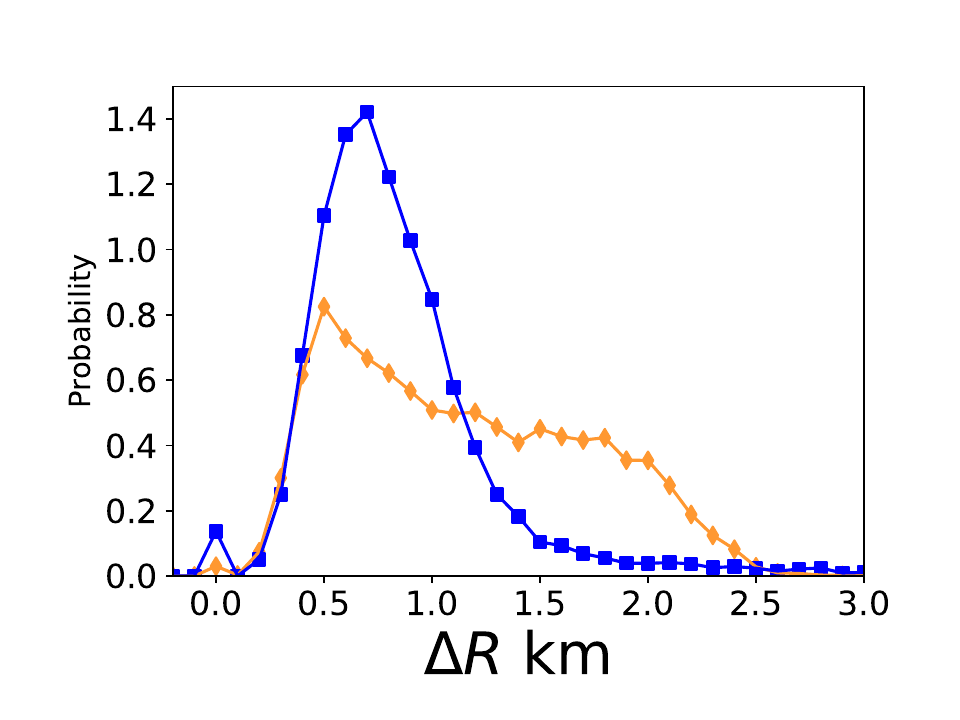}
    \end{tabular}
    \caption{Probability densities for how observable twin stars are. Top, the mass range over which twin stars are observable with the probability in log-scale. Bottom, the largest radius difference possible between twin stars.}
    \label{fig:probObs}
\end{figure*}
\begin{table*}[htbp]
    \addtolength{\tabcolsep}{0.5em}
    \begin{tabular}{lccccc}
        \hline
         & Absent & Both & Connected & Disconnected & Everything \\
        \hline
        $J_0$ MeV & -85.6$\pm$108 & -21.4$\pm$151 & -0.437$\pm$165 & 7.62$\pm$173 & -9.25$\pm$162\\
        $K_0$ MeV & 240.$\pm$11.5 & 240.$\pm$11.5 & 240.$\pm$11.5 & 240.$\pm$11.5 & 240.$\pm$11.5\\
        $J_{sym}$ MeV & 337$\pm$296 & 357$\pm$287 & 346$\pm$285 & 322$\pm$288 & 344$\pm$287\\
        $K_{sym}$ MeV & -80.3$\pm$83.7 & -109$\pm$120. & -123$\pm$113 & -135$\pm$137 & -119$\pm$115\\
        $L$ MeV & 63.4$\pm$15.2 & 65.4$\pm$15.8 & 61.7$\pm$15.4 & 71.3$\pm$13.4 & 63.2$\pm$15.5\\
        $E_{sym}(\rho_0)$ MeV & 31.8$\pm$1.85 & 31.7$\pm$1.85 & 31.7$\pm$1.85 & 31.7$\pm$1.85 & 31.7$\pm$1.85\\
        $\rho_t/\rho_0$ & 5.06$\pm$0.591 & 3.80$\pm$1.29 & 2.63$\pm$1.30 & 1.66$\pm$0.456 & 2.85$\pm$1.45\\
        $\Delta\varepsilon/\varepsilon_t$ & 0.760$\pm$0.170 & 0.560$\pm$0.194 & 0.472$\pm$0.195 & 0.799$\pm$0.110 & 0.540$\pm$0.223\\
        $c^2_{qm}$ & 0.591$\pm$0.244 & 0.361$\pm$0.363 & 0.667$\pm$0.229 & 0.796$\pm$0.158 & 0.651$\pm$0.254\\
        $\Delta M / M_\odot$ & N/A & 0.192$\pm$0.204 & N/A & 0.0141$\pm$0.0586 & N/A\\
        $\Delta R$ km & N/A & 1.23$\pm$2.75 & N/A & 1.14$\pm$0.965 & N/A\\
        $M_{\rm TOV} (M_\odot)$ & 2.16$\pm$0.102 & 2.14$\pm$0.127 & 2.24$\pm$0.220 & 2.18$\pm$0.149 & 2.22$\pm$0.202\\
        Max$_1 (M_\odot)$ & N/A & 1.83$\pm$0.565 & N/A & 0.685$\pm$0.332 & N/A\\
        Max$_2 (M_\odot)$ & N/A & 2.07$\pm$0.201 & N/A & 2.18$\pm$0.153 & N/A\\
    \hline
    \end{tabular}
    \caption{Mean and standard deviation for short.}
    \label{tab:shortMeanStd}
\end{table*}

\begin{table*}[htbp]
    \addtolength{\tabcolsep}{0.5em}
    \begin{tabular}{lccccc}
        \hline
         & Absent & Both & Connected & Disconnected & Everything \\
        \hline
        $J_0$ MeV & -85.0$\pm$109 & -22.8$\pm$151 & -0.105$\pm$165 & 11.6$\pm$172 & -9.09$\pm$162\\
        $K_0$ MeV & 241$\pm$11.6 & 241$\pm$11.5 & 240.$\pm$11.5 & 240.$\pm$11.5 & 240.$\pm$11.5\\
        $J_{sym}$ MeV & 337$\pm$297 & 357$\pm$286 & 346$\pm$285 & 331$\pm$287 & 345$\pm$287\\
        $K_{sym}$ MeV & -80.6$\pm$83.8 & -101$\pm$114 & -122$\pm$113 & -106$\pm$124 & -115$\pm$113\\
        $L$ MeV & 63.3$\pm$15.2 & 65.8$\pm$15.7 & 61.8$\pm$15.4 & 71.7$\pm$13.3 & 63.0$\pm$15.5\\
        $E_{sym}(\rho_0)$ MeV & 31.8$\pm$1.84 & 31.7$\pm$1.85 & 31.7$\pm$1.85 & 31.6$\pm$1.85 & 31.7$\pm$1.85\\
        $\rho_t/\rho_0$ & 5.06$\pm$0.592 & 3.84$\pm$1.26 & 2.65$\pm$1.30 & 1.77$\pm$0.435 & 2.89$\pm$1.45\\
        $\Delta\varepsilon/\varepsilon_t$ & 0.762$\pm$0.170 & 0.562$\pm$0.193 & 0.472$\pm$0.196 & 0.793$\pm$0.107 & 0.534$\pm$0.221\\
        $c^2_{qm}$ & 0.586$\pm$0.248 & 0.360$\pm$0.364 & 0.663$\pm$0.233 & 0.806$\pm$0.156 & 0.648$\pm$0.256\\
        $\Delta M / M_\odot$ & N/A & 0.199$\pm$0.195 & N/A & 0.0156$\pm$0.0561 & N/A\\
        $\Delta R$ km & N/A & 1.18$\pm$1.05 & N/A & 0.935$\pm$0.711 & N/A\\
        $M_{\rm TOV} (M_\odot)$ & 2.16$\pm$0.102 & 2.13$\pm$0.119 & 2.24$\pm$0.220 & 2.16$\pm$0.138 & 2.22$\pm$0.203\\
        Max$_1 (M_\odot)$ & N/A & 1.85$\pm$0.525 & N/A & 0.765$\pm$0.322 & N/A\\
        Max$_2 (M_\odot)$ & N/A & 2.07$\pm$0.188 & N/A & 2.16$\pm$0.143 & N/A\\
        \hline
    \end{tabular}
    \caption{Mean and standard deviation for mid-length.}
    \label{tab:midMeanStd}
\end{table*}

\begin{table*}[htbp]
    \addtolength{\tabcolsep}{0.5em}
    \begin{tabular}{lccccc}
        \hline
         & Absent & Both & Connected & Disconnected & Everything \\
        \hline
        $J_0$ MeV & -84.8$\pm$108 & -24.9$\pm$150. & -0.156$\pm$165 & 18.2$\pm$173 & -9.08$\pm$162\\
        $K_0$ MeV & 241$\pm$11.5 & 241$\pm$11.5 & 240.$\pm$11.5 & 241$\pm$11.6 & 240.$\pm$11.5\\
        $J_{sym}$ MeV & 335$\pm$297 & 362$\pm$284 & 346$\pm$286 & 342$\pm$286 & 345$\pm$287\\
        $K_{sym}$ MeV & -80.6$\pm$83.9 & -88.8$\pm$102 & -122$\pm$114 & -84.6$\pm$112 & -114$\pm$111\\
        $L$ MeV & 63.2$\pm$15.4 & 65.6$\pm$15.5 & 61.7$\pm$15.4 & 71.5$\pm$13.5 & 62.7$\pm$15.5\\
        $E_{sym}(\rho_0)$ MeV & 31.7$\pm$1.85 & 31.7$\pm$1.85 & 31.7$\pm$1.85 & 31.7$\pm$1.86 & 31.7$\pm$1.85\\
        $\rho_t/\rho_0$ & 5.05$\pm$0.593 & 3.93$\pm$1.22 & 2.65$\pm$1.30 & 1.88$\pm$0.434 & 2.92$\pm$1.45\\
        $\Delta\varepsilon/\varepsilon_t$ & 0.762$\pm$0.169 & 0.569$\pm$0.196 & 0.472$\pm$0.196 & 0.785$\pm$0.103 & 0.527$\pm$0.219\\
        $c^2_{qm}$ & 0.583$\pm$0.250 & 0.349$\pm$0.362 & 0.662$\pm$0.235 & 0.817$\pm$0.156 & 0.646$\pm$0.257\\
        $\Delta M / M_\odot$ & N/A & 0.215$\pm$0.189 & N/A & 0.0183$\pm$0.0650 & N/A\\
        $\Delta R$ km & N/A & 1.14$\pm$0.916 & N/A & 0.802$\pm$0.583 & N/A\\
        $M_{\rm TOV} (M_\odot)$ & 2.16$\pm$0.102 & 2.13$\pm$0.108 & 2.24$\pm$0.220 & 2.15$\pm$0.129 & 2.22$\pm$0.204\\
        Max$_1 (M_\odot)$ & N/A & 1.90$\pm$0.462 & N/A & 0.857$\pm$0.306 & N/A\\
        Max$_2 (M_\odot)$ & N/A & 2.06$\pm$0.183 & N/A & 2.14$\pm$0.134 & N/A\\
        \hline
    \end{tabular}
    \caption{Mean and standard deviation for long.}
    \label{tab:longMeanStd}
\end{table*}

\subsection{Existence and Observability of Twin Stars}
The most important question concerning twin stars is whether or not they can even exist. For the possible mechanisms and likelihoods to form twin stars from the dynamical point of view, we refer the readers to a recent study in Ref. \cite{Naseri:2024rby}.
Given current nuclear and astrophysics constraints, is there an EOS parameter space that yields twin star solutions from solving the TOV equations? In Tab. \ref{tab:twinCount}, we report the number of accepted EOSs in each category out of a total of 3.6 million. Also shown are the percent of accepted EOSs that yield twin star solutions. Unsurprisingly, as we require a longer second branch, the number of twin star solutions decreases, but even at the longest requirement considered, there is still a significant number of twin star EOSs that satisfy the astrophysical constraints used, and so the possibility of twin stars is worth exploring. These relative probabilities are admittedly found using just a single NS radius observation of GW170817. Including more radius and/or tidal deformation data would undoubtedly provide more stringent constraints on the nuclear EOS as shown, e.g., in Refs. \cite{Christian_2024,Li_2023,Verma_2025}. We chose, however, to use just this data point as a starting point because measures of NS tidal deformability still have large uncertainties and even such data are very limited. Also, NICER measurements of several NSs, while more accurate in some aspects, generally have two teams analyzing the same data that has been updated with slightly different results. Due to computational expense, we chose not to do multiple analyses to determine the effects of slightly different radius data, but rather focus on the relative probability of forming twin stars under the minimum constraint, see Ref. \cite{Li_NICERtwin_2025} for a use of various NICER analyses in constraining twin stars.

Next in importance to the existence of twin stars is our ability to actually observe them. In Fig. \ref{fig:probObs}, the probability distribution of the accepted twin star solutions are shown for two values that determine the observability of twin stars, depicted in the right panel of Fig. \ref{fig:cat}. The mean and standard deviation for these distributions are shown in Tabs. \ref{tab:shortMeanStd}--\ref{tab:longMeanStd}. In the top row of Fig. \ref{fig:probObs} is $\Delta M/M_\odot$, which is the difference in mass between the maximum mass of the first branch and the minimum mass of the second branch as defined in Ref. \cite{ZhangLi_twin}. This determines over what range of masses we could possibly observe twin stars. In all cases, the most probable value is zero, which indicates more of a sharp cusp, and less of a large dip, so there would not be very many twin stars to observe, although there are possibly many heavy NSs with relatively small radii on the second branch. In the Both category, however, there is a long tail at high $\Delta M/M_\odot$, albeit, with relatively low probability. We note also that the peak at zero for the Disconnected category decreases as we require a longer second branch. This indicates that allowing a short second branch to count as a twin star solution primarily increases the possibility of the scenario of almost no observable twin stars. Thus, if we wanted to focus on solutions that create more observable twin stars, we can require longer second branches.

\begin{figure*}
    \centering
    \addtolength{\tabcolsep}{-1em}
    \begin{tabular}{ccc}
        \includegraphics[width=0.35\linewidth]{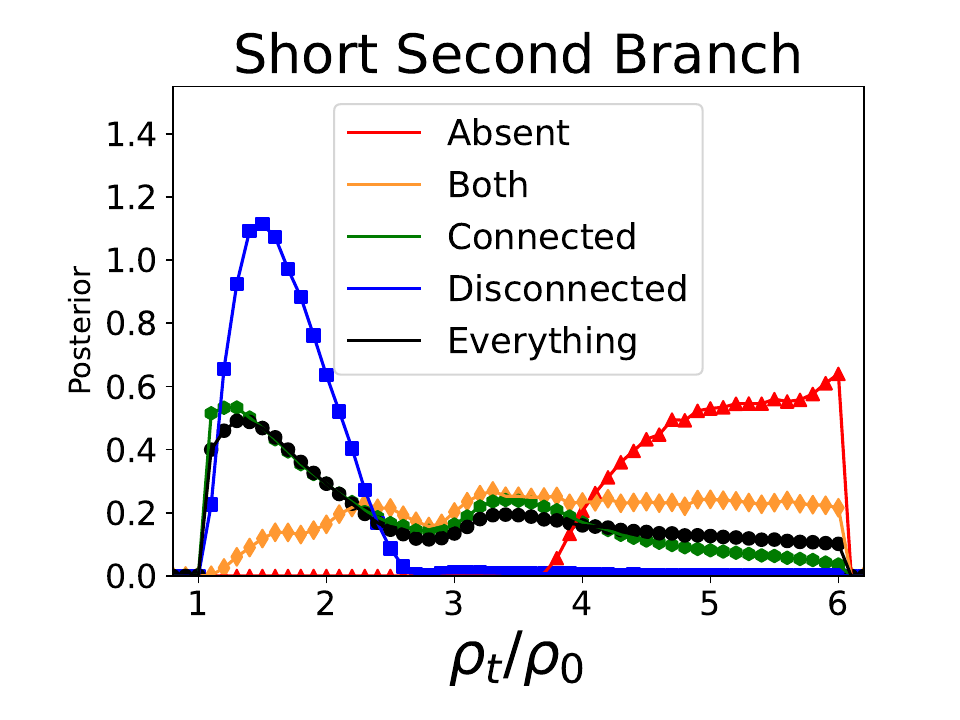} &
        \includegraphics[width=0.35\linewidth]{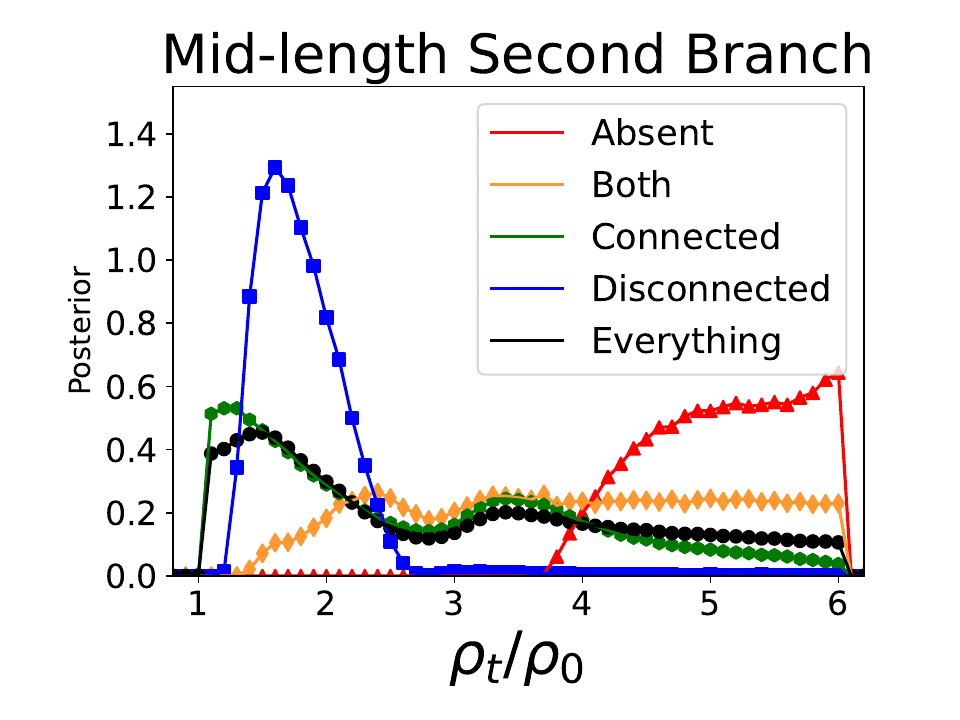} &
        \includegraphics[width=0.35\linewidth]{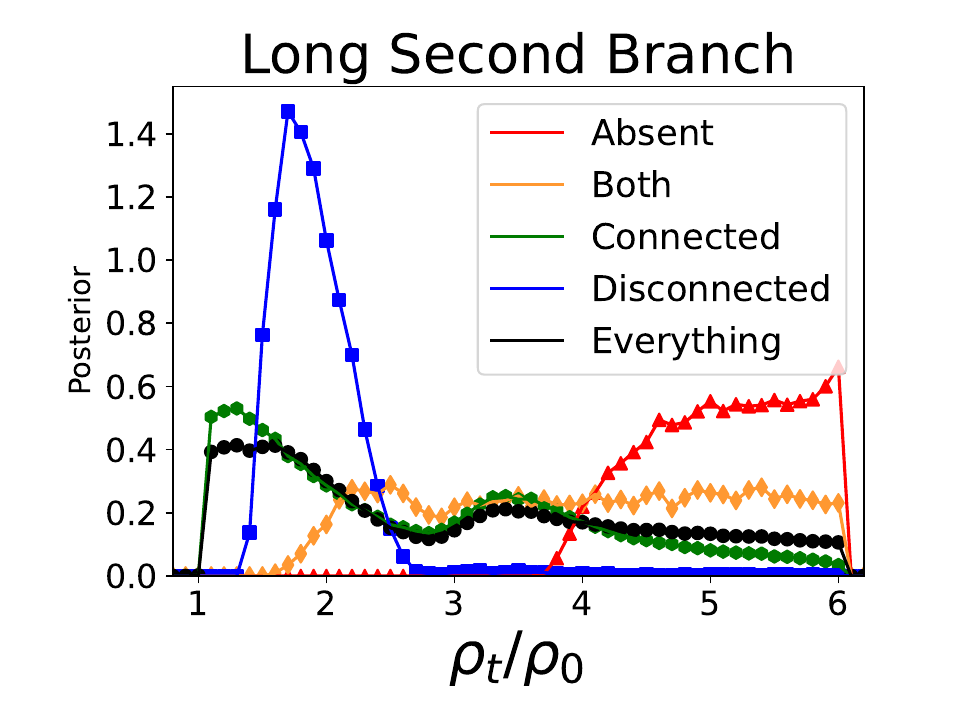} \\
        \includegraphics[width=0.35\linewidth]{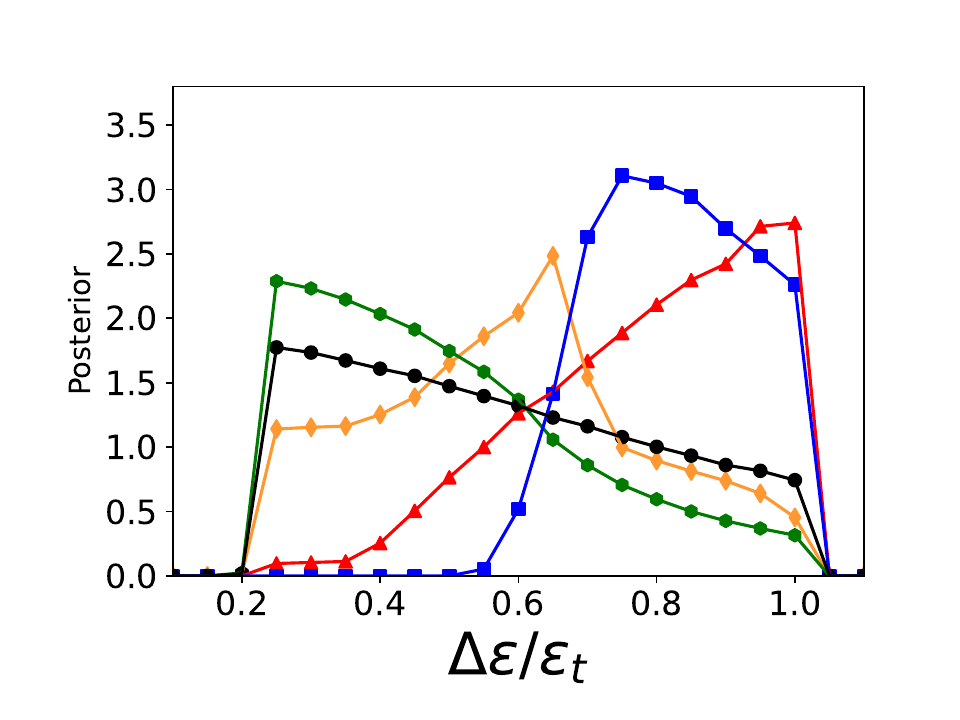} &
        \includegraphics[width=0.35\linewidth]{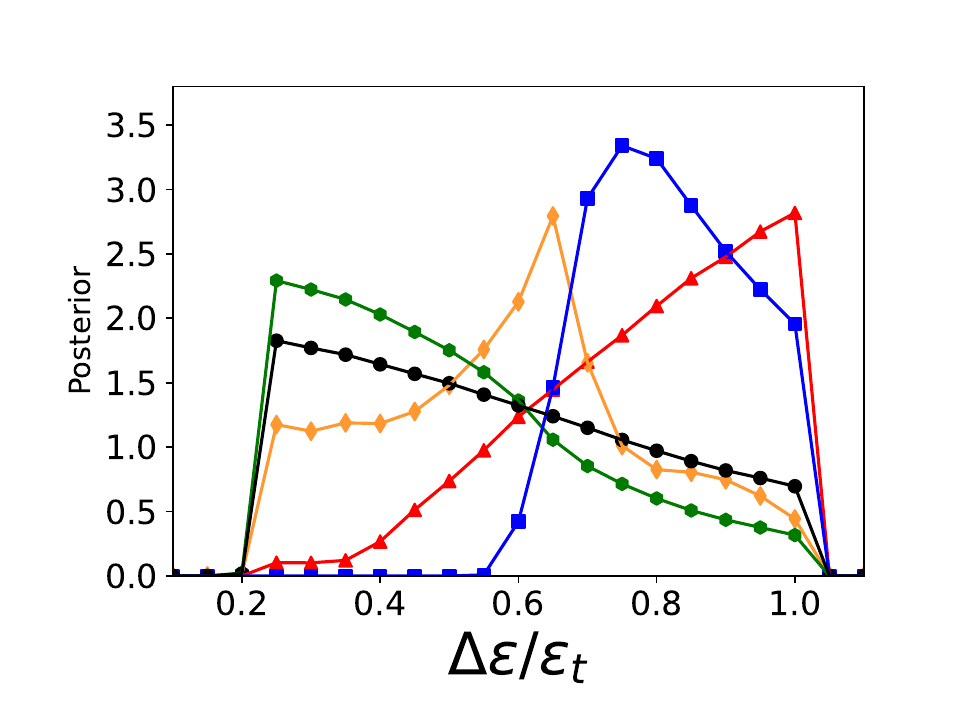} &
        \includegraphics[width=0.35\linewidth]{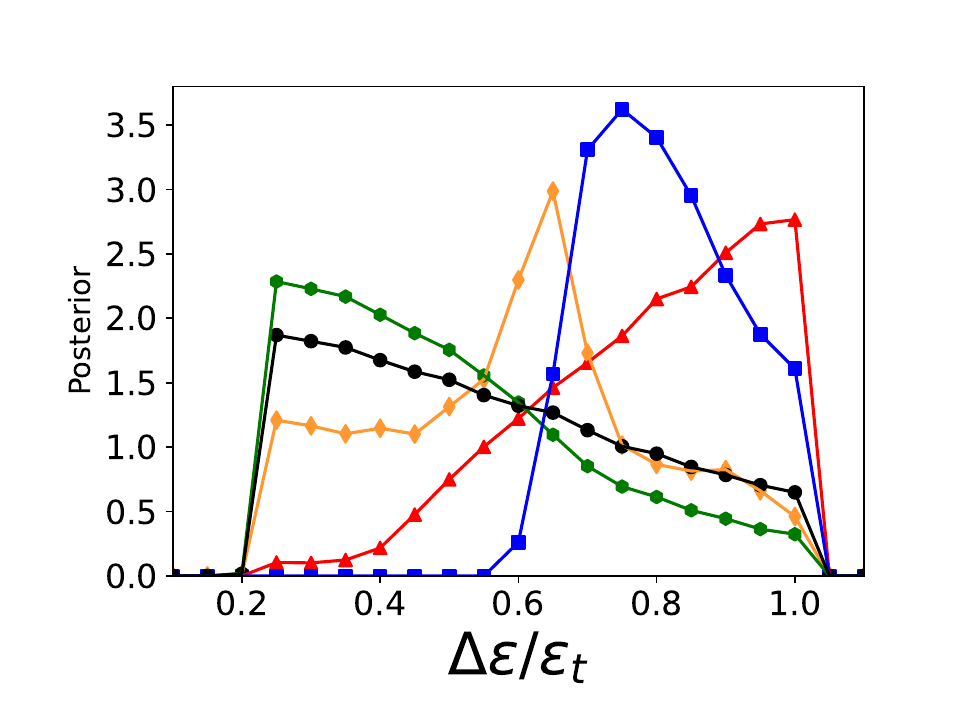} \\
        \includegraphics[width=0.35\linewidth]{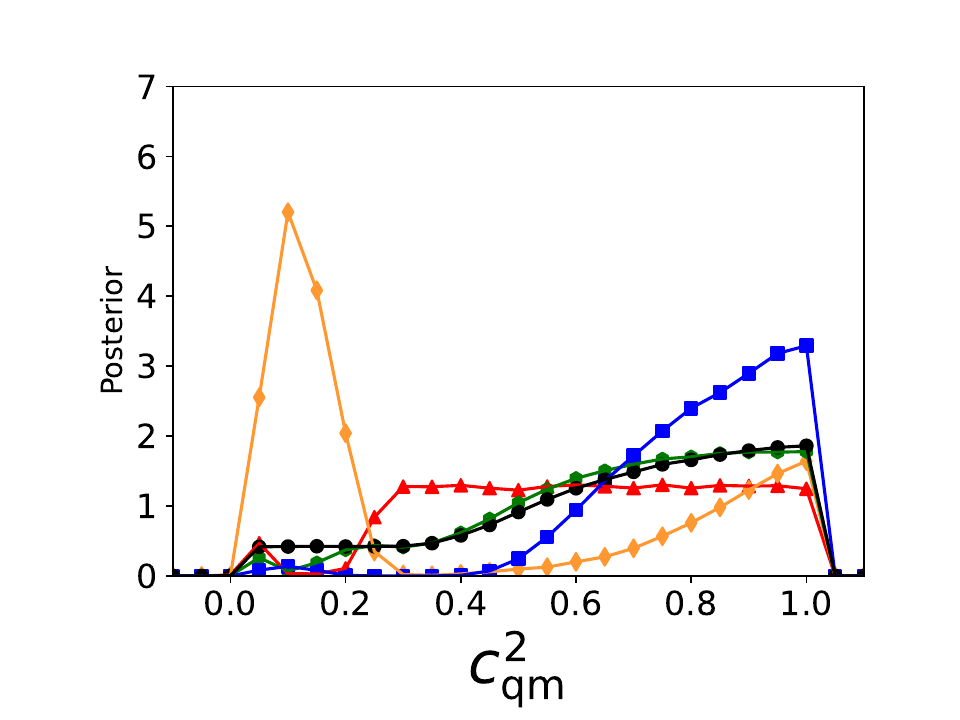} &
        \includegraphics[width=0.35\linewidth]{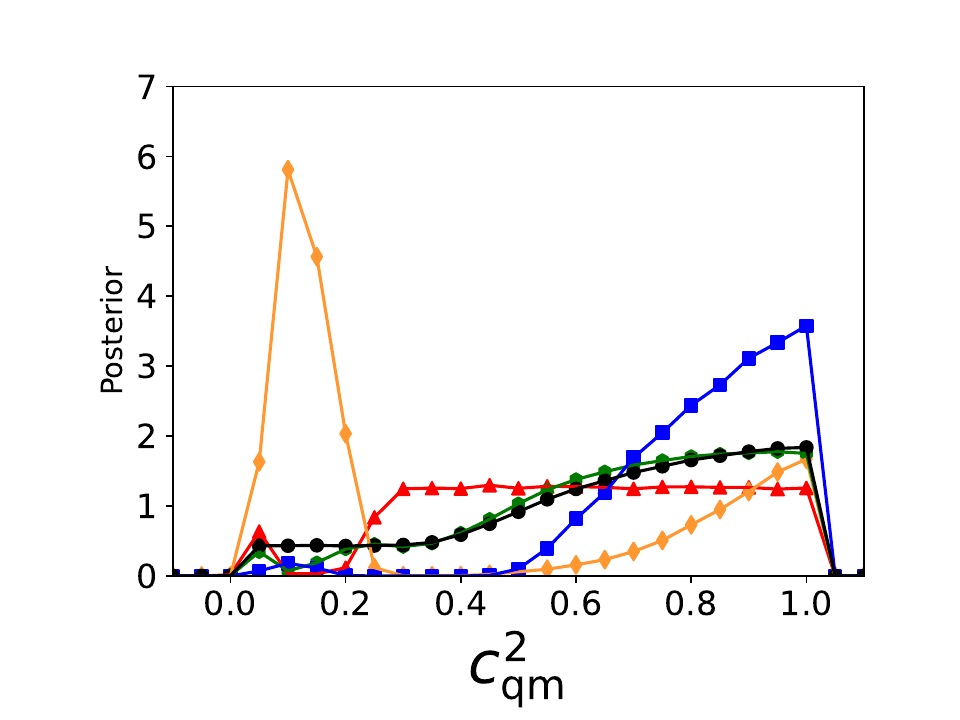} &
        \includegraphics[width=0.35\linewidth]{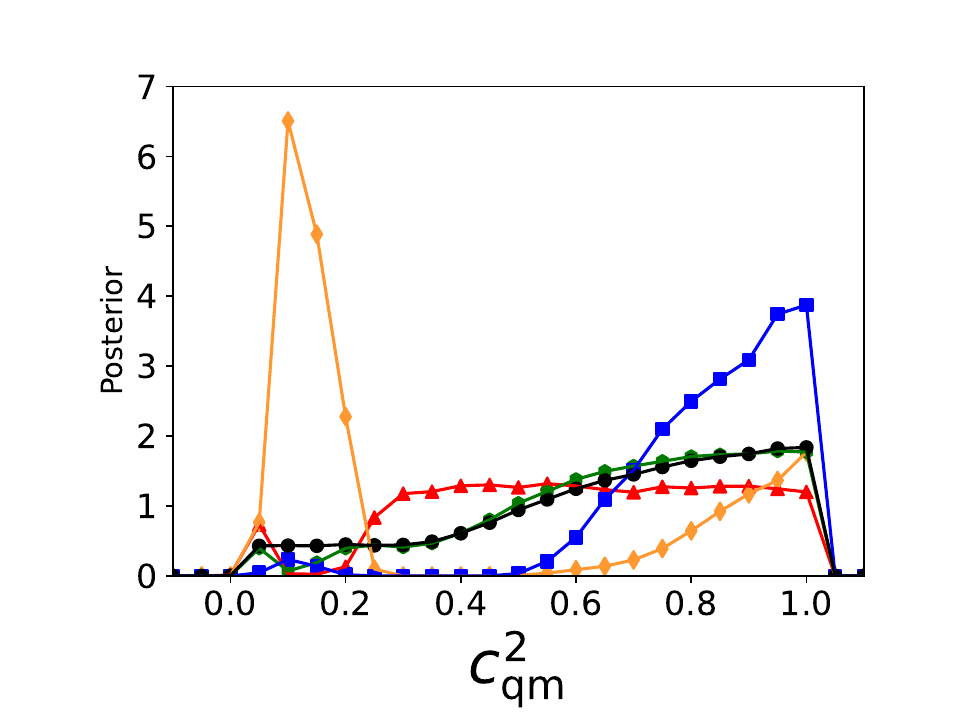}
    \end{tabular}
    \caption{PDFs of QM parameters.}
    \label{fig:pdfQM}
\end{figure*}

In the second row of Fig. \ref{fig:probObs}, we show the probability distribution of $\Delta R$, which is the largest separation in radii between twin stars, which are binned into 0.1 $M_\odot$ intervals before calculating the difference as done in Ref. \cite{Christian_2018}. In the Disconnected category, the MaP is about 0.8 km, near the edge of current observing capabilities, but many studies suggest that accuracies of 0.2 km may be possible within a decade \cite{Chatziioannou:2021tdi, Pacilio:2021jmq, Bandopadhyay:2024zrr, Finstad:2022oni, Walker:2024loo}. The Both category again has a long tail with values up to 2.0 km being probable, which is well within the range of our current observation power. Looking at the difference between requiring different lengths of the second branch, we notice that $\Delta R$ seems to saturate around 0.8 km as the length increases and the second peak at 0 km, which are twin stars that would be impossible to observe since they are nearly identical, gradually disappears. We see again that by requiring a longer second branch, we remove more of the hard to observe twin star solutions.

\subsection{NS EOS Constraints}

\begin{figure*}
    \centering
    \addtolength{\tabcolsep}{-1em}
    \begin{tabular}{ccc}
        \includegraphics[width=0.35\linewidth]{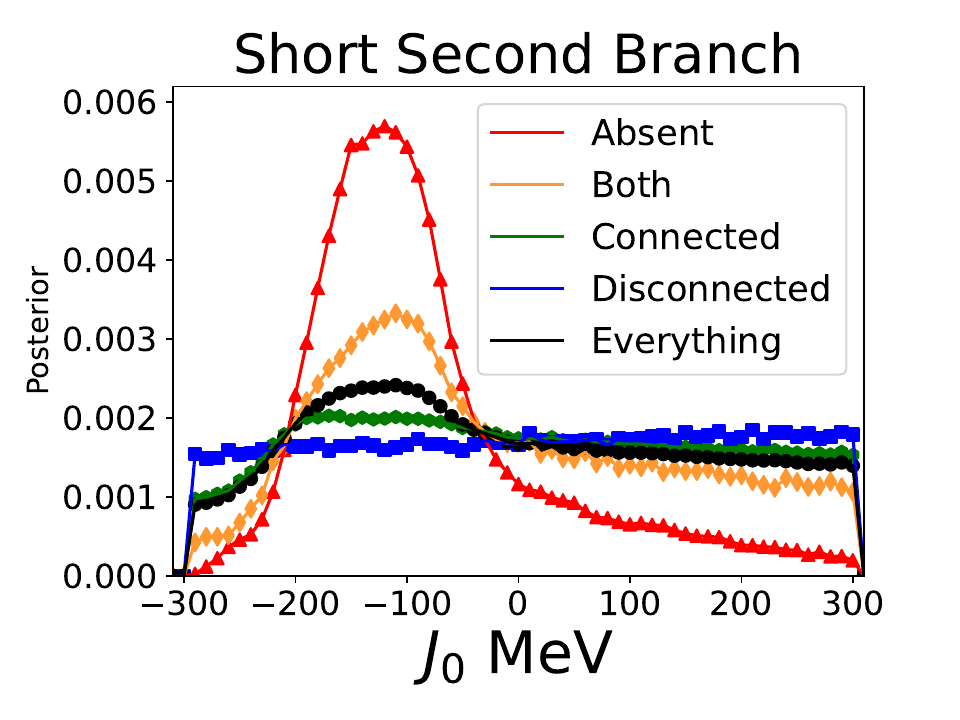} &
        \includegraphics[width=0.35\linewidth]{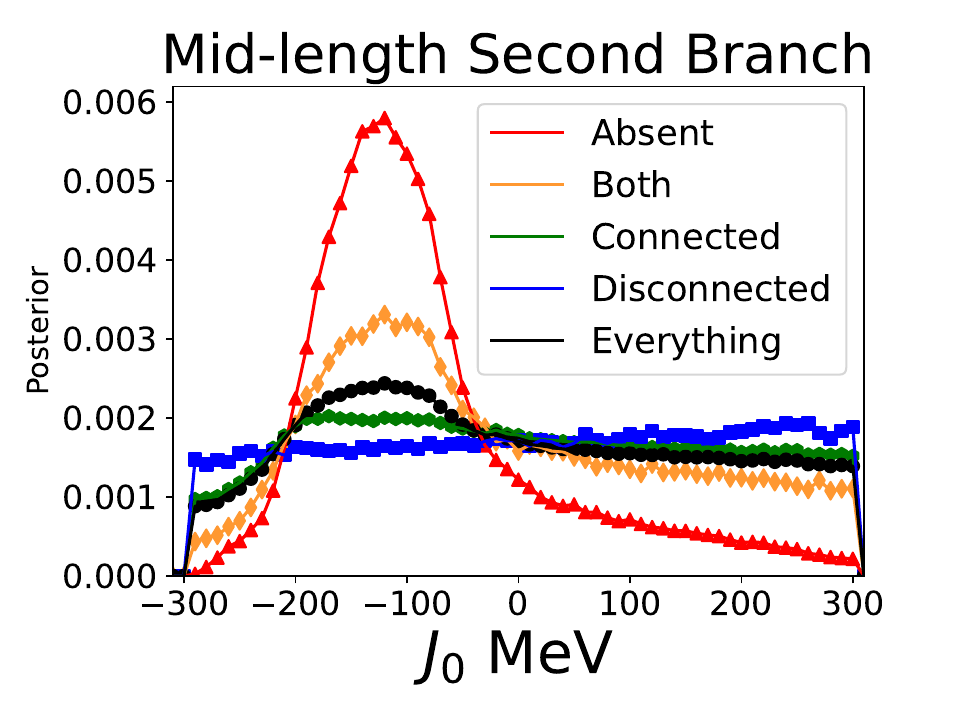} &
        \includegraphics[width=0.35\linewidth]{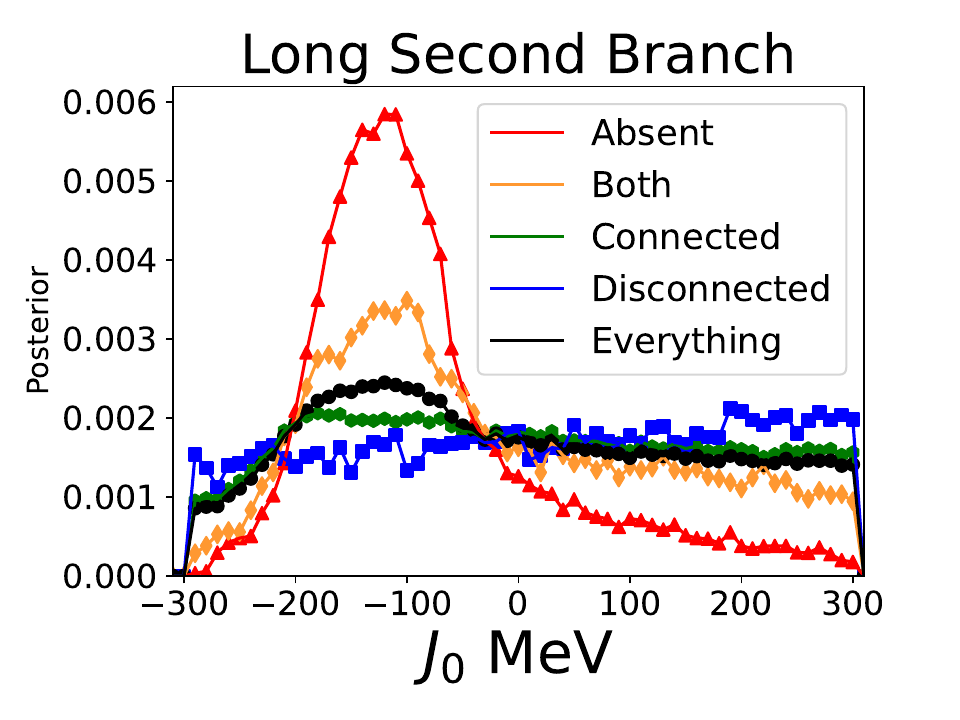} \\
        \includegraphics[width=0.35\linewidth]{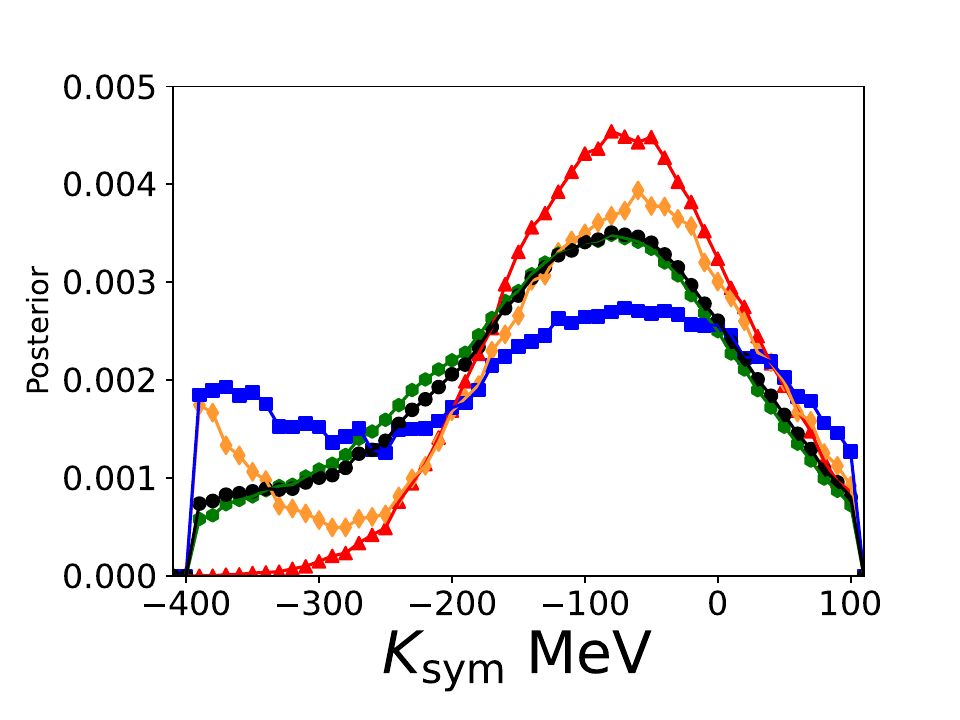} &
        \includegraphics[width=0.35\linewidth]{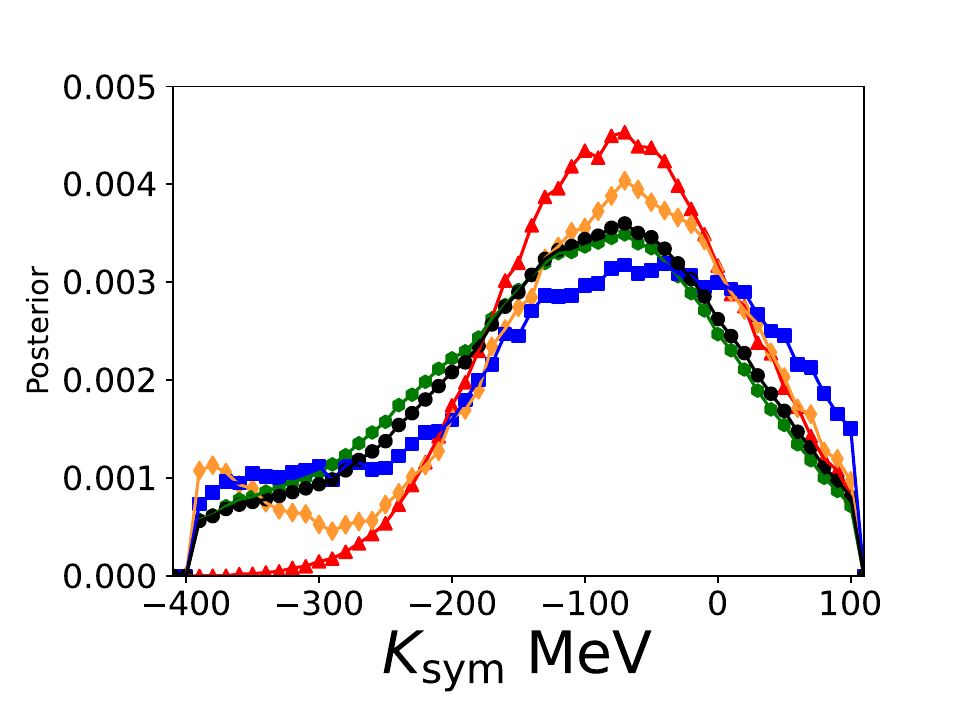} &
        \includegraphics[width=0.35\linewidth]{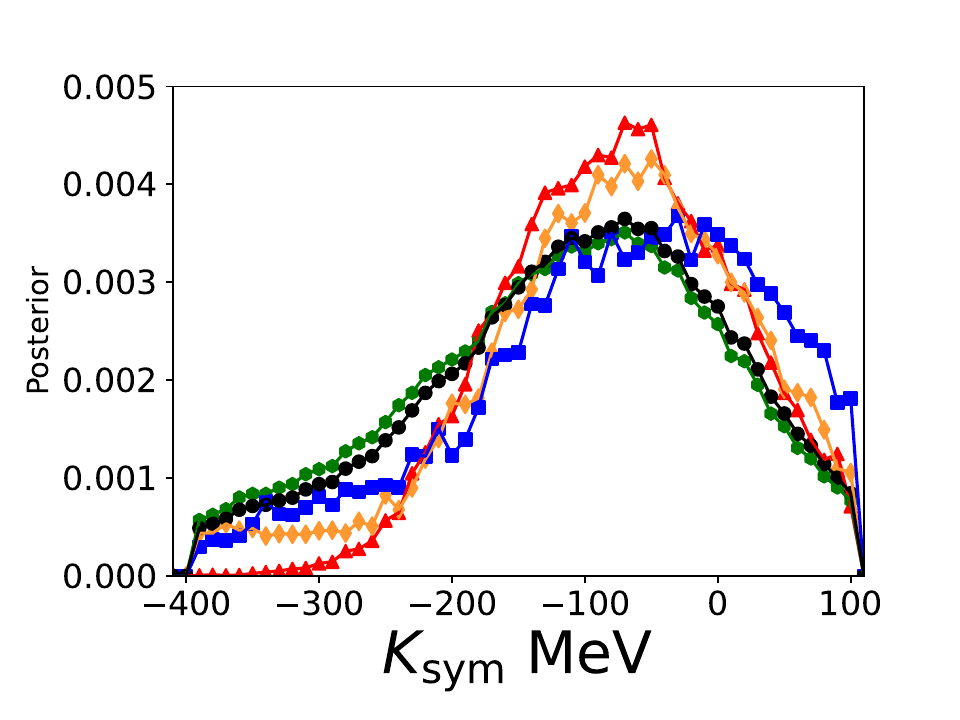} \\
        \includegraphics[width=0.35\linewidth]{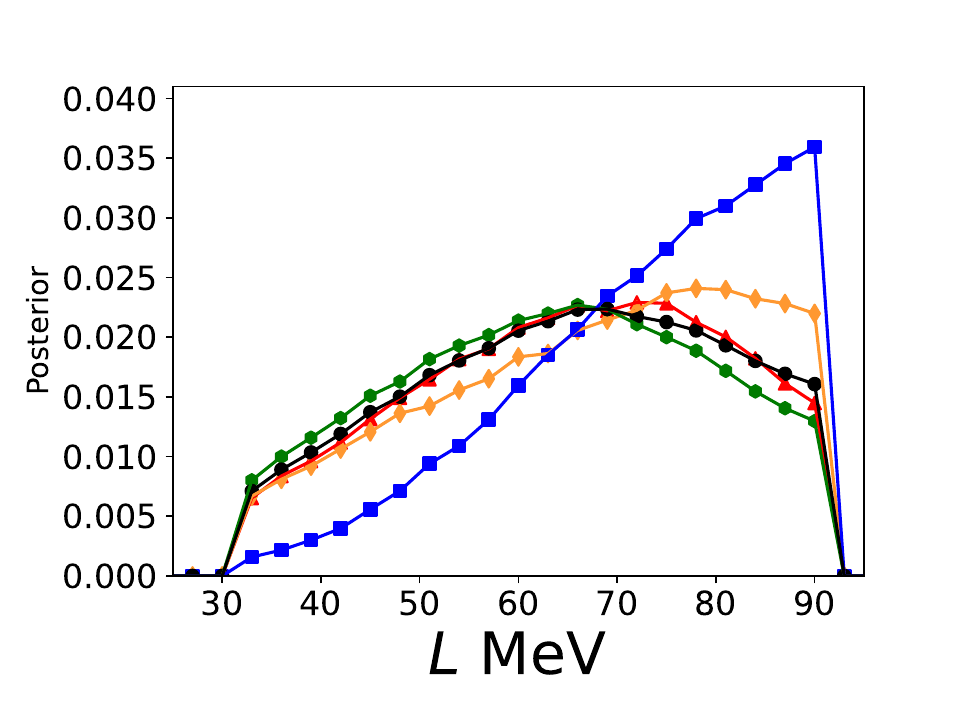} &
        \includegraphics[width=0.35\linewidth]{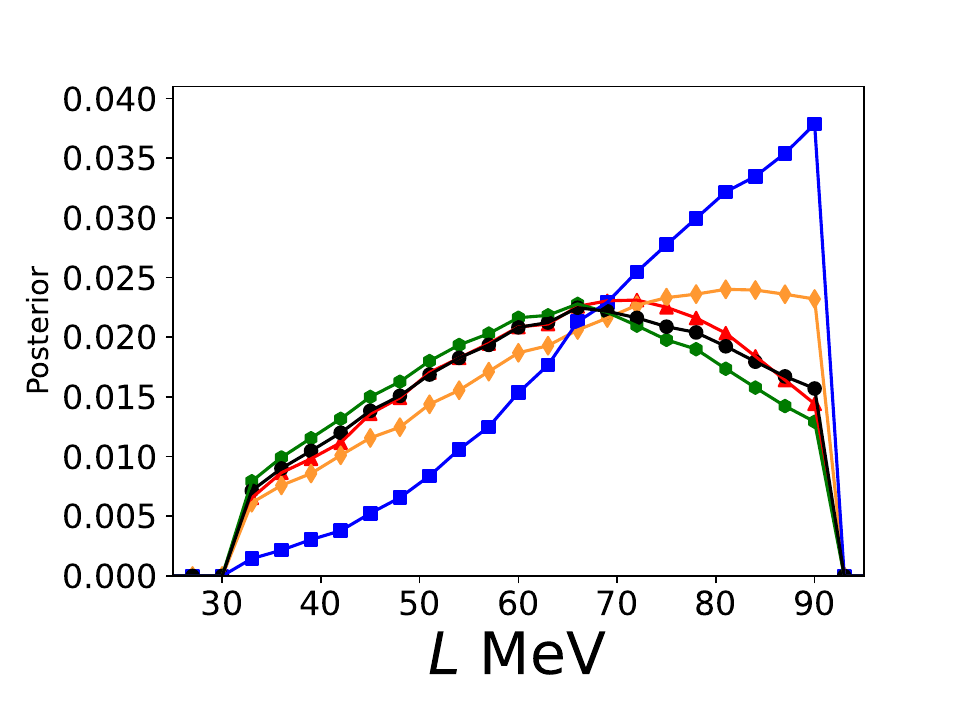} &
        \includegraphics[width=0.35\linewidth]{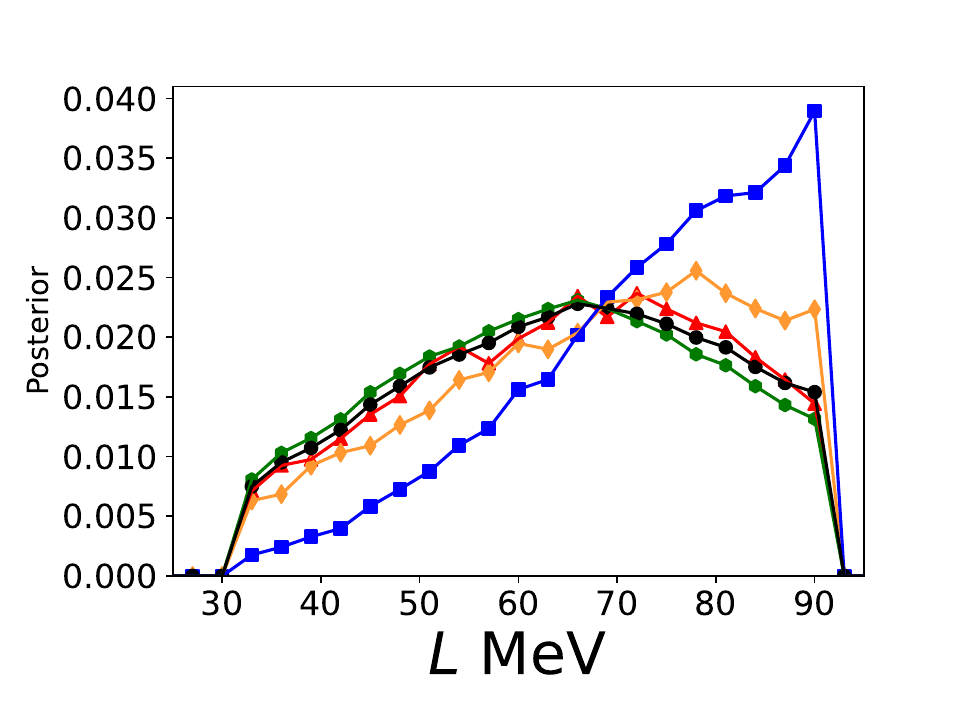}
    \end{tabular}
    \caption{PDFs of high-density HM EOS parameters}
    \label{fig:pdfHM}
\end{figure*}

We now look at the parameter space most likely to yield NSs in each category. In Tabs. \ref{tab:shortMeanStd}-\ref{tab:longMeanStd}, we show the mean and standard deviation of the posteriors of all nine parameters. While the PDFs are non-Gaussian and highly asymmetric, these values demonstrate some general trends. The MaP values are clear from the marginalized PDFs shown in Figs. \ref{fig:pdfQM} and \ref{fig:pdfHM}. They are therefore not listed.

Beginning with the high-density, QM EOS, we immediately see large differences in the PDFs for each EOS category. Categories Connected and Everything are very similar throughout because the majority of EOSs have a connected hybrid branch as shown in Tab. \ref{tab:twinCount}. The results for the Everything category are consistent with our earlier work \cite{XieLi_phasetrans}, and the transition density is discussed in detail in our most recent paper \cite{hybridPrecision}. It is more interesting to examine the other categories whose parameter space is usually overshadowed by the larger Connected category. The transition density of EOSs where NSs have no quark matter (Absent) is very high, peaking at $\rho_t/\rho_0 = 6$ at the upper bound of the prior, indicating this parameter is unconstrained in this category and could go higher if the upper limit of its prior is enlarged. This is reasonable since such a large density may not be reached even in the cores of the most massive NS, so no QM would be present. The Disconnected category peaks at very low transition densities around $\rho_t/\rho_0 \approx 1.5$, which matches other recent studies' findings of very low transition densities, see e.g., Refs. \cite{Mendes:2024hbn, Yuan:2025njl, hybridPrecision, Laskos_Patkos_2024, Ayriyan:2024zfw, Ayriyan:2025rub}. Because an EOS must have a large energy-density discontinuities at the phase transition to produce twin stars in this category, which significantly softens the EOS, the phase transition must occur early on. The QM EOS is then required to be very stiff with a speed of sound near the speed of light to produce the minimum M$_{\rm{TOV}}$ of 1.97 $M_\odot$. These trends, widely noted in the literature, are again shown in all three rows of Fig. \ref{fig:pdfQM} for the Disconnected category. Lastly for the transition density, the PDF is still nearly uniform for the Both category, indicating that twin stars are possible for a transition at any density depending on the other EOS parameters. This is good because the results for the Disconnected category would be ruled out by findings of recent Beam Energy Scan (BES) experiments at the Relativistic Heavy-Ion Collider (RHIC) \cite{bes2022_1,bes2022_2,bes2022_3,Yong22}, which currently place a lower limit on the hadron-quark transition density in cold NS matter to around $(3-4)\rho_0$ \cite{hybridPrecision}.

In the second row of Fig. \ref{fig:pdfQM}, we show the PDFs of the energy density discontinuity, and our results match those of Alford, Han, and Prakash in Ref. \cite{Alford:2013aca}. The Connected category favors low values in order for the hybrid branch to be connected; the Disconnected category favors high values so that the phase transition immediately destabilizes the star. The Both category is a middle ground that just satisfies the Seidov condition in Eq. (\ref{eq:seidov}), so that the star destabilizes shortly after the phase transition. For the Absent category, $\Delta \varepsilon / \varepsilon_t$ is unconstrained, peaking at the upper boundary again. This is because if the energy density jump is too large, all hybrid stars will be unstable. The length requirement for the second branch does play a role in determining this parameter. The right shoulder for the Disconnected category at high values of $\Delta \varepsilon/\varepsilon_t$ drops as we require a longer second branch. This is due to the instability caused by a larger jump, which if too large, will prevent a long branch of stable twin stars.

The last row of Fig. \ref{fig:pdfQM} has the PDFs for the speed of sound squared in QM. Several interesting observations can be made. Firstly, as mentioned earlier, Disconnected EOSs require a very high speed of sound in order to produce the minimum M$_{\rm{TOV}}$ of 1.97 $M_\odot$ required by astrophysical observations, and so this PDF peaks at $c^2_{\rm qm} = 1$. The Connected category is less peaked, but favors high values, see our previous work for more detailed discussions \cite{XieLi_phasetrans}. This comes from the softening of even a relatively weak phase transition, requiring a stiff QM EOS to compensate.

This high value for the speed of sound is in contrast to the conformal limit ($c^2_{\rm qm} < 1/3$) predicted by pQCD. However, the pQCD regime only applies when quarks become asymptotically free, around 40 $\rho_0$, far beyond the densities in NS considered here \cite{Somasundaram:2022ztm,Zhou:2023zrm}. Many studies have examined a peak in the speed of sound profile within NS densities, before approaching the conformal limit at very high densities, see, e.g., Refs. \cite{Cai:2023pkt, Cai:2023ajw, Cai:2024oom, Tajima:2024qzj, Xia_2021, Jimenez:2024hib} and references therein. Obviously, at high densities the approximation that the speed of sound is constant fails; the CSS model is only useful at NS densities. Thus, the high speed of sound found probable in our analysis can still be consistent with pQCD because the two apply at different densities.

Secondly, of course, NSs without QM again do not constrain the QM parameter for the speed of sound, whose PDF is nearly unchanged from its uniform prior except for the lack of low values. This is because (1) the stiffness of QM is irrelevant to MR data if QM is not present, and (2) the lack of low values is because this combination yields EOSs in the Both category. 

Thirdly and most interestingly, the PDF($c^2_{\rm qm}$) for Both EOSs has a minor peak at $c^2_{\rm qm} = 1$, but the main peak is at $\approx 0.1$. To the authors' best knowledge, such a two-peaked feature has not been previously reported in the literature. This is likely due to most studies fixing the speed of sound squared at $1/3$ or $1$, or focusing only on Disconnected EOSs where the Seidov condition is violated. It is interesting to note that the main peak is located below the conformal limit of $c^2_{\rm qm}=1/3$ predicted by pQCD, valid at densities above about 40$\rho_0$. Qualitatively, it is not surprising to see two peaks in the PDF($c^2_{\rm qm}$) in this category where there are two branches of hybrid stars having QM cores at very different energy densities. The hybrid stars on the second branch have smaller radii but the same or larger masses, thus higher energy densities, compared to the ones on the first branch. The QM at high energy densities can afford to be softer (having $c^2_{\rm qm}$ much less than 1/3) to produce similar or even higher pressure to support massive hybrid stars on the second branch, compared to hybrid stars having relatively low-density QM cores (having $c^2_{\rm qm}$ close to 1) on the first branch. This is because the pressure in QM is proportional to the product of $c^2_{\rm qm}$ and its energy density $\varepsilon_{\rm{qm}}$.

Ironically, the above findings are from applying the so-called Constant Sound Speed (CSS) model where the stiffness $c^2_{\rm qm}$ of QM was introduced as a constant coefficient of its EOS in Eq. (\ref{css}). However, both mathematically and physically, nothing prevents the $c^2_{\rm qm}$ from varying when the EOSs are randomly generated and then selected in the MCMC process according to the likelihood function of Eq. (\ref{ll}) in our Bayesian framework. Namely, the CSS model itself does not require the data-preferred $c^2_{\rm qm}$ value to stay the same (as its name may imply) in all hybrid stars on the two branches where the QM energy densities (thus the QM pressure $(p-p_t)$ values) are rather different. Therefore, the observed two-peaked PDF($c^2_{\rm qm}$) in the Both category indicates that the $c^2_{\rm qm}$ decreases with increasing density as the MR curve jumps (reflected by the gap in PDF($c^2_{\rm qm}$) for $c^2_{\rm qm}$ between about 0.2 and 0.6) from the first to second hybrid branch. This information may have significant ramifications in understanding the density profile of speed of sound $dp/d\varepsilon$ and the associated trace anomaly $\Delta\equiv 1/3-p/\varepsilon$ in massive neutron stars, see, e.g., Refs. \cite{Jimenez:2024hib,Tan:2021ahl, Tan:2021nat,Altiparmak:2022bke,Fujimoto:2022ohj,Cai:2023pkt,Cai:2023ajw}. Thus, the dual-peaked feature in PDF($c^2_{\rm qm}$) warrants further investigations more quantitatively especially when more accurate NS radius data become available. Indeed, there are many unresolved issues regarding the density dependence of $c^2_{\rm qm}$ in massive NSs, see, e.g., Refs. \cite{Cai:2024kfs,Cai:2025nxn} for recent reviews. Investigations of hybrid twins may help address some of these issues.

\begin{figure*}
    \centering
    \addtolength{\tabcolsep}{-0.4em}
    \begin{tabular}{ccc}
        \includegraphics[width=0.33\linewidth]{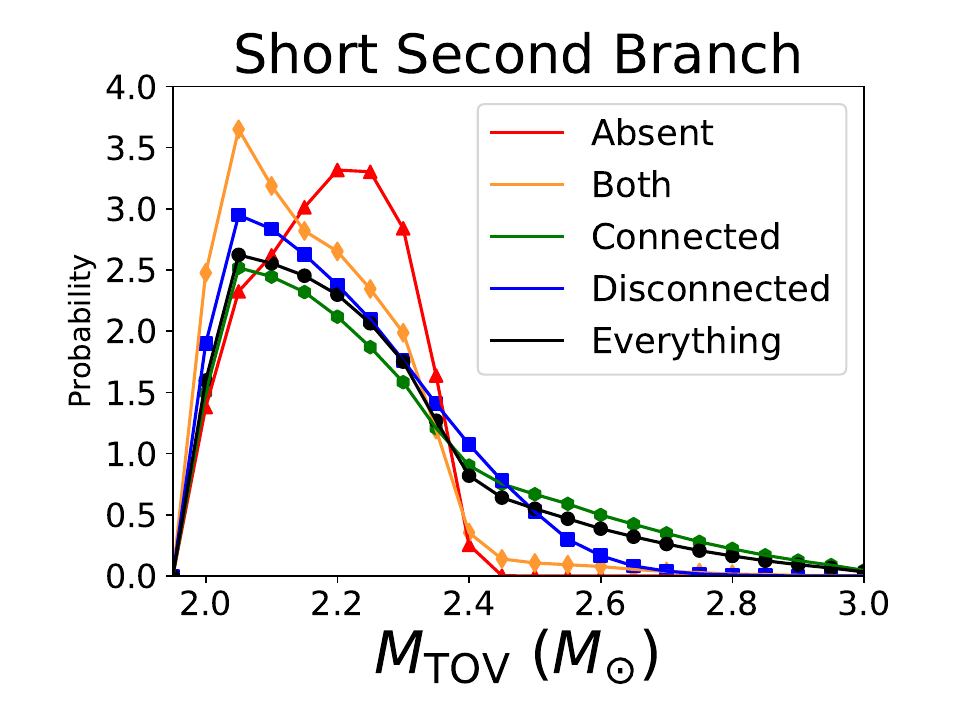} &
        \includegraphics[width=0.33\linewidth]{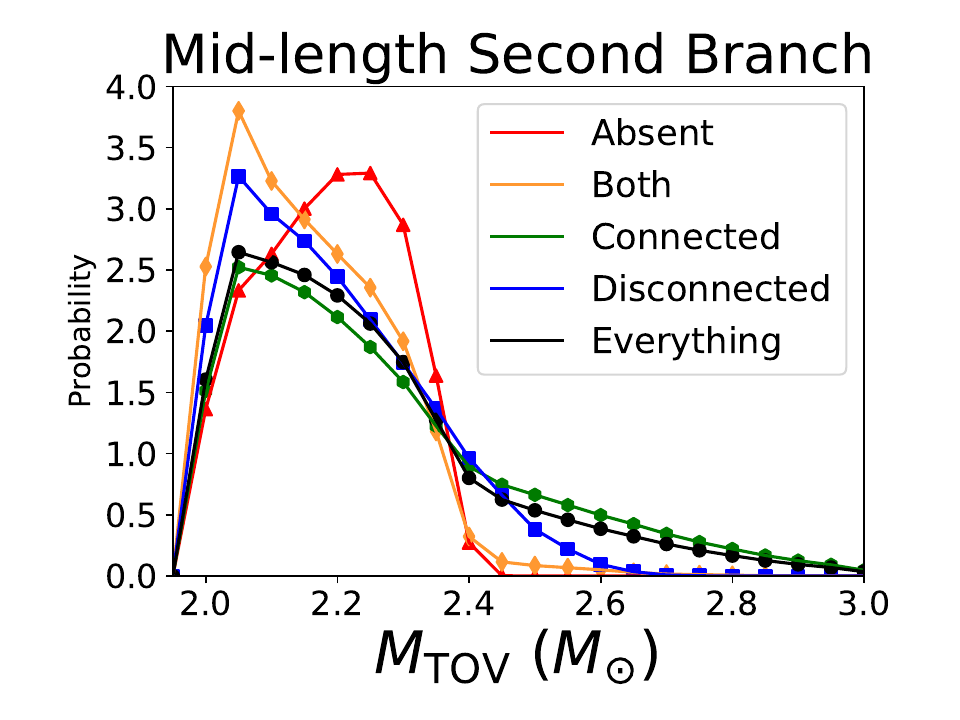} &
        \includegraphics[width=0.33\linewidth]{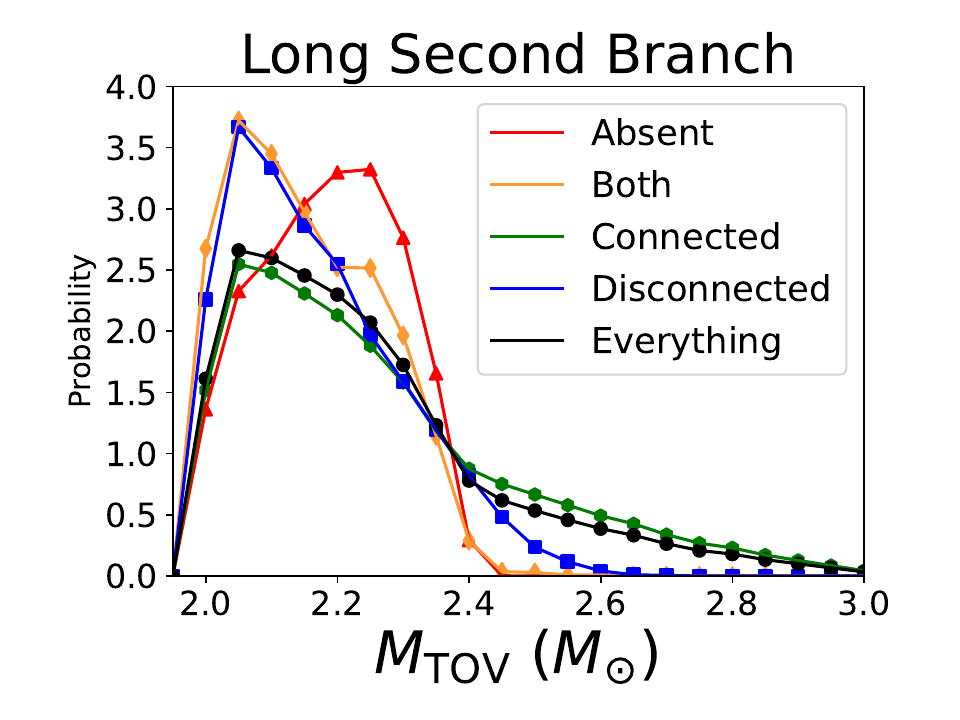} \\
        \includegraphics[width=0.33\linewidth]{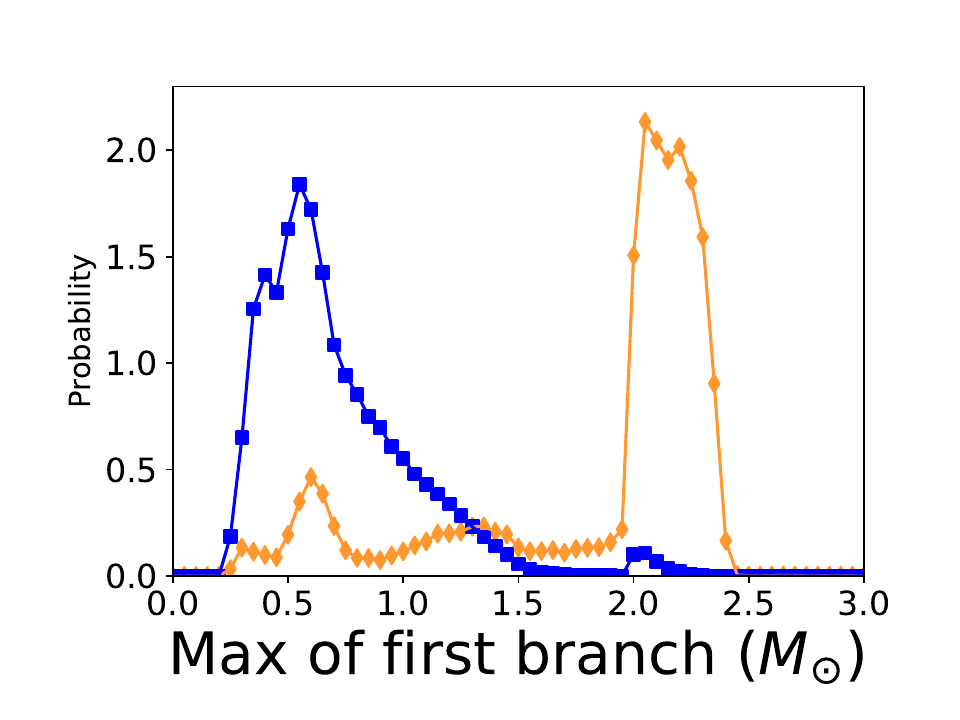} &
        \includegraphics[width=0.33\linewidth]{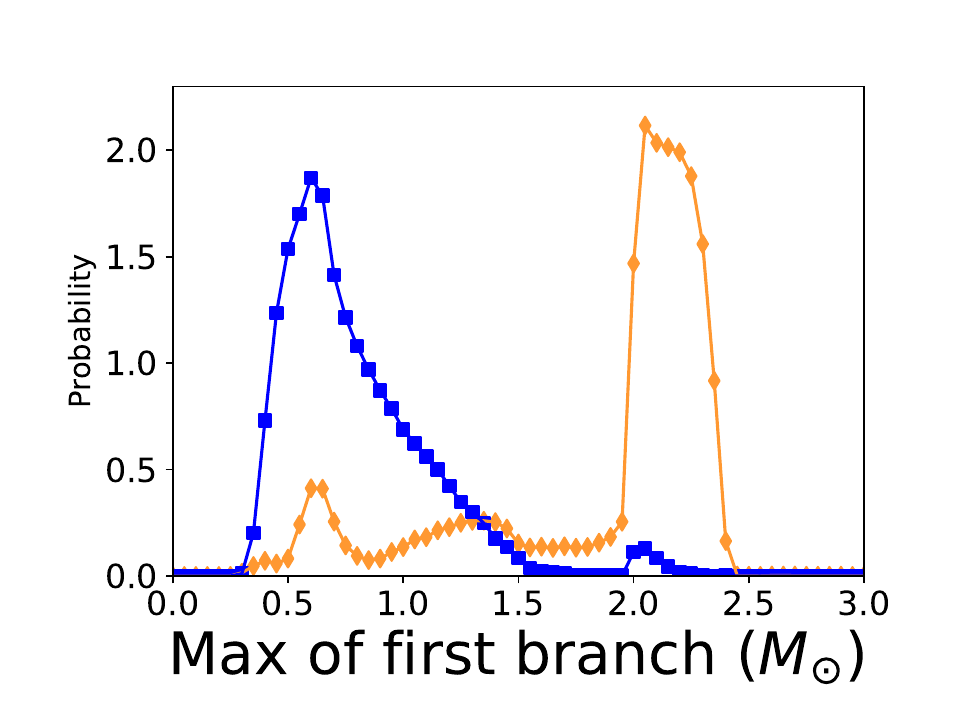} &
        \includegraphics[width=0.33\linewidth]{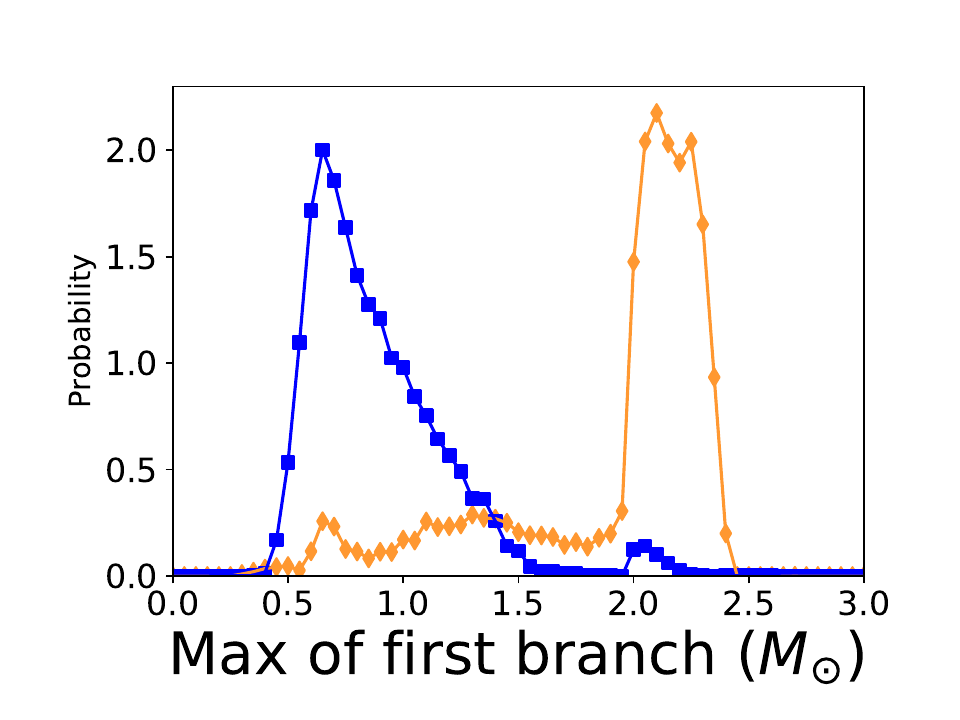} \\
        \includegraphics[width=0.33\linewidth]{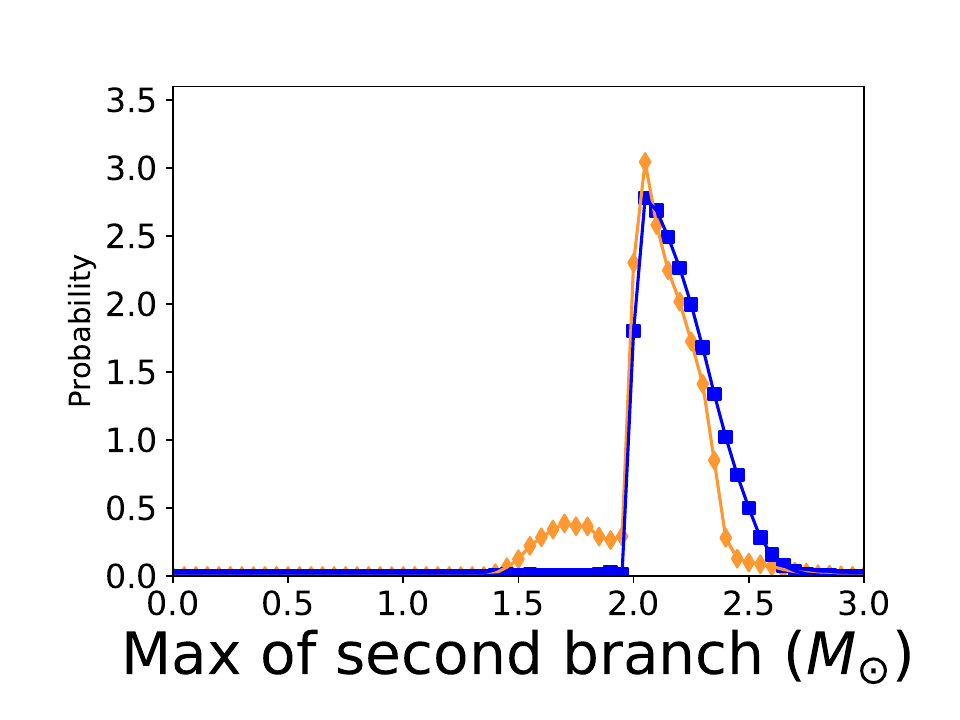} &
        \includegraphics[width=0.33\linewidth]{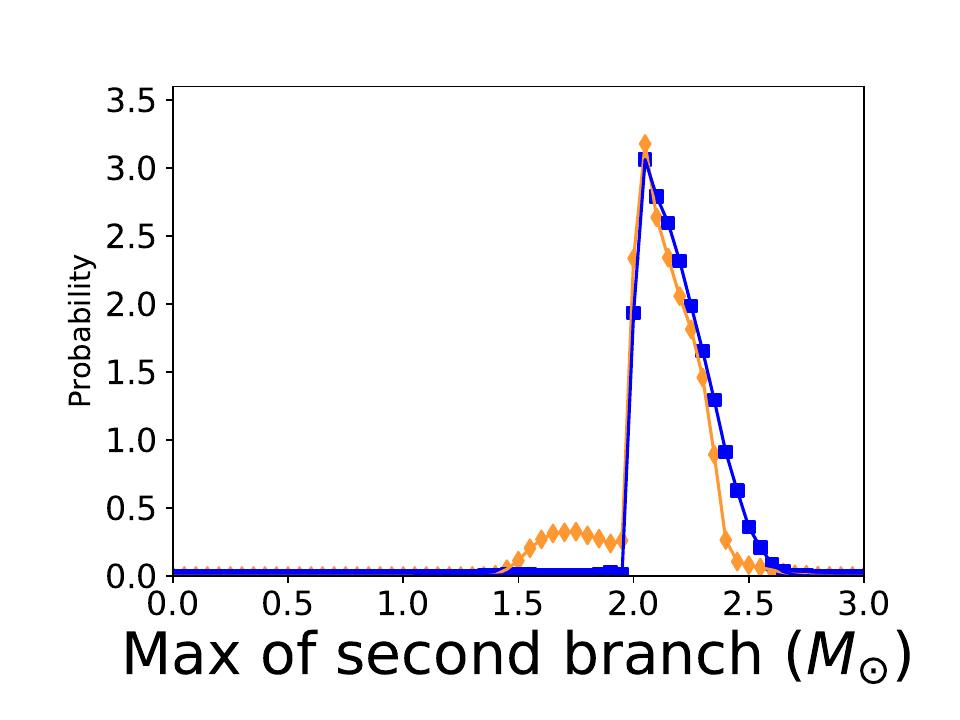} &
        \includegraphics[width=0.33\linewidth]{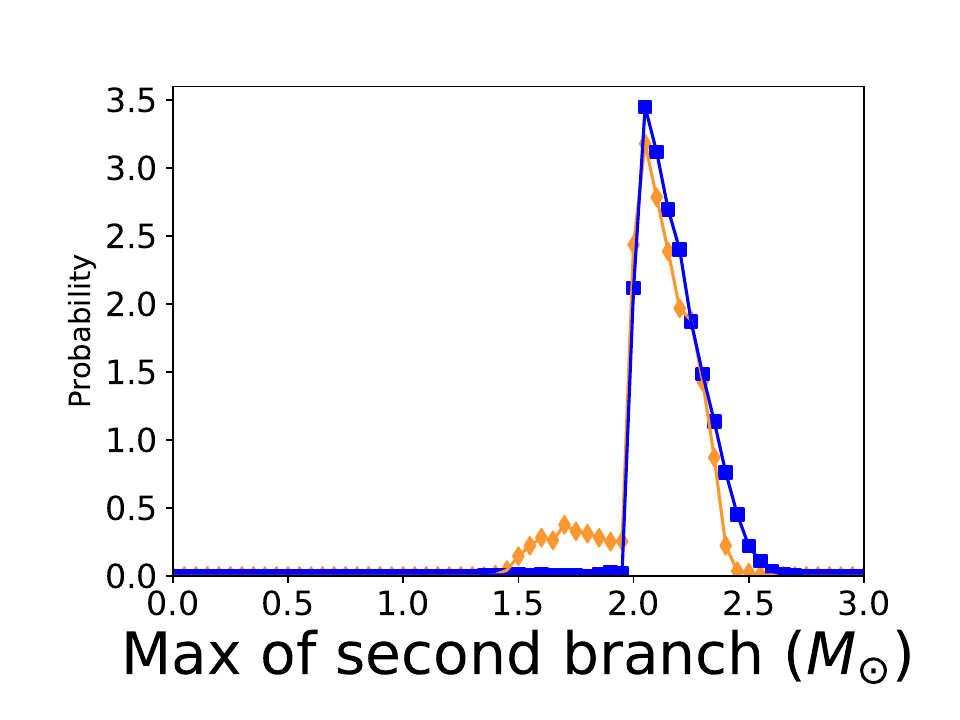}
    \end{tabular}
    \caption{Probability distribution of M$_{\rm{TOV}}$ (top), the maximum mass reached on the first branch (middle), and the maximum mass reached on the second branch (bottom) of all accepted EOSs in their respective categories.}
    \label{fig:probMass}
\end{figure*}

Shown in Fig. \ref{fig:pdfHM} are the HM EOS parameters that are actually constrained by the astrophysical data used. The parameters $E_{\rm sym}(\rho_0)$ and $K_0$ are already well constrained, and the skewness of symmetry energy, $J_{\rm sym}$, has proven elusive to any existing astrophysical constraint \cite{xie2019bayesian,xie2020bayesian}, so their PDFs are not much changed from their priors. Beginning with the SNM skewness parameter, $J_0$, we see that its PDF is less and less constrained as we go from Absent to Both to Connected to Disconnected. For the first three categories, $J_0$ peaks around $-120$ MeV, but only the Absent EOSs show a significant peak. The Disconnected category on the other hand remains completely unconstrained with almost no change from the uniform prior. This is because $J_0$ controls the stiffness of HM EOS at densities three to four times the saturation density. If, however, at these densities, there is QM, $J_0$ becomes irrelevant in determining NS properties. This idea can be checked with Tabs. \ref{tab:shortMeanStd}-\ref{tab:longMeanStd}. The standard deviation of $J_0$ is related to how spread out its posterior distribution is and increases as the average value of $\rho_t/\rho_0$ decreases.

$K_{\rm sym}$ shows a similar pattern. While now every category has a peak around $-50$ MeV, the width of that peak increases as the average transition density decreases. The PDFs of Both and Disconnected of $K_{\rm sym}$ are also sensitive to the length requirement of the second branch, with a shorter requirement allowing softer values near this parameter's lower bound. From Tabs. \ref{tab:shortMeanStd}-\ref{tab:longMeanStd} we see that this can also be explained by a change in the average transition density. If the transition density is low enough, even $K_{\rm sym}$ becomes unimportant.

Finally, the PDFs of the slope $L$ of symmetry energy is shown at the bottom of Fig. \ref{fig:pdfHM}. Both categories Absent and Connected peak around $65-70$ MeV, consistent with its fiducial value of about $57.7 \pm 19$ MeV \cite{Li:2013ola}. The Both PDF peaks at a slightly higher value, stiffening the symmetry energy around $(1-2)\rho_0$ in response to the increased softening of the EOS from a larger $\Delta \varepsilon/\varepsilon_t$. Lastly, the Disconnected category peaks at 90 MeV, at the upper bound of the prior, indicating higher values would also satisfy the astrophysical data. This large value may be justified by the results of PREX-II \cite{prexData,prexAnalysis}, but its tension with other experiments has so far been unresolved, see, e.g., Ref. \cite{Reed2024-prexcrex} for a recent discussion.

\subsection{NS maximum Mass}
We report lastly the maximum mass M$_{\rm{TOV}}$ of all NSs, as well as the maximum masses of the first and second branches for twin stars. Fig. \ref{fig:probMass} shows the probability distributions of these maximum masses for the accepted EOSs. For most categories, the distributions peak around 2.05 $M_\odot$ just above the required minimum M$_{\rm{TOV}}$ of 1.97 $M_\odot$. Only the Absent category has a peak at about 2.2 $M_\odot$, but it has a sharp drop, so that almost no NS without QM can exceed 2.4 $M_\odot$. If the secondary object with a mass around 2.5 $M_\odot$ measured by LIGO/VIRGO collaborations in GW190814 was a NS \cite{Abbott_2020}, then that may require a phase transition to very stiff quark matter, with a speed of sound near one, in order to reach this mass since only hybrid star categories have solutions in that range.

The first and second maxima are only slightly affected by the different length requirements on the second branch. We see that for most EOSs in the Both category, the maximum mass on the first branch meets the 1.97 $M_\odot$ requirement. The maximum on the second branch is also usually above this limit, although some are below, indicating that with a strong enough softening in the EOS the densest hybrid stars are not the most massive. With their low transition density, the Disconnected category of NSs almost always undergo a phase transition before reaching 1.5 $M_\odot$, so a high speed of sound is required for the quark matter to support a 1.97 $M_\odot$ star on the second branch. We mention for comparison that in terms of the categories used in Ref. \cite{Christian_2018}, Both EOSs are usually Category I or II, and Disconnected EOSs are usually Category III or IV.

\subsection{A Closer Look At EOSs in the Both Category}
The unusual results of EOSs in the Both Category deserve more attention. It will be useful to split our analysis into two categories. First, if we have a low transition density, $\rho_t \lesssim 2\rho_0$, then the QM EOS must be very stiff in order to meet the 1.97 $M_\odot$ constraint, as shown most clearly by the PDFs for Disconnected EOSs. Categorization into Both, Connected, or Disconnected then depends on the parameter $\Delta \varepsilon / \varepsilon_t$, whether it is low (Connected), high (Disconnected), or in-between (Both).

Now, if we instead had a high transition density, then the NS will meet the two solar mass requirement on the first branch, as shown clearly for Both EOSs in Fig. \ref{fig:probMass} in the previous subsection. It is likely that the radius constraint has also been met by the first branch. Therefore, there is no constraint on the QM stiffness. Zhou \& Huang in Ref. \cite{chunhuang_2025twin} found that requiring the existence of twin stars did not constrain the speed of sound in QM, but it was their choice of four MR measurements from NICER that forced the speed of sound toward the causal limit. In this work, we chose to focus on just the single radius data from GW1701817, so we do not have that constraint. This means that the large peak in the PDF of $c_{\rm qm}^2$ at 0.1 for EOSs in the Both category is due to the categorization. That is, we see this peak because it is not covered up by the more common Connected category, and such a speed of sound yields Both EOSs and not other categories. This can be justified by the comments in Refs. \cite{Alford:2013aca} and \cite{Zdunik_2006}. In Sec. III (A) of Ref. \cite{Alford:2013aca}, Alford, Han, and Prakash discuss the nearly horizontal boundary on their phase diagram between categories Both and Connected, which is a small distance below the $\Delta \varepsilon_{crit}$-line in the $\Delta \varepsilon$\textendash$p_t$ plane that marks the boundary for Seidov stability, see Fig. 3 of Ref. \cite{Alford:2013aca}. Zdunik et al. note in Ref. \cite{Zdunik_2006} Sec. 4.2 that the separation between the boundary line between Both and Connected and the line of Seidov stability is determined by the stiffness of the QM EOS, namely, a softer QM EOS will increase the separation between these two lines, which increases the allowable range of $\Delta \varepsilon / \varepsilon_t$ for the Both category. Thus, having a very soft QM EOS lends itself well to the formation of twin stars in the Both category, and so we see the large peak in the PDF($c_{\rm qm}^2$) at very low values.

\begin{figure}
    \centering
    \includegraphics[width=0.9\linewidth]{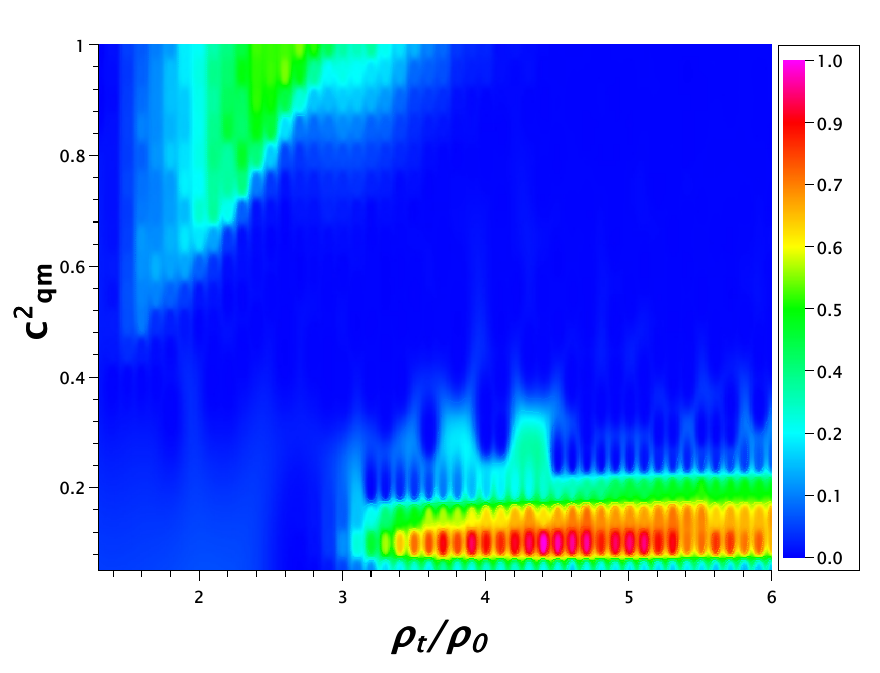}

    \includegraphics[width=0.9\linewidth]{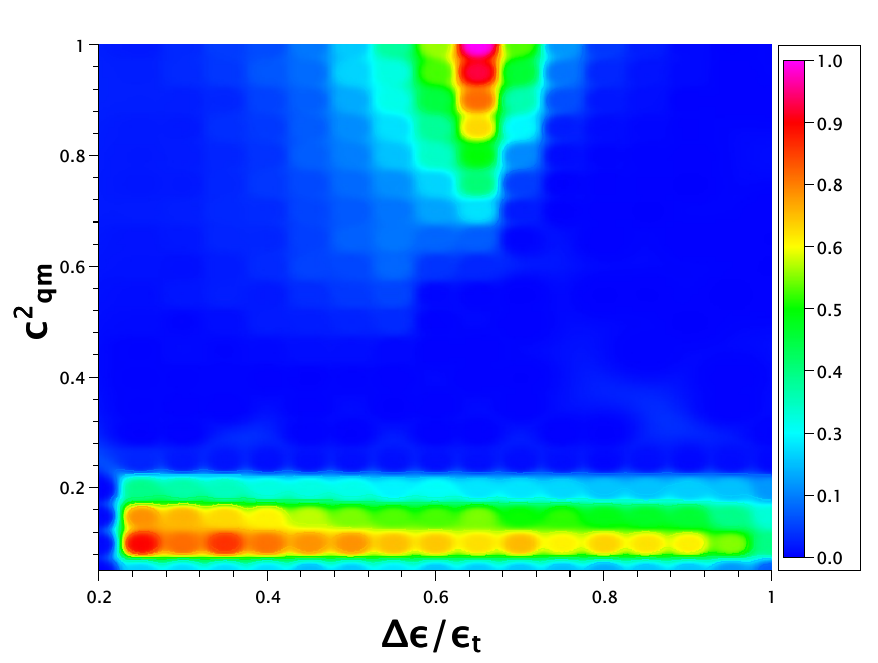}
    \caption{The 2D posterior probability distribution functions for the QM parameters in the Both category, which shows their correlation.}
    \label{fig:cscor}
\end{figure}
In Fig. \ref{fig:cscor}, we show the two-dimensional posterior distribution functions for $\rho_t/\rho_0$-$c_{\rm qm}^2$ and $\Delta \varepsilon / \varepsilon_t$-$c_{\rm qm}^2$, which shows these parameters' correlations. These plots are for the Both category under the mid-length branch requirement as a demonstration of the observations above. As can be seen, for low values of the transition density, the speed of sound in QM is very high. For these high values of $c_{\rm qm}^2$, the energy density discontinuity is moderate. On the other hand, if $\rho_t/\rho_0$ is high, then the $c_{\rm qm}^2$ is typically below 0.2, and $\Delta \varepsilon / \varepsilon_t$ is relatively uncorrelated with $c_{\rm qm}^2$.

\section{Effects of Recent Measurements}\label{nicer}
\begin{table}[htbp]
    \centering
    \begin{tabular}{|c|c|c|c|}
        \hline
         & Run 4 & Run 5 & Run 6 \\
        \hline
        Absent & 211,968 & 87,448 & 86,276\\
        Both & 105,317 & 45,377 & 57,449\\
        Connected & 1,319,405 & 643,233 & 731,049\\
        Disconnected & 51,600 & 63,960 & 110,909\\
        Everything & 1,688,291 & 840,018 & 985,683\\
        \% Twin & 9.29 & 13.02 & 17.08\\
        \hline
    \end{tabular}
    \caption{The number of accepted EOS in each category from the Bayesian analyses with recent observational constraints: $R_{2.1} = 12.92_{-1.13}^{2.09}$ km, $M_{\rm TOV} \ge 2.08~M_\odot$, and $R_{1.4} = 11.36_{-0.63}^{+0.95}$ km for runs 4, 5, and 6, respectively.}
    \label{tab:twinCount_nicer}
\end{table}

\begin{figure*}
    \centering
    \addtolength{\tabcolsep}{-0.8em}
    \begin{tabular}{ccc}
        \includegraphics[width=0.35\linewidth]{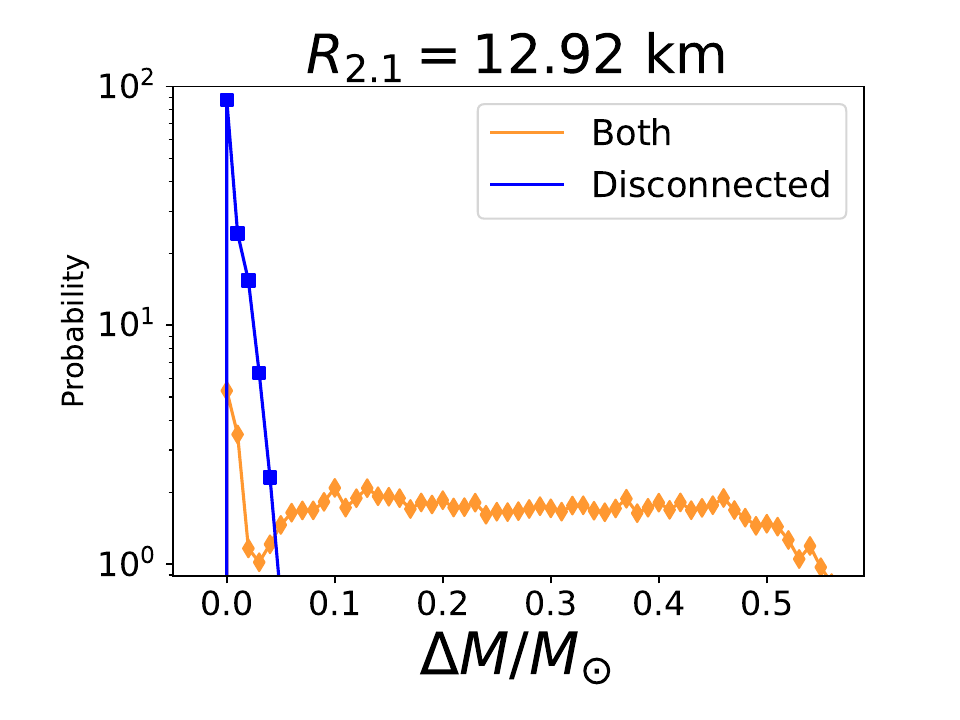} &
        \includegraphics[width=0.35\linewidth]{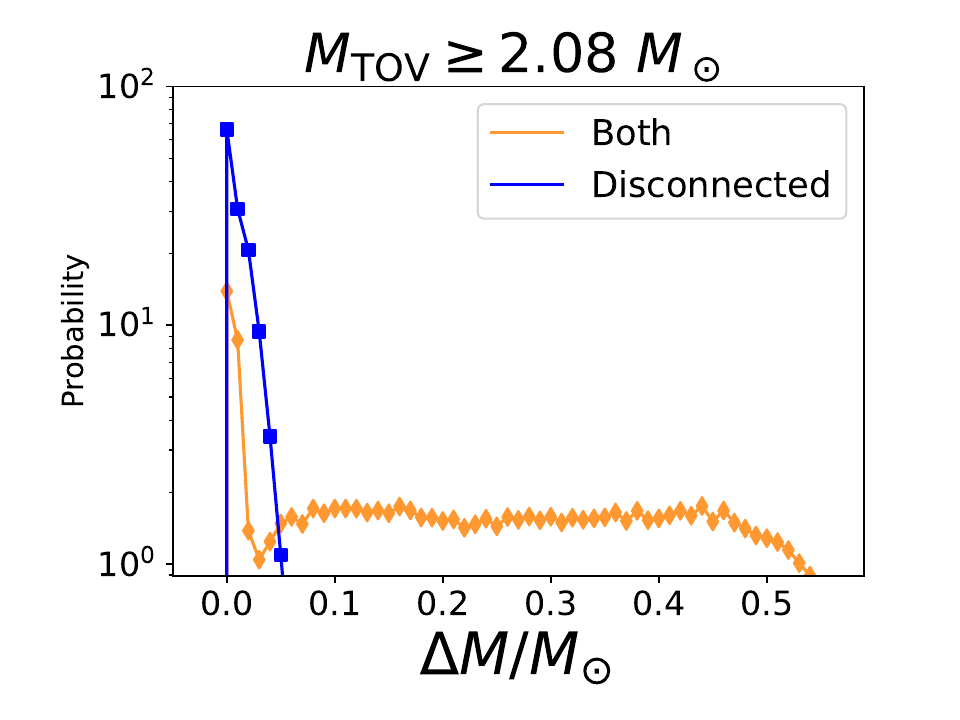} &
        \includegraphics[width=0.35\linewidth]{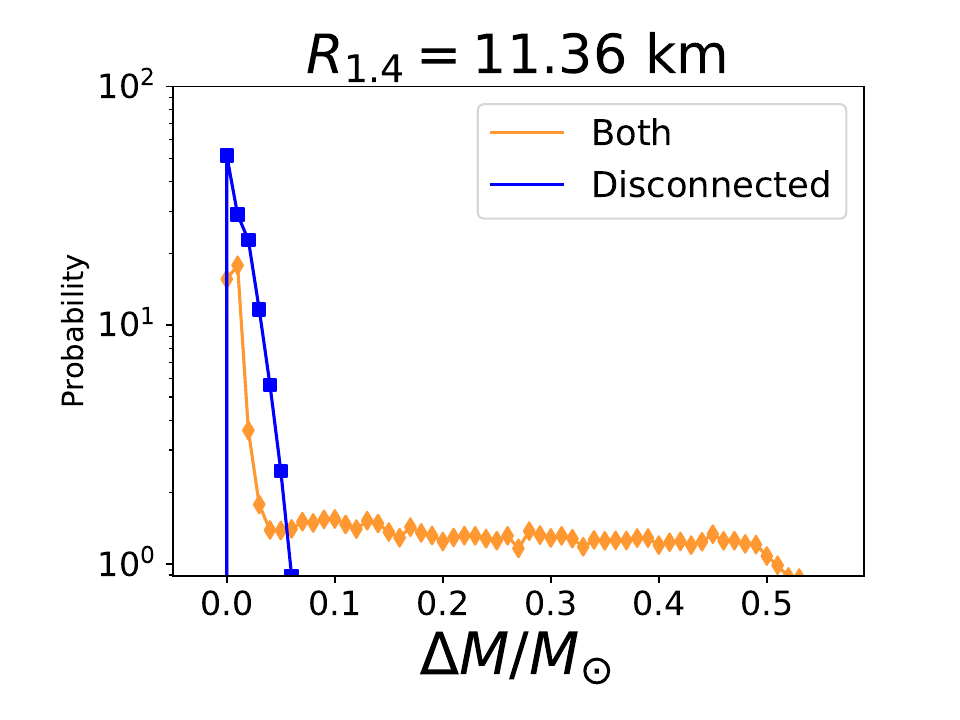} \\
        \includegraphics[width=0.35\linewidth]{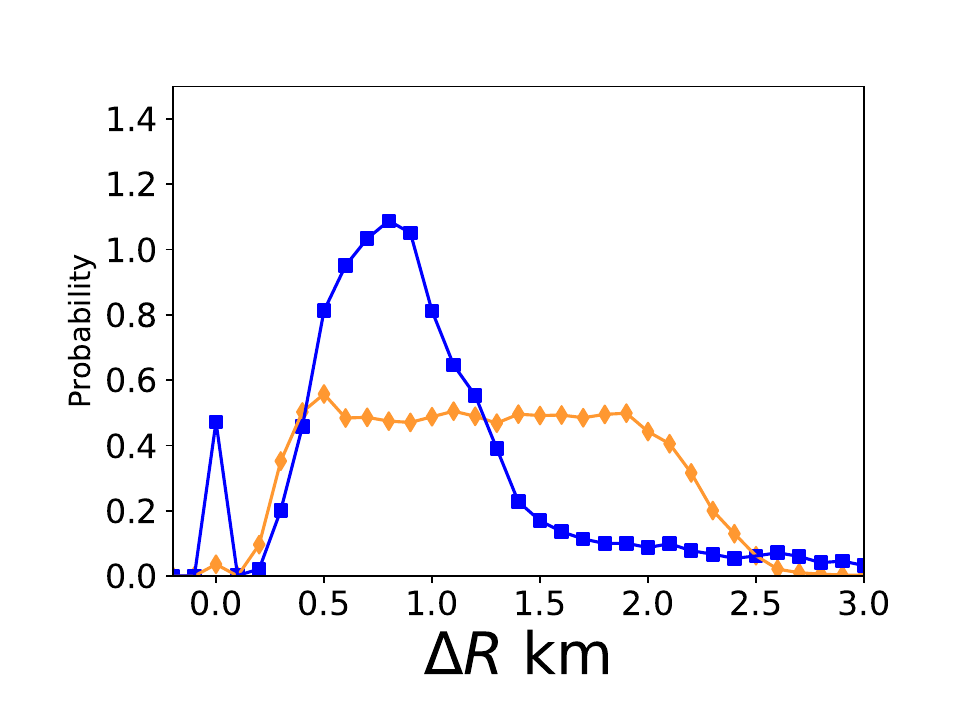} &
        \includegraphics[width=0.35\linewidth]{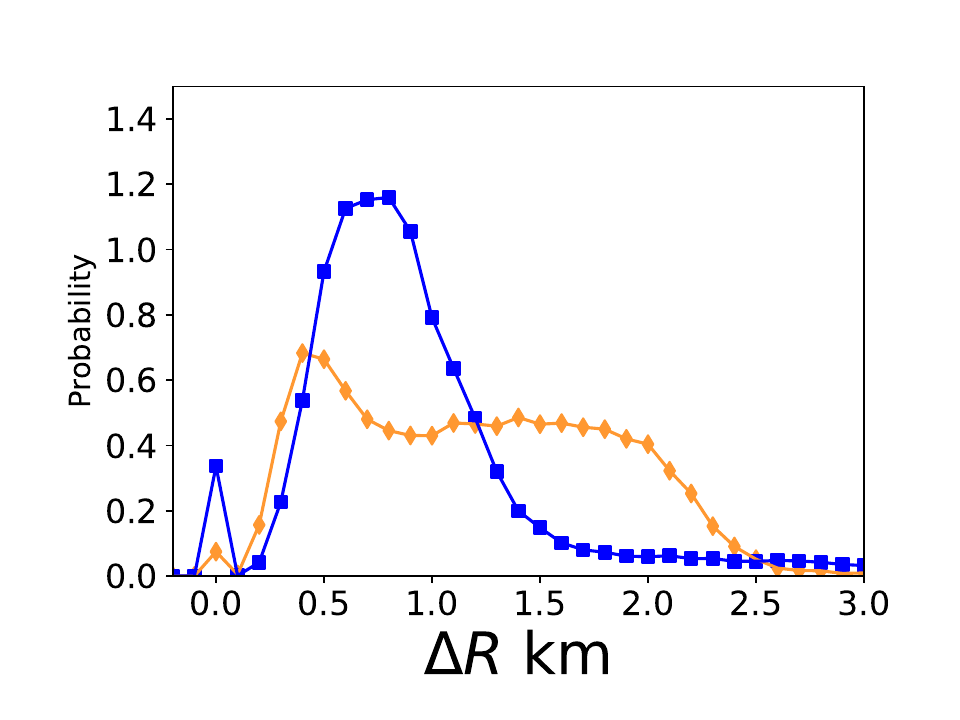} &
        \includegraphics[width=0.35\linewidth]{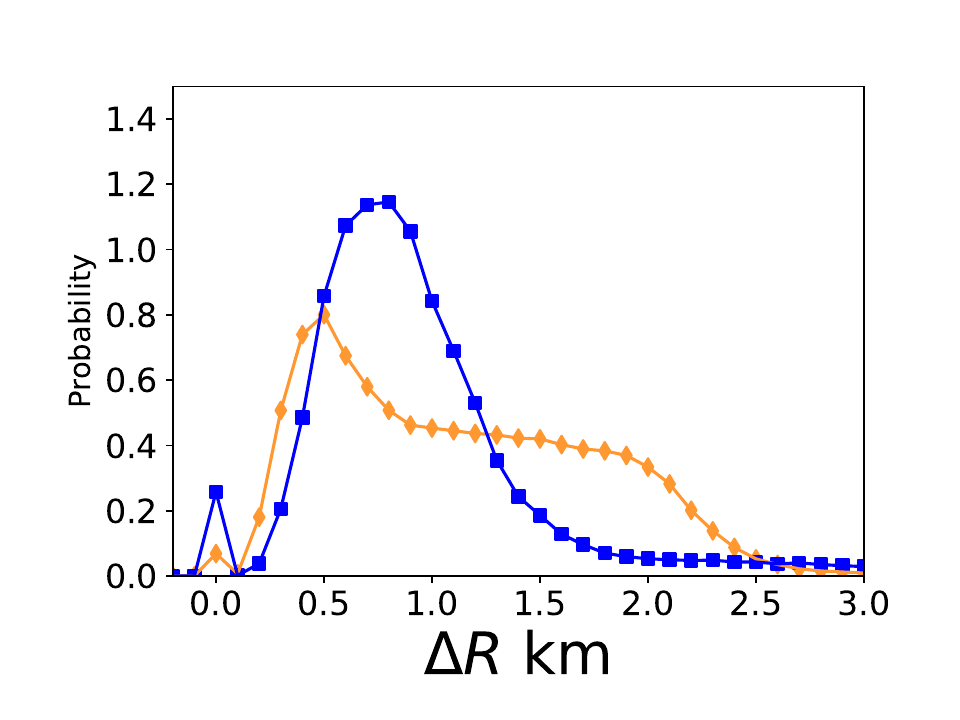}
    \end{tabular}
    \caption{Same as Fig. \ref{fig:probObs} but with more recent observations.}
    \label{fig:probObs_nicer}
\end{figure*}

We now examine the effects of using more recent observational data of NSs from NICER. Run 4 specified a single mass-radius observation based on the analysis in Ref. \cite{Dittmann:2024mbo} of PSR J0740+6620, $R_{2.1}=12.92_{-1.13}^{2.09}$ km. From our binning process, this effectively increased the minimum maximum mass to 2.05 $M_{\rm TOV}$. As seen in Tab. \ref{tab:twinCount_nicer}, while a very large number of EOS were accepted including twin star configurations, the percentage of twin stars dropped, indicating the current value for high-mass NS mass radius data is less favorable to twin stars than the data from canonical NS. Additionally, this is the only time the number of Both EOS accepted was greater than the number of Disconnected EOS. This is likely due to Disconnected EOS reaching 2 $M_\odot$ only on the second branch, which is often more compact than the radius data from PSR J0740+6620. Run 5 again used the LIGO/VIRGO data for $R_{1.4} = 11.9 \pm 0.875$ km, but this time set the mass cut-off for the minimum $M_{\rm TOV}$ at 2.08 $M_\odot$, up from 1.97 $M_\odot$. This increased mass constraint reduced the total number of accepted EOS, but the twin percentage was only slightly lower than earlier, 13.02\% to 14.75\%. Run 6 was the most similar to the previous set of calculations replacing the radius constraint from GW170817 with that of PSR J0437+4715, $R_{1.4} = 11.36_{-0.86}^{+0.95}$ \cite{Choudhury:2024xbk}. In this case, the percentage of twin star configurations that were accepted increased, so smaller radii of canonical NS are more favorable to twin star EOS. Recently, an even smaller radius was reported by Mauviard et al. (2025) for PSR J0614+3329 with a mass of $1.44_{-0.07}^{+0.06}\;M_\odot$ and radius of $10.29_{-0.86}^{+1.01}$. Due to computational expense, we are unable to explore every combination of pulsar data, which is left for future work, but the three runs here work as a starting place to examine their individual effects.

\begin{figure*}
    \centering
    \addtolength{\tabcolsep}{-1em}
    \begin{tabular}{ccc}
        \includegraphics[width=0.35\linewidth]{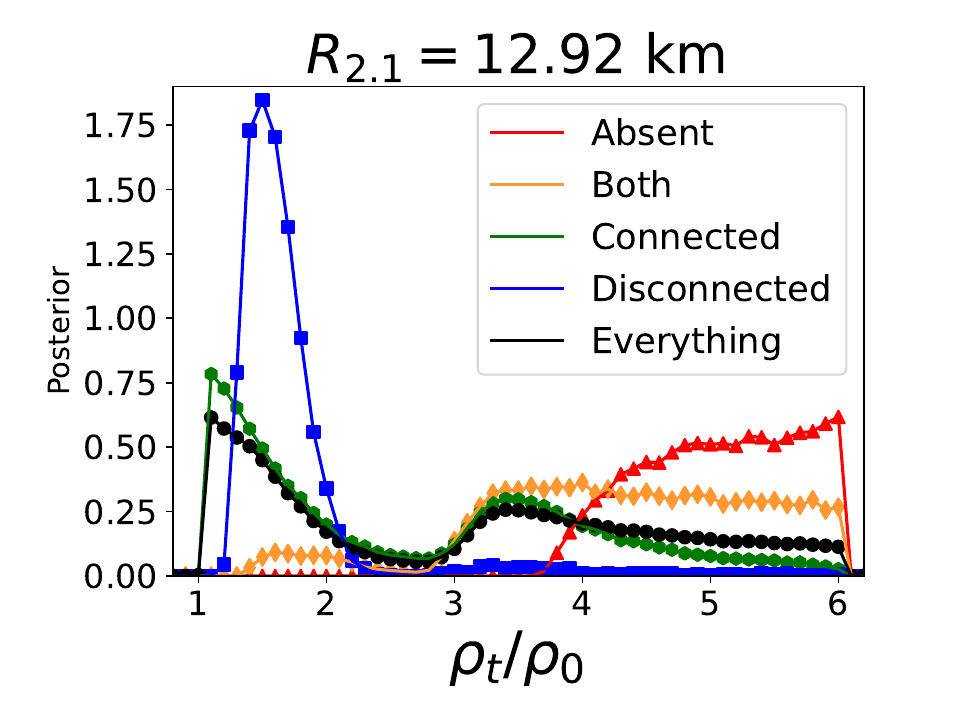} &
        \includegraphics[width=0.35\linewidth]{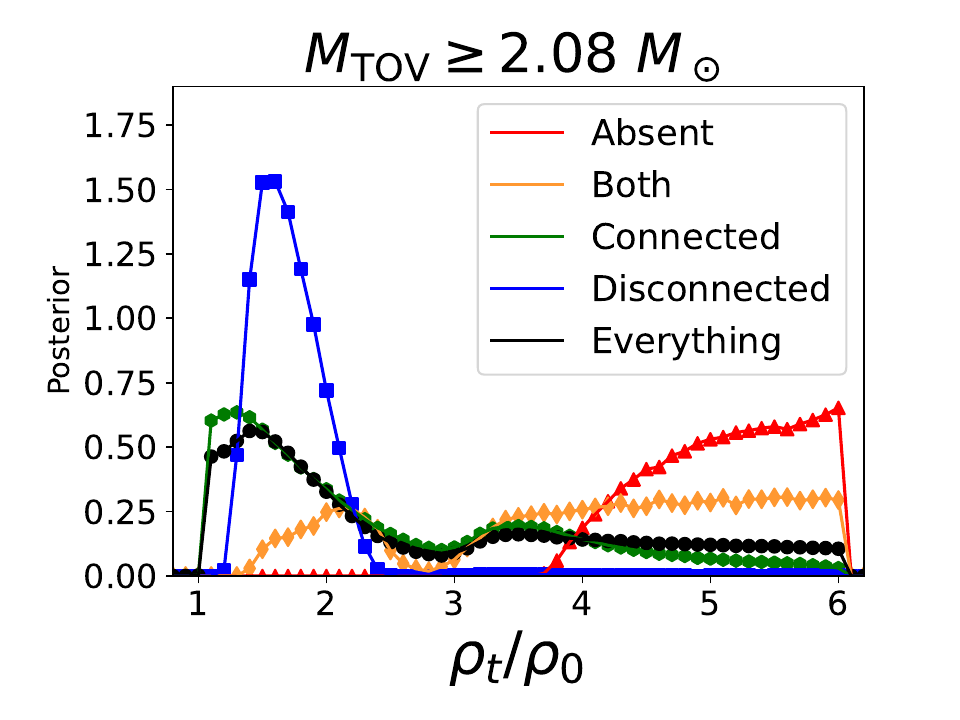} &
        \includegraphics[width=0.35\linewidth]{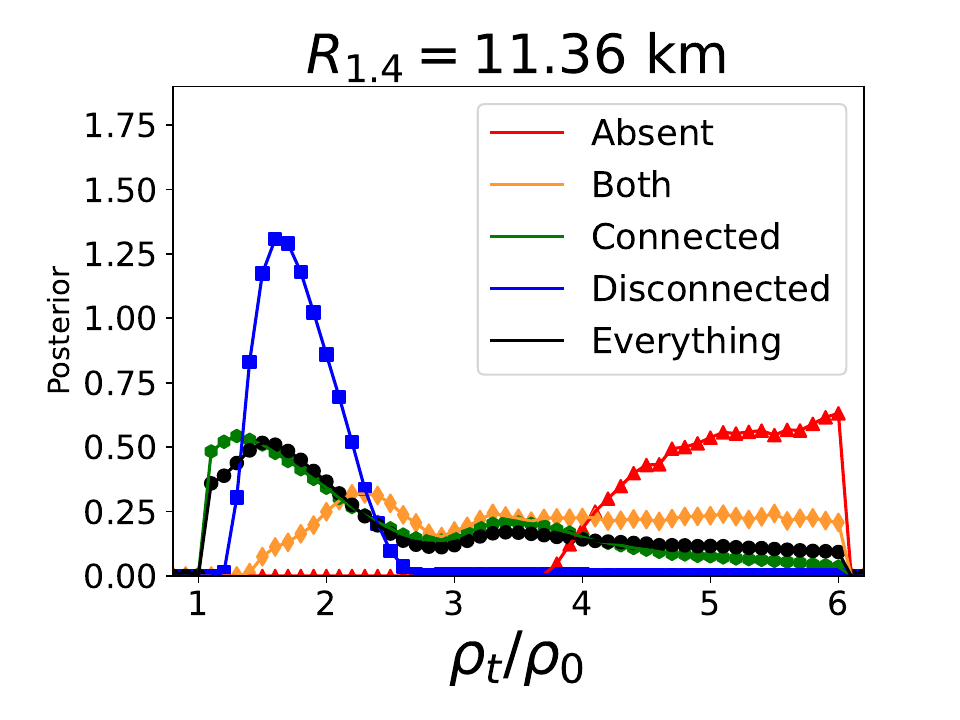} \\
        \includegraphics[width=0.35\linewidth]{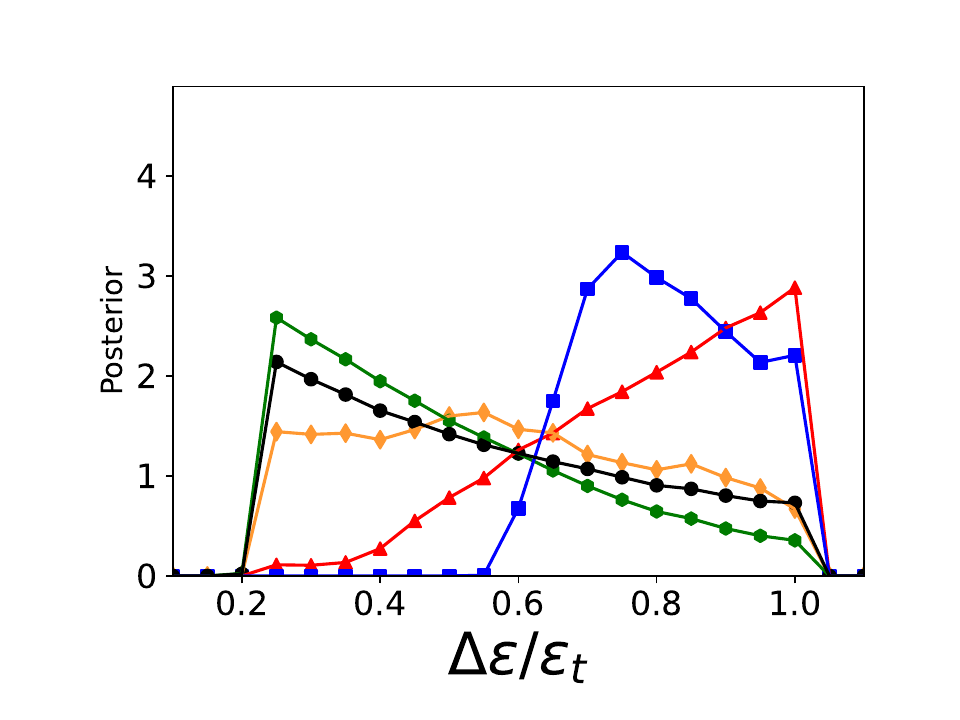} &
        \includegraphics[width=0.35\linewidth]{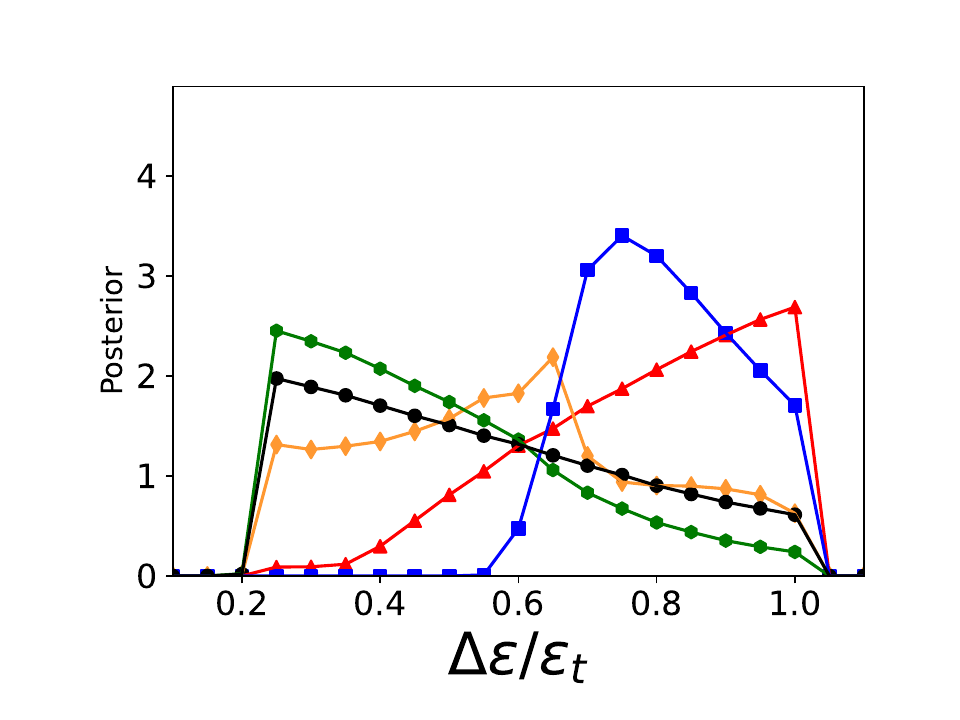} &
        \includegraphics[width=0.35\linewidth]{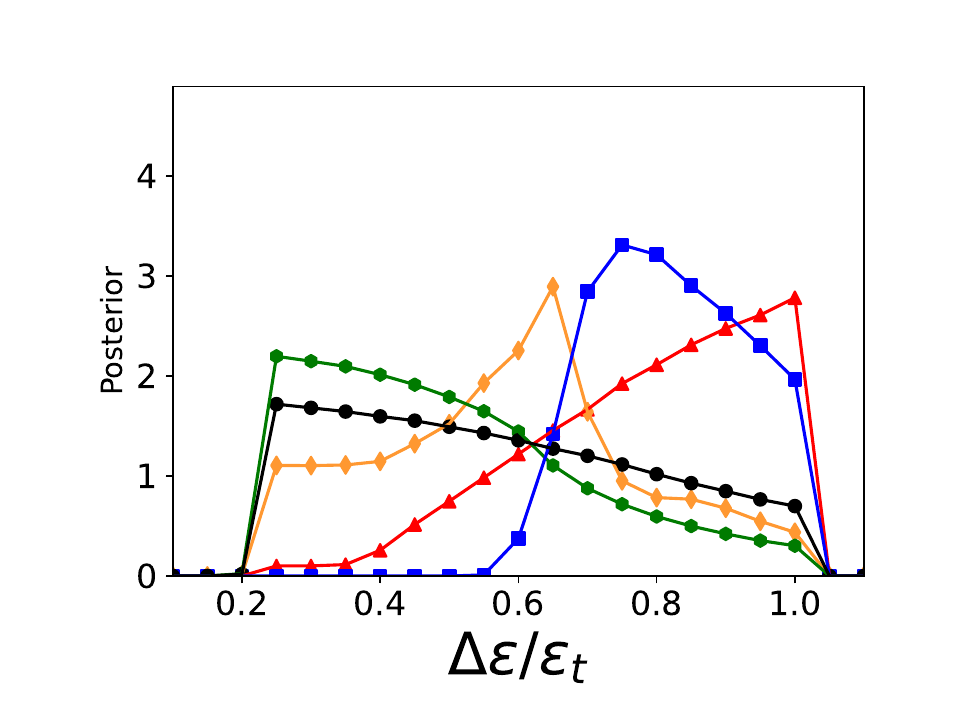} \\
        \includegraphics[width=0.35\linewidth]{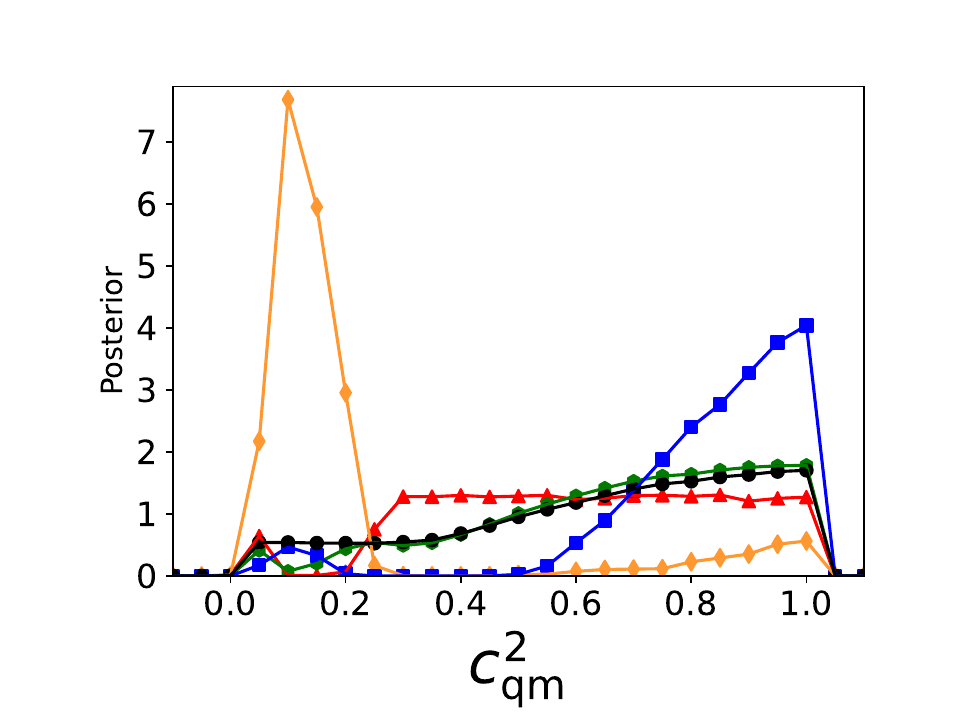} &
        \includegraphics[width=0.35\linewidth]{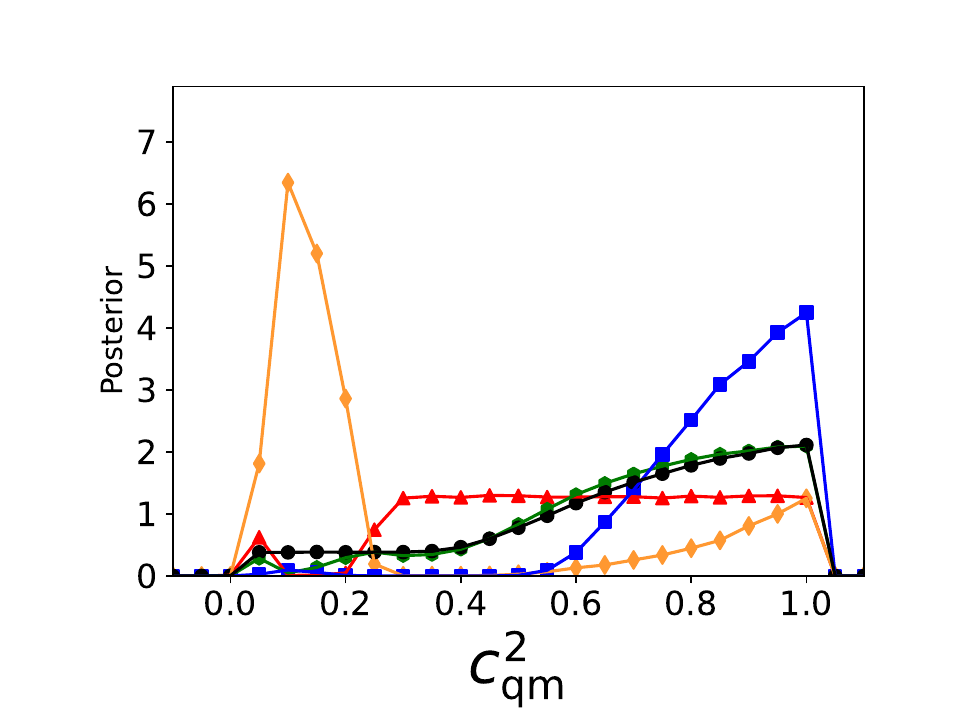} &
        \includegraphics[width=0.35\linewidth]{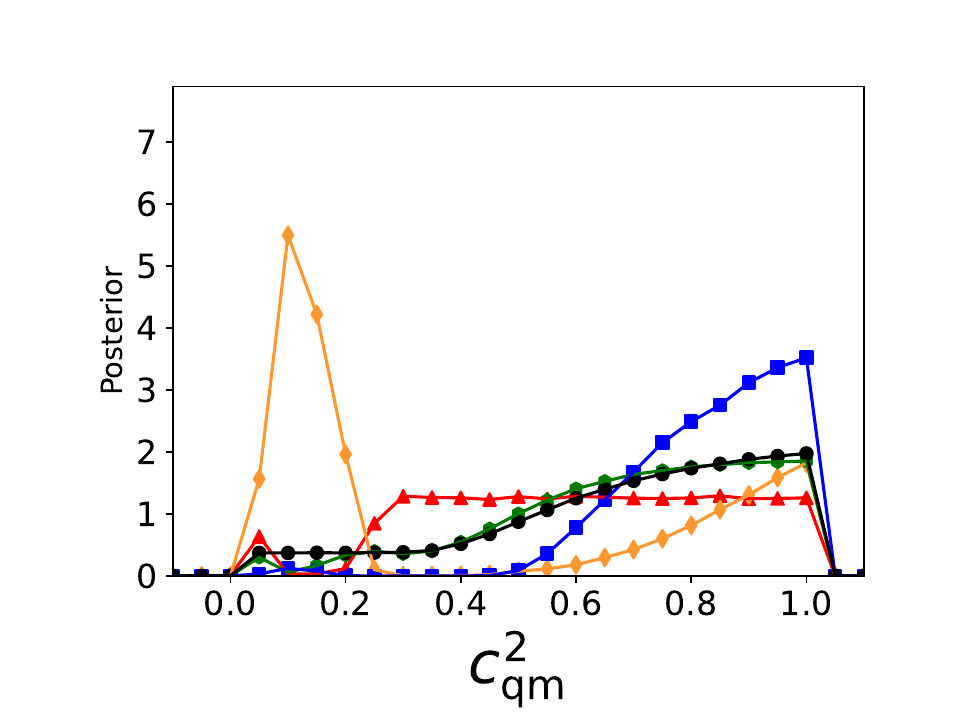}
    \end{tabular}
    \caption{Same as Fig. \ref{fig:pdfQM} but with more recent observations.}
    \label{fig:pdfQM_nicer}
\end{figure*}

\begin{figure*}
    \centering
    \addtolength{\tabcolsep}{-1em}
    \begin{tabular}{ccc}
        \includegraphics[width=0.35\linewidth]{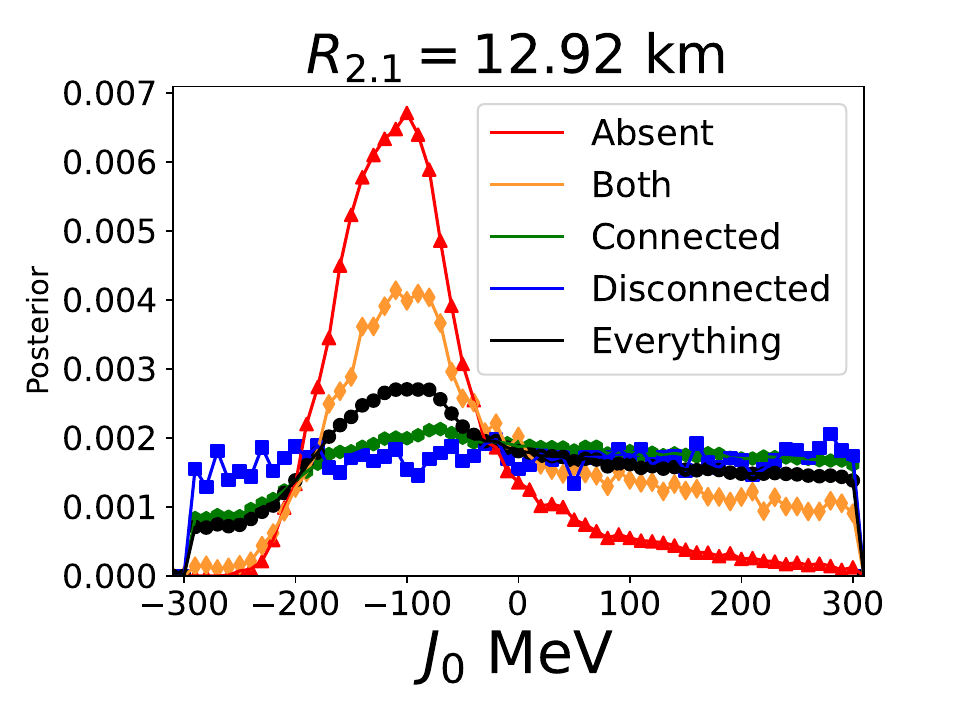} &
        \includegraphics[width=0.35\linewidth]{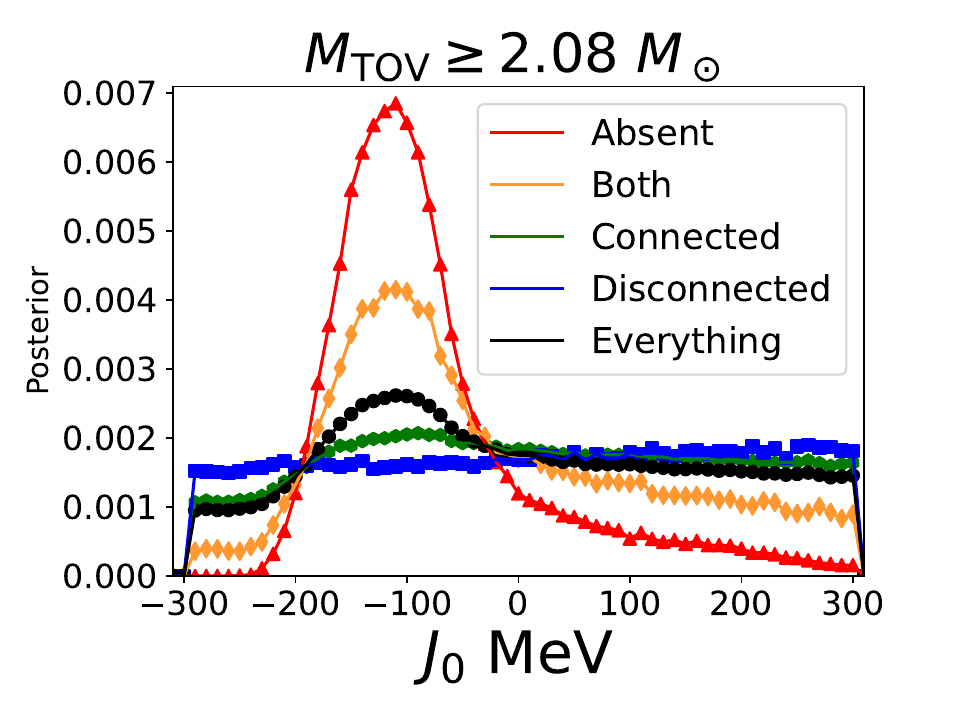} &
        \includegraphics[width=0.35\linewidth]{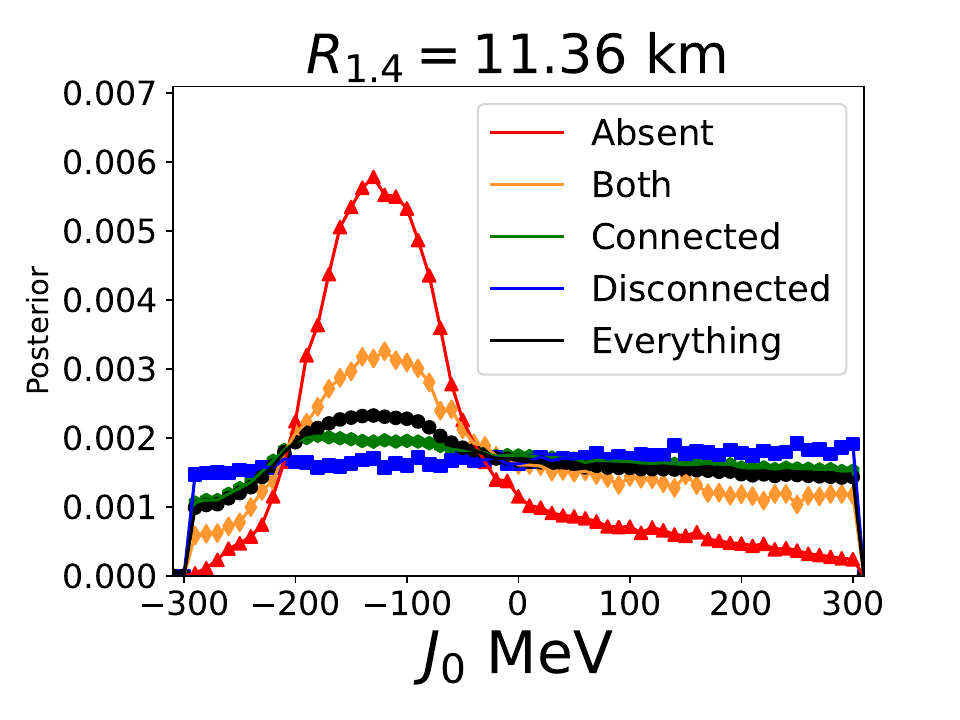} \\
        \includegraphics[width=0.35\linewidth]{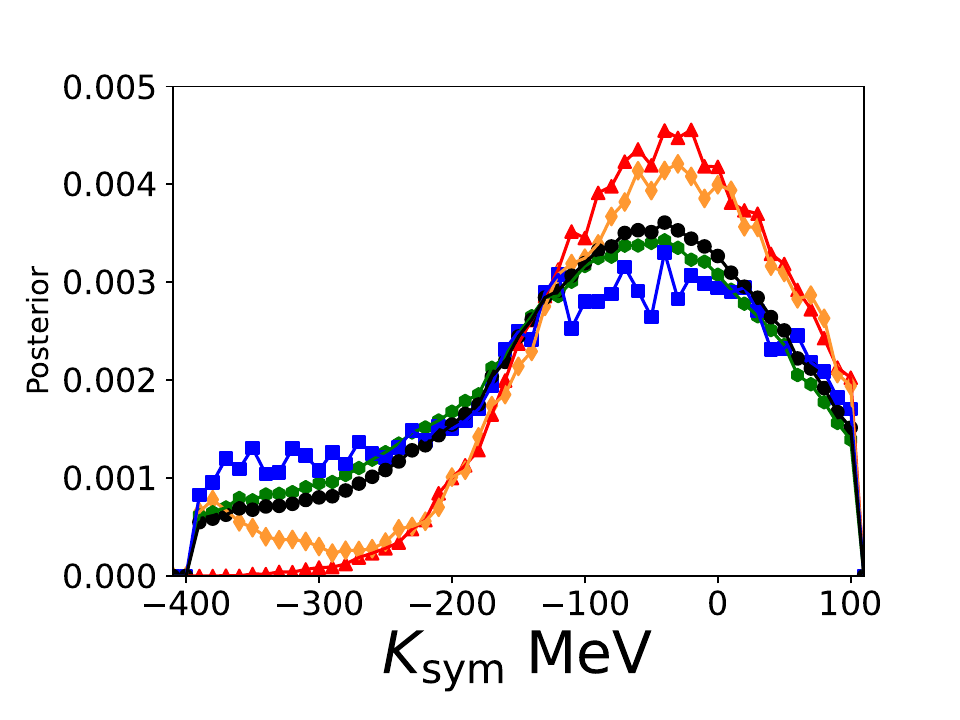} &
        \includegraphics[width=0.35\linewidth]{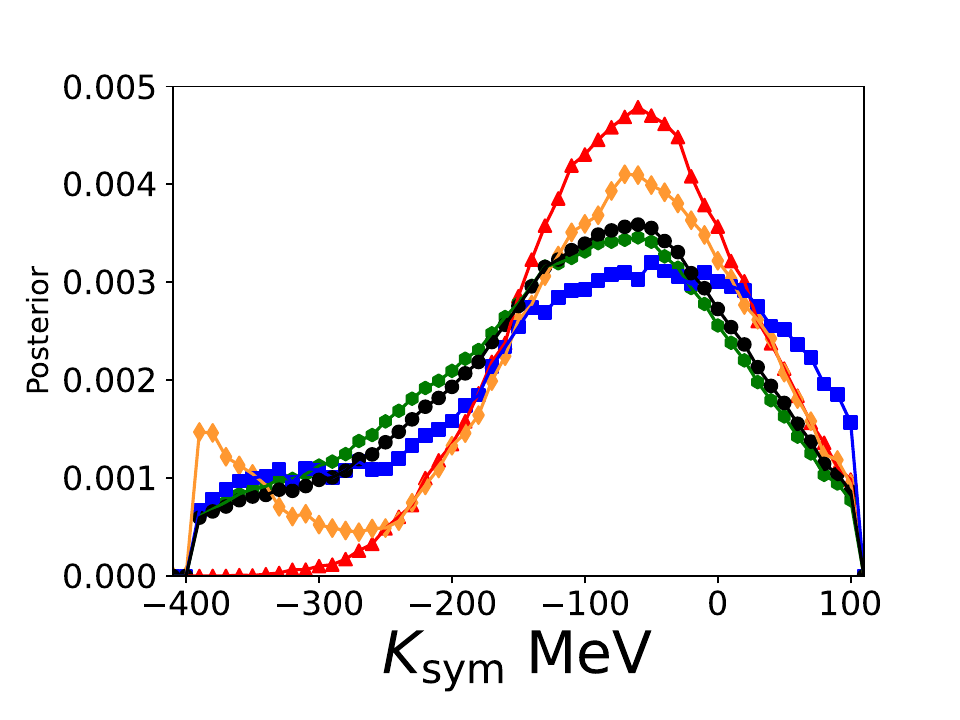} &
        \includegraphics[width=0.35\linewidth]{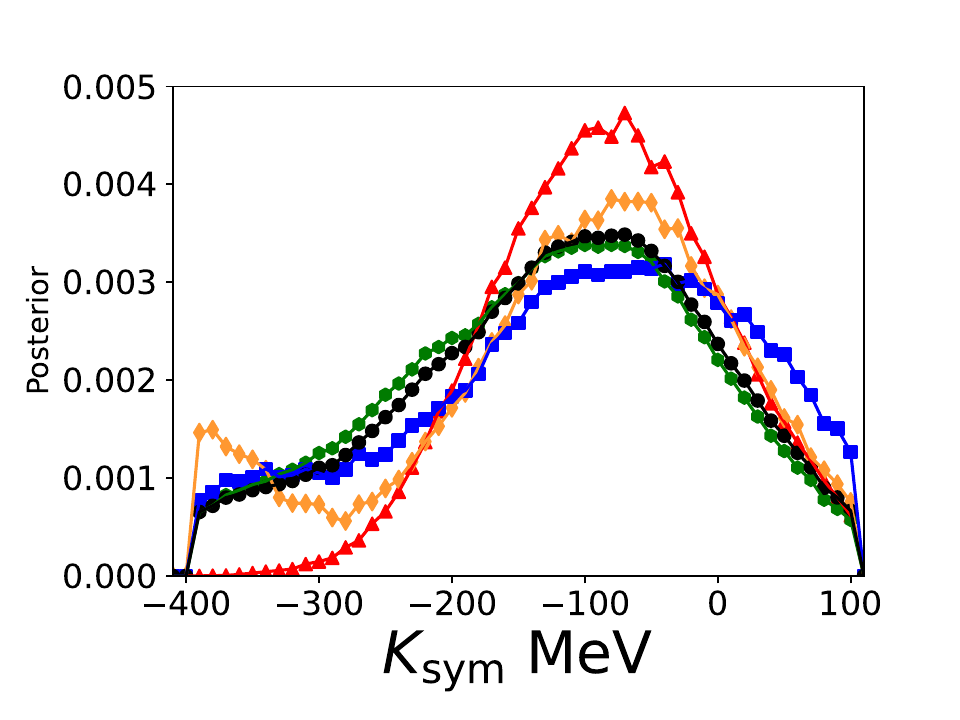} \\
        \includegraphics[width=0.35\linewidth]{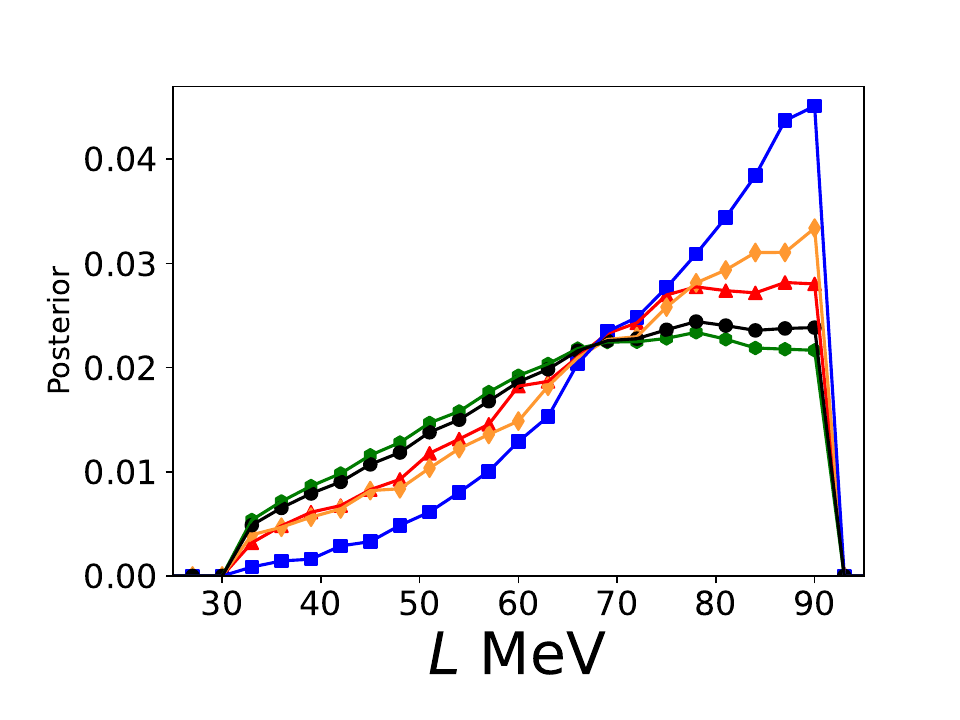} &
        \includegraphics[width=0.35\linewidth]{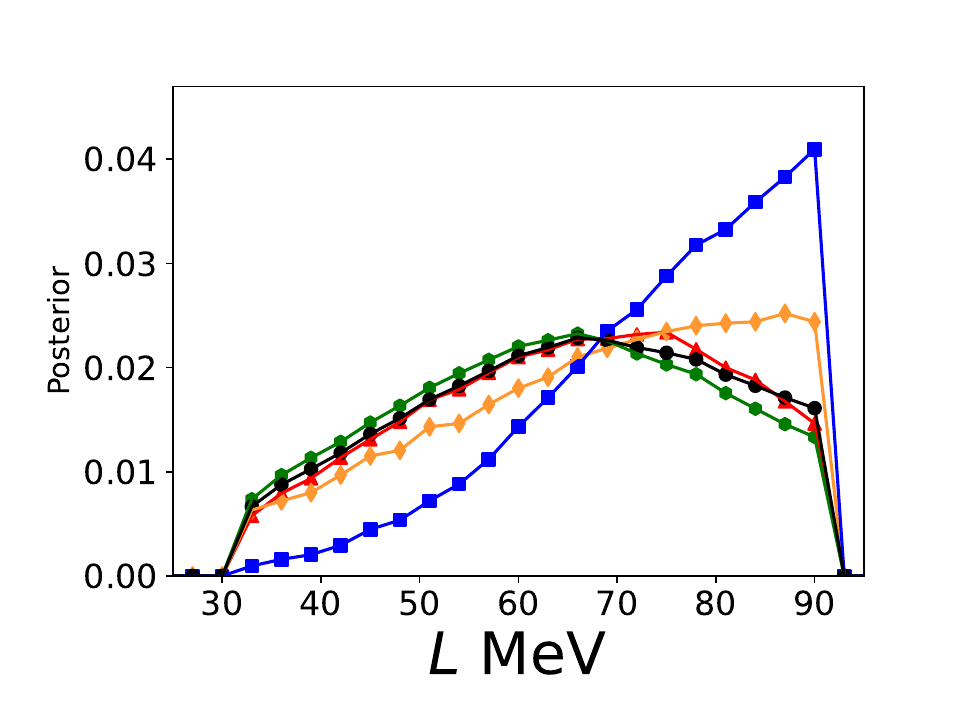} &
        \includegraphics[width=0.35\linewidth]{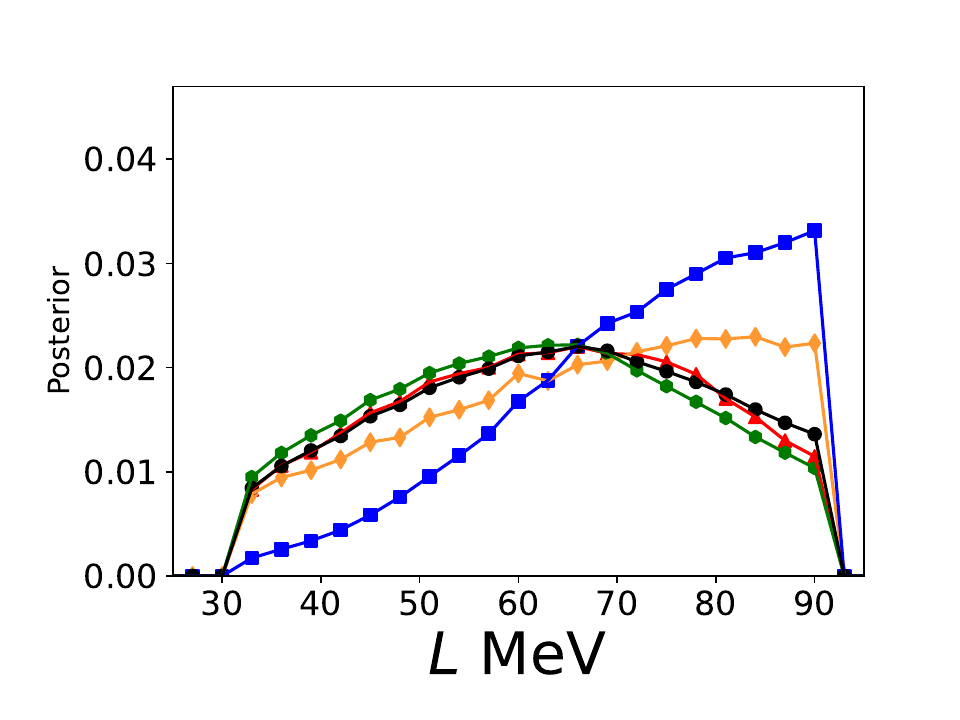}
    \end{tabular}
    \caption{Same as Fig. \ref{fig:pdfHM} but with more recent observations.}
    \label{fig:pdfHM_nicer}
\end{figure*}

\begin{figure*}
    \centering
    \addtolength{\tabcolsep}{-0.4em}
    \begin{tabular}{ccc}
        \includegraphics[width=0.33\linewidth]{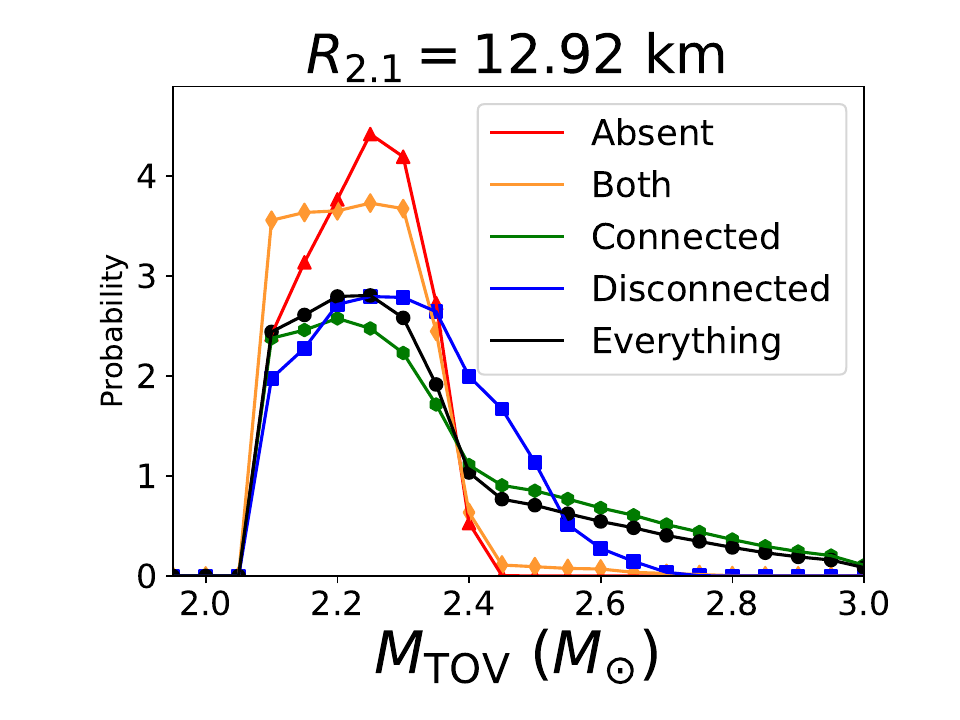} &
        \includegraphics[width=0.33\linewidth]{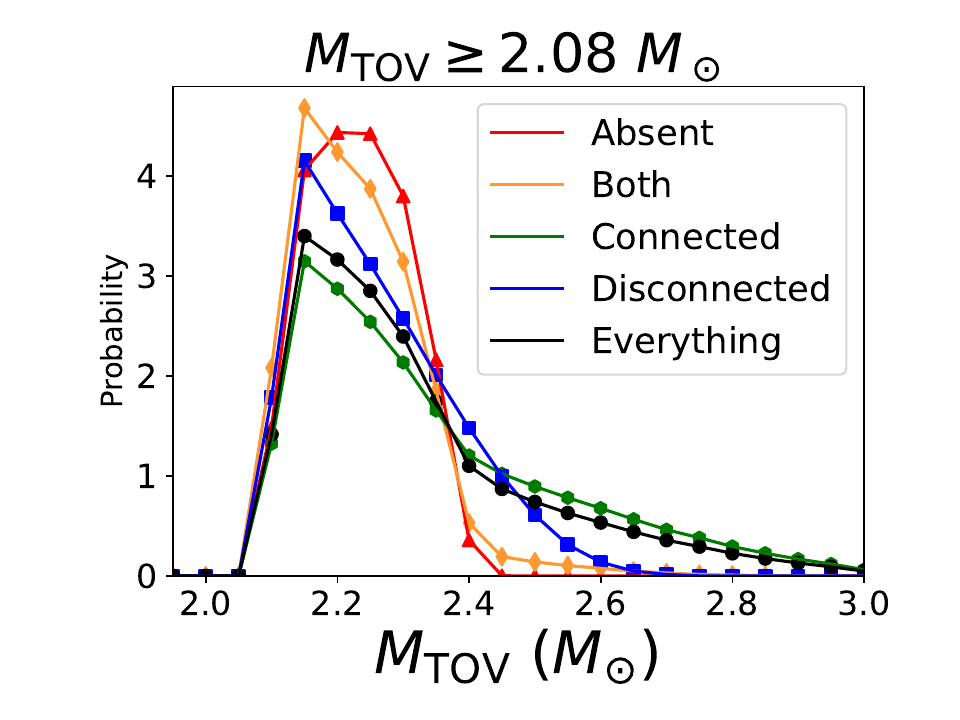} &
        \includegraphics[width=0.33\linewidth]{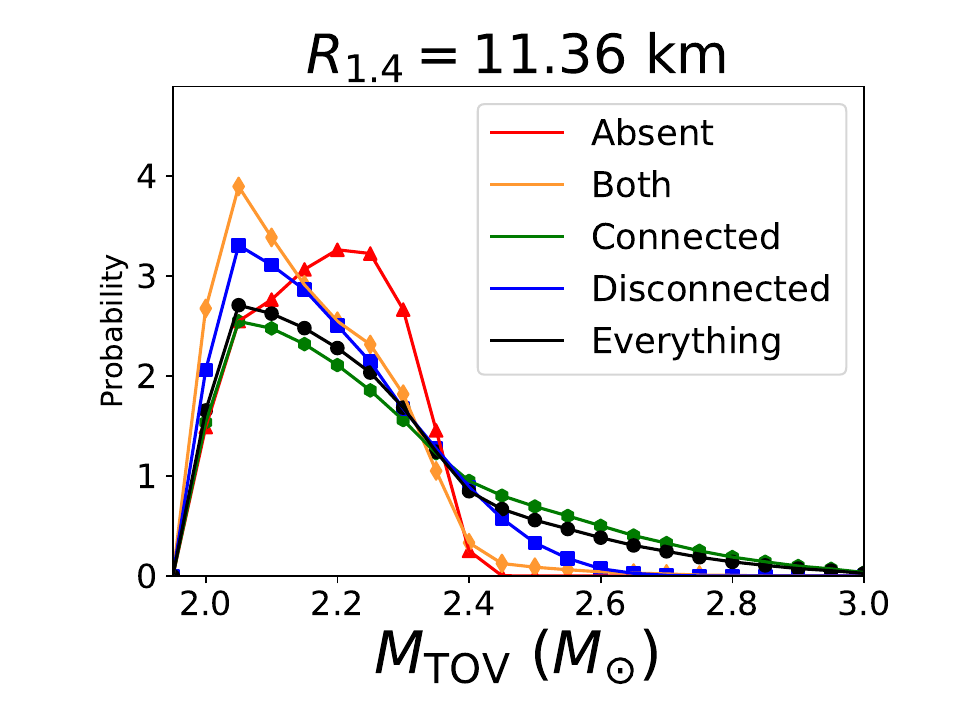} \\
        \includegraphics[width=0.33\linewidth]{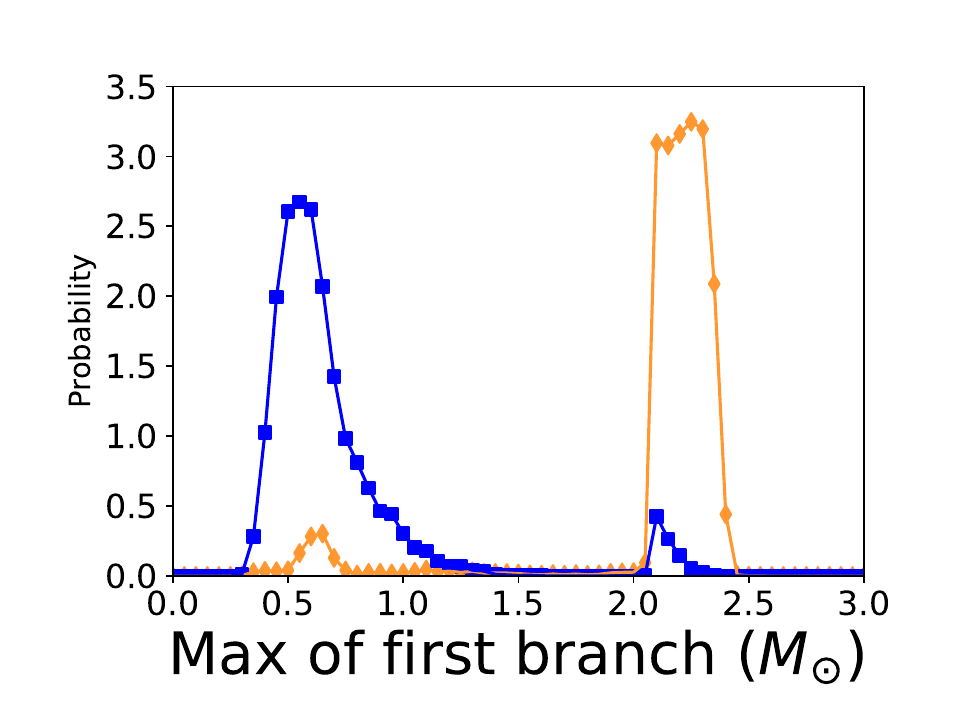} &
        \includegraphics[width=0.33\linewidth]{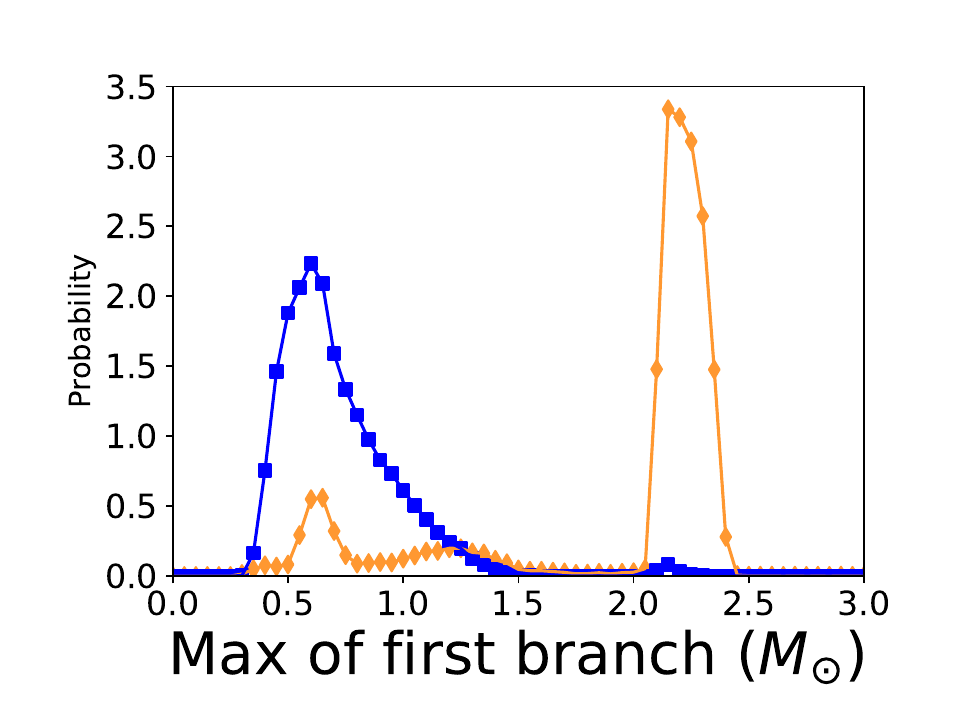} &
        \includegraphics[width=0.33\linewidth]{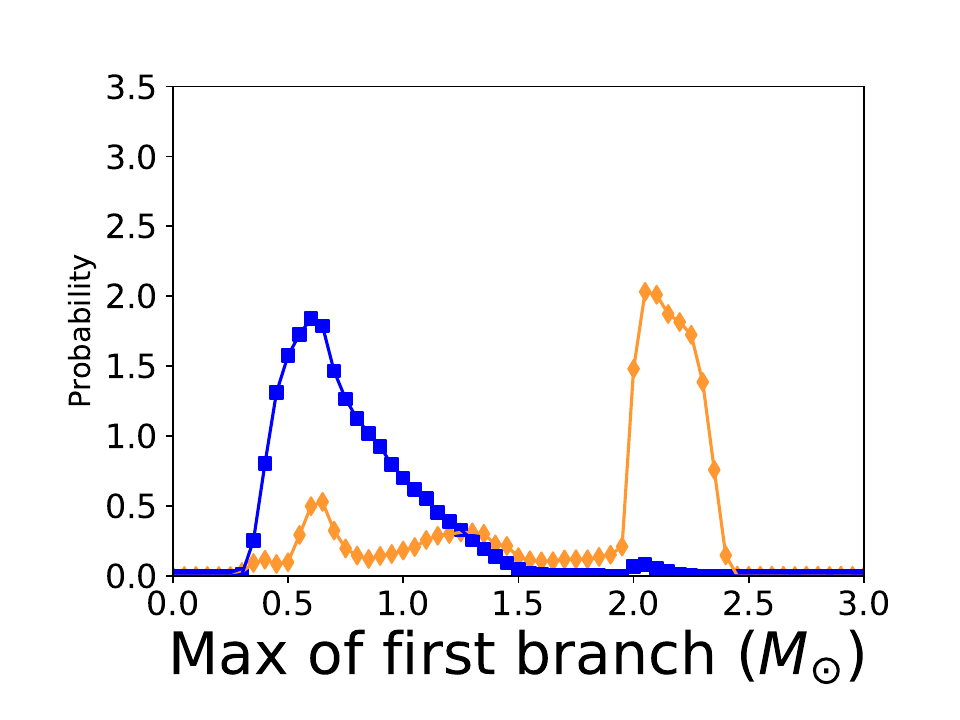} \\
        \includegraphics[width=0.33\linewidth]{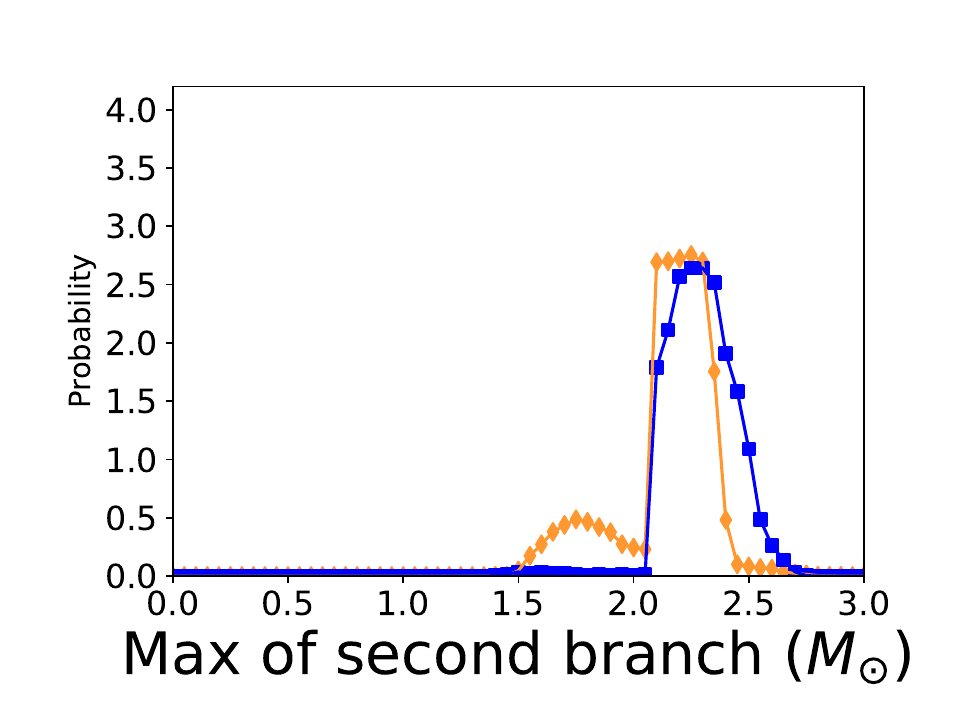} &
        \includegraphics[width=0.33\linewidth]{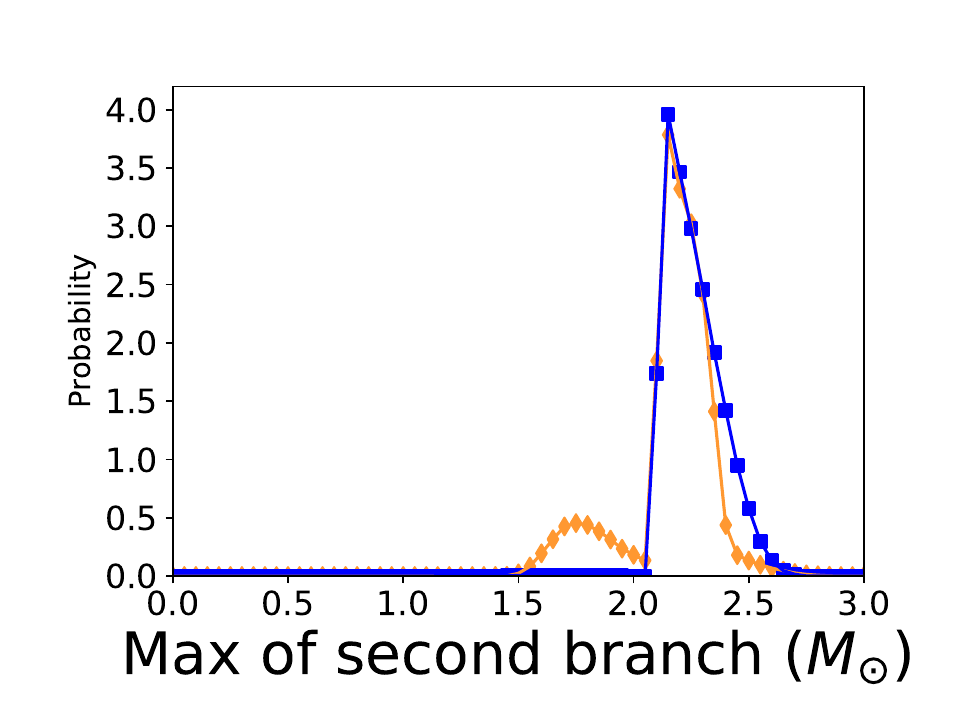} &
        \includegraphics[width=0.33\linewidth]{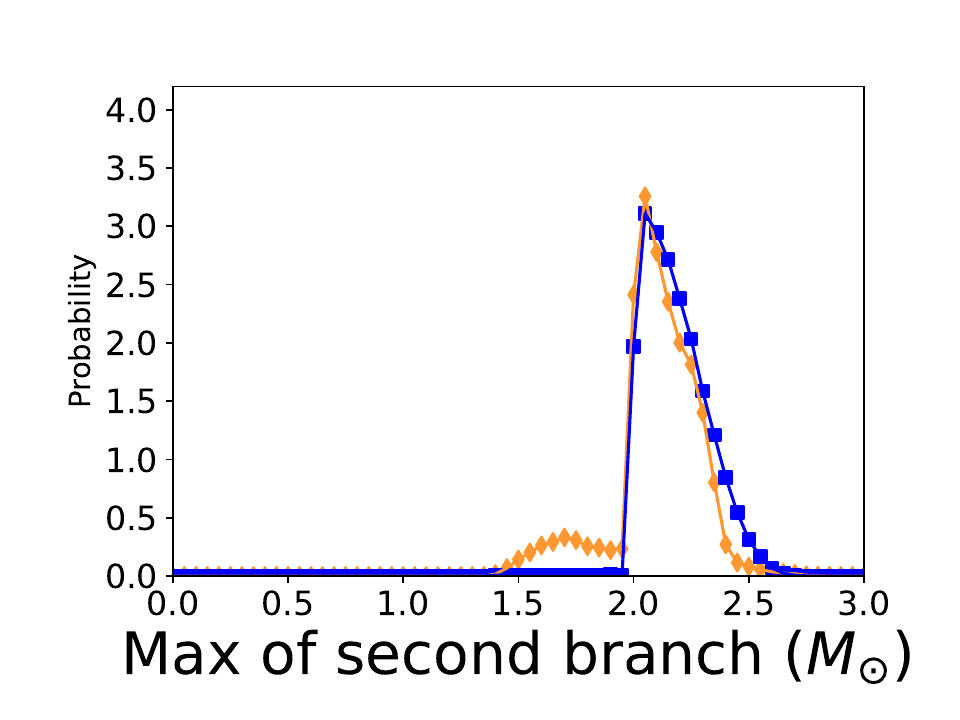}
    \end{tabular}
    \caption{Same as Fig. \ref{fig:probMass} but with more recent observations.}
    \label{fig:probMass_nicer}
\end{figure*}

The probability distributions for observables and EOS parameters are shown in Figs. \ref{fig:probObs_nicer}--\ref{fig:probMass_nicer}. For the observability of twin stars, $\Delta M$ and $\Delta R$, the pattern in the distribution is similar. Only the analysis done with $R_{2.1}$ shows some difference in the Both category, with the peak in $\Delta M$ at zero significantly reduced, and the small peak in $\Delta R = 0.5$ km flattening out. This increases the chances of EOS in the Both category being observable.

Fig. \ref{fig:pdfQM_nicer} shows that using $R_{1.4} = 11.36$ km again closely matches the earlier results, c.f. middle column of Fig. \ref{fig:pdfQM}. Using $M_{\rm TOV} \ge 2.08~M_\odot$ differs only slightly. For the analysis using the mass and radius of a high-mass NS, however, we see significant changes. Across all categories, it is now very unlikely for the phase transition to occur between 2--3 $\rho_0$. It will occur either near saturation density, or above $\sim3.5~\rho_0$. This shows a first-order phase transition around 2 $\rho_0$ is inconsistent with massive NS measurements. For the energy density discontinuity, $\Delta \varepsilon/\varepsilon_t$, this parameter actually becomes less constrained. Lastly, for the squared speed of sound in QM, the secondary peak in the PDF near 1 for the Both category has nearly disappeared. Related to this, in Fig. \ref{fig:probMass_nicer}, the maximum mass of the first branch is now, except for a very few EOS, always above 2 $M_\odot$. Previously, there were many configurations that had low transition densities that required a very stiff QM EOS in order to meet the 1.97 $M_\odot$ constraint. The mass-radius measurements of PSR J0740+6620 disfavors these configurations, so only high-mass twins exist in the Both category, which corresponds to a single peak in $c^2_{\rm qm}$ at low values.

The HM EOS (PDFs shown in Fig. \ref{fig:pdfHM_nicer}) is not much changed by the various NICER constraints. Only the slope of symmetry energy when $R_{2.1}$ is considered has much change, with a shift toward stiffer values. This is because the contribution of $L$ around 1--2 $\rho_0$ mainly affects the radius of canonical NS, whose radius is not specified in this scenario. Thus, the parameter is free to take stiffer values in order to support the more massive NS required by NICER observations. This likely explains in part the large number of accepted EOS despite a \textit{de facto} requirement of a larger minimum maximum mass.

Finally, shown in Fig. \ref{fig:probMass_nicer}, are the maximum masses of the NS, and the first and second branches. The analysis with radius data from PSR J0437+4715, as expected, is unchanged from using the LIGO/VIRGO data. For the fifth run, using a higher $M_{\rm TOV}$, the cutoff is now at 2.08 $M_\odot$, but the general pattern of which EOS category allows more massive NS is unchanged. Absent and Both categories still drop off at 2.4 $M_\odot$, while the others have a longer tail at high masses. The maximum of the first and second branches in this analysis is also little affected. The probability of the maximum mass of the first branch occurring between 1.5--2 $M_\odot$ is suppressed for Both EOS. The use of $R_{2.1}$ again shows the most change. The maximum of the first branch was already discussed---the Both category no longer has many EOS with a first branch ending before 2 $M_\odot$. For the second branch, the peak for Both and Disconnected EOS is slightly broader, making masses even above 2--2.1 $M_\odot$ likely. The maximum mass across all categories is nearly uniform between 2.1--2.4 $M_\odot$ before a sharp drop off for Absent and Both and a longer decline for the rest.

\section{Summary and Conclusions}\label{conclusion}
In this work, we have investigated the relative probability of forming twin stars given astrophysical constraints using a nine-dimensional meta-model capable of mimicking most HM and QM EOSs. By categorizing the EOSs based on their MR curve topologies, it is possible to see the most probable values of EOS parameters that yield twin star solutions. The corresponding EOS parameter subspace is significant and deserves careful study due to the important advances in nuclear physics this could lead to.

First, if we can observationally confirm the existence of twin stars, then we will be able to narrow the possible EOS to this subspace. The accuracy in radius measurements is already available to observe twin stars predicted by certain EOSs, and may be soon for most of this subspace. Of course, the existence of twin stars only indicates some new form of star, and this may not be the neutron star+hybrid star or hybrid star+hybrid star explored here. Others have explored the possibility of other NS core properties that could also create similar observations. This is an important caveat to our work, that there are other NS compositions possible, see, e.g., Refs. \cite{Grippa_2025, Kain_2021, Lopes_2025, kcl2-qgxh} for recent discussions on dark matter admixed NS or Refs. \cite{zdunik2013, Lai2018, Zhou2018, Xia_2021, Vidana:2018bdi} for discussions on strange quark stars. The scenario in this work is but one possible, although in the authors' opinion reasonable, configuration, and remains an important question for the community.

Second, with the categories' respective posteriors, future terrestrial experiments, astrophysical observations, and progress in nuclear theory may help update the prior bounds here such that certain categories' subspace of high probability is removed. Transition density is an excellent example. Investigating the QCD phase diagram, including the first-order hadron-quark phase transition boundary and its critical endpoint as functions of temperature as well as baryon and isospin chemical potentials, has long been at the forefront of both experimental and theoretical high energy physics. Indeed, significant progress has been made about the QCD phase diagram on the temperature-baryon chemical potential plane \cite{Chen24, QCD}, see, e.g., the recent BES experiments at RHIC by the STAR Collaboration \cite{bes2022_1,bes2022_2,bes2022_3,STAR:2025owm}. However, to our best knowledge, there are few broadly accepted conclusions about the isospin dependence of various boundaries and the critical end point on the QCD phase diagram, although there have been serious efforts and some stimulating results, see, e.g., Refs. \cite{Sag09,David10,Kent,Pavlo}. Given the diverse predictions about the transition density in the literature, it will be interesting to study in the future how our results may be affected by the prior range and PDF of the transition density as we did in Ref. \cite{hybridPrecision} on a different topic using a similar model. 

The most interesting, but unsurprising, result presented here is the appearance of two peaks in the PDF for $c_{\rm qm}^2$, one below while the other above the conformal limit, when considering NSs in the Both category where two hybrid stars having different densities coexist. It indicates a decreasing speed of sound squared $c_{\rm qm}^2$ with increasing density in QM toward its conformal limit.
\\

\noindent{\bf Acknowledgement:} We thank Bao-Jun Cai, Wen-Jie Xie and Nai-Bo Zhang for helpful feedback on the code and useful discussions. XG was supported in part by the Texas Space Grant Consortium. XG and BAL were supported in part by the U.S. Department of Energy, Office of Science, under Award Number DE-SC0013702.
\\

\section*{DATA AVAILABILITY} All data used in this work are publicly available \cite{dataset}.

\bibliographystyle{nst}
\bibliography{references}

\clearpage
\end{document}